\documentclass[a4paper,aps,prd,twocolumn,groupaddress]{revtex4}
\usepackage[dvips]{graphicx}
\usepackage{float}
\usepackage{amsmath}
\usepackage{amssymb}
\usepackage{enumerate}
\usepackage{verbatim}

\begin{document}

\title{Constraints on cosmic string tension imposed by the limit on the stochastic gravitational wave background from the European Pulsar Timing Array}

\author{Sotirios A. Sanidas}
\email{sotiris.sanidas@gmail.com}
\affiliation{Jodrell Bank Centre for Astrophysics, University of Manchester, Manchester, M13 9PL, United Kingdom}

\author{Richard A. Battye}
\email{rbattye@jb.man.ac.uk}
\affiliation{Jodrell Bank Centre for Astrophysics, University of Manchester, Manchester, M13 9PL, United Kingdom}

\author{Benjamin W. Stappers}
\email{Ben.Stappers@manchester.ac.uk}
\affiliation{Jodrell Bank Centre for Astrophysics, University of Manchester, Manchester, M13 9PL, United Kingdom}

\date{\today}


\begin{abstract}
We investigate the constraints that can be placed on the cosmic string tension by using the current Pulsar Timing Array (PTA) limits on the gravitational wave background. We have developed a code to compute the spectrum of gravitational waves (GWs) based on the widely accepted one-scale model. In its simplest form the one-scale model for cosmic strings allows one to vary: (i) the string tension, $G\mu/c^2$; (ii) the size of cosmic string loops relative to the horizon at birth,  $\alpha$;  (iii) the spectral index of the emission spectrum, $q$; (iv) the cut-off in the emission spectrum, $n_{*}$; and (v) the intercommutation probability, $p$. The amplitude and slope of the spectrum in the ${\rm nHz}$  frequency range is very sensitive to these unknown parameters. We have also investigated the impact of more complicated scenarios with multiple initial loop sizes $\alpha$, in particular the 2-$\alpha$ models proposed in the literature and a log-normal distribution for $\alpha$. We have computed the constraint on $G\mu/c^2$ due to the limit on a stochastic background of GWs imposed by the European Pulsar Timing Array (EPTA). Taking into account all the possible uncertainties in the parameters we find a conservative upper limit of $G\mu/c^2<5.3\times 10^{-7}$ which typically occurs when the loop production scale is close to the gravitational backreaction scale, $\alpha\approx\Gamma G\mu/c^2$. Stronger limits are possible for specific values of the parameters which typically correspond to the extremal cases $\alpha\ll \Gamma G\mu/c^2$ and  $\alpha\gg \Gamma G\mu/c^2$. This limit is less stringent than the previously published limits which are based on cusp emission, an approach which does not necessarily model all the possible uncertainties. We discuss the prospects for lowering this limit by two orders of magnitude, or even a detection of the GW background,  in the very near future in the context of the Large European Array for Pulsars (LEAP) and the Square Kilometre Array (SKA).
\end{abstract}


\maketitle


\begin{section}{INTRODUCTION}

Cosmic strings are one-dimensional topological defects \cite{vs94} which may have formed in the early Universe during various phase transitions expected in Grand Unified Theories (GUTs). Kibble \cite{kib76} first proposed and investigated the production of cosmological scale topological defects in the framework of spontaneous symmetry breaking in gauge theories. Subsequently, they attracted the interest of many cosmologists, because cosmic strings born during a GUT phase transition are a possible source for the density fluctuations which eventually led to galaxy formation \cite{vil81a,sv84,tur84,bb90b}. The initial enthusiasm diminished somewhat after the analysis of the Cosmic Background Explorer (COBE) satellite data. It was realized that the amplitude of the measured Cosmic Microwave Background (CMB) anisotropies on large-scales and the amplitude of density fluctuations measured in the galaxy distribution on smaller scales could not be reconciled in cosmic string models. This ruled them out as the primary source for the large-scale structure of the Universe (see, for example, \cite{abr97,abr99}). These results were confirmed by other experiments \cite{pog01}, but nonetheless, cosmic strings may still contribute to the anisotropy seen in the CMB temperature, but with less than $10\%$ contribution \cite{bgm06,bhku08,bm10}.

While appearing to be distinctly non-minimal in terms of structure formation - two mechanisms giving rise to similar amplitude fluctuations - such ideas are natural within a number of well-motivated inflationary models. It has been shown that cosmic strings are present in most modern inflationary scenarios, be they standard field theory strings, such as in supersymmetric hybrid inflation \cite{jea96,jrs03}, or cosmic superstring models, as in the case of brane inflation \cite{kkl+03,st02,dv04a}. The term cosmic superstrings is used to describe cosmic strings which are also fundamental strings but with their tension, or mass per unit length, $\mu$ is reduced from $\sim M_{\rm pl}^2$ where $M_{\rm pl}$ is the Planck mass, to a lower value, capable of evading constraints on the string tension, by a warp factor. See, for example, \cite{cmp04,jst03}.

Cosmic strings have a wide range of astrophysical signatures including: ultra high energy cosmic rays \cite{bs00,vac10}, gamma ray bursts \cite{bps87,bhv01}, radio bursts and synchrotron radiation \cite{vac08,cfsv86}, Aharonov-Bohm radiation \cite{cmv10,jmv10}, gravitational lensing (strong/micro) \cite{vil81c,gmtm+08,ksv08}, CMB imprints (non-gaussianity, small/large-scale anisotropies, B-mode polarization) \cite{ks84,tps98,bat98,bhku07,frsb08, pw08,tnsy+09,gdf+11}, effects on matter power spectra in $21$-cm surveys \cite{bdhh10,kw08}. So far no detection has been possible, but an interesting opportunity lies in their imprint on the stochastic gravitational wave background (SGWB) \cite{vil81d,hr84,vv85,bat86,ac89,bb90a,ca92,cbs96}. The SGWB created by a cosmic string network has a very broad spectrum with frequencies ranging from below the nHz scale to beyond the GHz scale, making them a potential source for every present or future GW detection experiment. Such broad GW spectra are expected from only some primordial sources, such as inflation \cite{all88,skc06}, global phase transitions \cite{jkm08} and self-ordering scalar fields \cite{ffdg09}.

A SGWB created by cosmic strings is a possible primordial source for detection by Pulsar Timing Arrays (PTAs) operating in the nHz frequency range. A PTA \cite{det79,fb90} consists of an ensemble of millisecond pulsars (MSPs) that are observed periodically over an extended period of time, usually a number of years, which sets the frequency probed with maximum sensitivity by the array. The existence of a SGWB changes the Earth-MSP distance and therefore it manifests itself as noise in the time of arrival of pulses. These timing irregularities will have a specific signature which allows us to distinguish them  from other types of noise \cite{hd83}. The usage of MSPs is necessary since this category of  neutron stars combines a series of characteristics which makes them very stable ``clocks'', capable of providing high quality timing measurements. In the case of a SGWB generated by cosmic strings, the quantity we can constrain is their linear energy density (or tension in the Nambu-Goto approximation), $\mu$, usually expressed through the dimensionless quantity $G\mu/c^2$, where $G$ is Newton's constant and $c$ the speed of light. The value of $\mu\sim \eta^2$ in natural units, where $\hbar=c=1$, is typically related to  the energy scale of the phase transition, $\eta$, at which they are formed.

The precise details of the SGWB due to cosmic strings are very sensitive to the nature of the string evolution and the spectrum of the radiation emitted by the string loops that are created due to the intercommutation of the long strings. In particular, the spectrum in the band probed by PTAs depends on the distribution of loops formed and the amount of radiation emitted into high frequency modes due to loops formed in the matter era. This can be quantified in terms of a spectral index and cut-off in the spectrum \cite{vs94}. Since these details are not well understood, this has led to some confusion in the literature with a number of inconsistent constraints being published, some of them based on the same data \cite{dv00,dv01,dv05,hog06,scm+06,jhv+06,smc07,dh07,oms10,vlj+11}. This is due to the estimates of the amplitude of the SWGB being based on different assumptions; some of which we believe are too strong for our current level understanding of string evolution. In this paper we will use the limit on the SGWB imposed by the European Pulsar Timing Array \cite{vlj+11} (EPTA) to compute constraints on the cosmic string tension as a function of the parameters which describe the details of the string evolution and radiation. Taking into account all possible uncertainties, we define the most reliable, as opposed to the tightest, constraint as that which corresponds to the set of parameters with the highest upper bound. By parameterizing our ignorance of string evolution we should be able to impose a constraint on $G\mu/c^2$ which will never be violated.

The structure of the paper is as follows. In Sec.~II we give a description of the one-scale model of cosmic strings which we used to construct the GW spectra and we present a model for the GW emission mechanism from cosmic string loops. In Sec.~III we present the results describing the effect of the various cosmic string model parameters on the GW power spectrum. The results of multiple scale models will also be presented there, in which we extend the applicability of the one-scale model to more realistic possibilities. Finally, in Sec.~IV we present robust limits on the cosmic string tension using the recently published European Pulsar Timing Array limit on the SGWB. We conclude with a discussion of the results in Sec.~V.

\end{section}


\begin{section}{Modelling the spectrum of gravitational wave emission from a cosmic string network}\label{sec:onescale}

Our calculation of the SGWB expected from a cosmic string network is based on the one-scale model \cite{kib85,ben86a,ben86b}. Caldwell and Allen \cite{ca92} (hereafter, CA92), and DePies and Hogan \cite{dh07} (hereafter, DH07) have previously considered this model and our implementation is a combination of both their approaches (see also, \cite{bbcd12} for a recent investigation). Before going into the details of the one-scale model we will give a very brief picture of a cosmic string network.

The basic constituents of a cosmic string network are loops and ``infinite'' (or long) strings; loops so large that only a part of them lies within our horizon radius, appearing as extremely long strings with no ends. These ``infinite" cosmic strings stretch along with the expansion of the Universe and oscillate at relativistic speeds. A fundamental dynamical process which impacts the evolution of the cosmic string network is intercommutation, whereby strings (self-)intersect, exchange partners and form new loops \cite{she87}. Cosmic string loops have significant tension, equal to their linear energy density in the Nambu-Goto approximation, so after their formation they start to oscillate relativistically and decay by emitting their energy into the ``preferred channel'' which in the case of local strings is thought to be GWs \cite{vil81d,vv85}. An alternative picture is suggested by Abelian-Higgs simulations \cite{vhs97,vah98} (see also, \cite{hin11} for a recent discussion), where we have the creation of microscopic loops from the string network which immediately decay via gauge boson emission. In that case, the emission of GWs from the cosmic string network is significantly suppressed and the dominant energy loss mechanism is field quanta emission, implying that the constraints derived here would be invalid. This point remains as a caveat to our analysis.

This mechanism for loop creation and their subsequent decay is important since otherwise the cosmic strings would dominate the energy density of the Universe rather quickly. Instead the loop production allows the network to achieve a scaling regime \cite{kib85} with all the properties of the network being related to  the horizon radius $\propto t$, known as the one-scale model.

The important features of the one-scale model are:

\begin{enumerate}[i.]
 \item The evolution of the cosmic string network is considered to take place in a homogeneous, flat FRW Universe and all the lengthscales of the network are linear multiples of the particle horizon radius. The particle horizon radius in such a Universe is
     \begin{equation}
     d_{\rm H}(t)=a(t)c\int_0^t {dt^{\prime}\over a(t^{\prime})}\,,
     \label{horizonradius}
     \end{equation}
     where $a(t)$ is the scale factor. The network evolution is considered to take place in a co-moving volume of size $V(t)=a^3(t)D^3$, where $D$ is an arbitrary length scale which will cancel out from our final equations.
     \item The energy density of the infinite cosmic strings is
     \begin{equation}
     \rho_{\infty}(t)=\frac{A \mu c^2}{d_{\rm H}^2(t)}\,,
     \label{rhoinfinite}
     \end{equation}
     where $A$ quantifies the number of infinite strings present in a horizon volume. The value of $A$ was determined in numerical simulations \cite{bb89a} which suggest a value of $A\approx 52$ for the radiation-dominated era and $A\approx 31$ for the matter-dominated era (more recent simulations \cite{bos11} give values within $10\%$ of these ). The $\Lambda$-dominated epoch is believed to be only a very small fraction of the cosmic history, so we used the same value for $A$ as in the matter-dominated era presuming that our results will not be radically different in the case of the SGWB calculation.
 \item Every cosmic string loop is born with a size $\ell_{\rm b}(t_{\rm b})$ which is a constant fraction of the horizon radius at the birth time $t_{\rm{b}}$
     \begin{equation}
     \ell_{\rm{b}}(t_{\rm{b}})=\alpha d_{\rm{H}}(t_{\rm{b}})\,.
     \label{one_scale}
     \end{equation}
     A correction has to be applied in Eq.~\eqref{one_scale} because cosmic string loops are born with relativistic peculiar velocities which are very quickly reduced due to the expansion of the Universe, resulting in an energy loss from the loops almost immediately after their birth. By taking into account this redshifting of their initial velocity we can write the initial length (energy) of a newborn cosmic string loop as $\ell_{\rm{b}}(t_{\rm{b}})=f_{\rm{r}}\alpha d_{\rm H}(t_{\rm{b}})$, where $f_{\rm{r}}$ incorporates this energy reduction and is $f_{\rm{r}}\approx0.71$ \cite{bb89b}.
\end{enumerate}


\begin{subsection}{Number density $n(\ell,t)$ of cosmic string loops}
\label{sec:nlt}

In the one-scale model the ``infinite" string network needs to lose sufficient energy so as to maintain the scaling regime which allows us to calculate $n(\ell,t)d\ell$, the number density of cosmic string loops with lengths between $\ell$ and $\ell +d\ell$ at time $t$ after the creation of the cosmic string network. This quantity gives us a full ``map'' of the number and size of cosmic string loops present throughout the cosmic history.
In order to calculate the number density of cosmic string loops we need information about their birth rate. Allen and Caldwell \cite{ac91a,ca92} calculated in detail the formation rate of cosmic string loops from the conservation of the cosmic string stress-energy tensor. Briefly, they did this by first calculating the equation of energy conservation of the cosmic string network
\begin{equation}
\frac{d}{dt}\left[a^3(t)(\rho_{\rm{tot}}(t)+p_{\rm{tot}}(t))\right]=\dot{p}_{\rm{tot}}(t)a^3(t)\,,
\end{equation}
where $\rho_{\rm{tot}}$ and $p_{\rm{tot}}$ are, respectively, the energy density and pressure of the whole system (infinite cosmic strings, cosmic string loops and GWs emitted). Combining this with the equation of state of infinite strings
\begin{equation}
p_{\infty}(t)=\frac{1}{3}\rho_{\rm{\infty}}(t)\left[2\langle\upsilon^2\rangle/c^2-1\right]\,,
\end{equation}
where $\langle\upsilon^2\rangle$ is the mean squared velocity of infinite cosmic strings, they derived the amount of energy lost by the network to create new cosmic string loops per unit time.
\begin{equation}
\frac{dE_{\rm{loop,cr}}}{dt}=-V(t)\left[\dot{\rho}_{\infty}(t)+2\frac{\dot{a}(t)}{a(t)}\rho_{\infty}(t)\left(1+\langle\upsilon^2\rangle/c^2\right)\right]\,.
\label{dEdt}
\end{equation}
The mean squared velocity of infinite strings is also determined by numerical simulations \cite{bb89a} and is $\langle\upsilon^2\rangle/c^2=0.43$ for the radiation-dominated era and $\langle\upsilon^2\rangle/c^2=0.37$ for the matter-dominated era. The latter value will also be used in our calculations for the $\Lambda$-dominated era. The most recent evolution simulations \cite{bos11} suggest similar values. In the one-scale model, since we know the size of the newborn loops, we can write
\begin{equation}
\frac{dE_{\rm{loop,cr}}}{dt}=\mu\alpha d_{\rm H}(t)c^2\frac{dN_{\rm{loop}}}{dt}\,,
\label{dE-dN}
\end{equation}
where $N_{\rm{loop}}$ is the total number of loops created since the creation of the network within the volume $V(t)$ and $dN_{\rm{loop}}/dt$ is the corresponding formation rate. Combining Eqs.~\eqref{dEdt},\eqref{dE-dN} we get
\begin{align}
\frac{dN_{\rm{loop}}}{dt}&=-\frac{V(t)}{\mu \alpha d_{\rm{H}}(t)c^2}\nonumber\\
&\times\left[\dot{\rho}_{\infty}(t)+2\frac{\dot{a}(t)}{a(t)}\rho_{\infty}(t)\left(1+\langle\upsilon^2\rangle/c^2\right)\right]\,,
\label{dNdt1}
\end{align}
and by using Eq.~\eqref{rhoinfinite} we can bring it to the simpler form
\begin{equation}
\frac{dN_{\rm{loop}}}{dt}=\frac{2V(t)\rho_{\infty}(t)}{\mu\alpha d_{\rm{H}}(t)c^2}\left[\frac{c}{d_{\rm{H}}(t)}-\frac{\dot{a}(t)\langle\upsilon^2\rangle}{a(t)c^2}\right]\,.
\label{dNdt2}
\end{equation}
Knowing the formation rate, we can calculate the number of loops born at any instant in cosmic history. Note, that this rate was actually calculated from the energy lost by the network in order to maintain scaling, which means that our results are automatically normalized. This is not always the case in other methods.

Each cosmic string loop decays by GW emission with a constant rate
\begin{equation}
\frac{dE_{\rm{loop,em}}}{dt}=-\Gamma G\mu^2c\,,
\end{equation}
where $E_{\rm{loop,em}}$ is the energy emitted by a cosmic string loop in the form of GWs and $\Gamma$ is a constant which describes the efficiency of the emission mechanism. The value of $\Gamma$ depends on the shape of the cosmic string loops. Throughout this work, a value of $\Gamma\approx 50$ will be used which has been calculated as the average for representative loops through numerical simulations \cite{ca95}. Analytic calculations of $\Gamma$ have also been performed \cite{ac94,aco94} but only for specific cases. Assuming that GW emission is the dominant energy loss mechanism of cosmic string loops, the length $\ell(t,t_{\rm b})$ of a cosmic string loop at time $t$, if this loop was born at time $t_{\rm{b}}$, can be written as
\begin{equation}
\ell(t,t_{\rm{b}})=f_{\rm{r}}\alpha d_{\rm{H}}(t_{\rm{b}})-\frac{\Gamma G\mu}{c}(t-t_{\rm{b}})\,.
\label{cslooplength}
\end{equation}
This relation for the variation length of a cosmic string loop with the cosmic time and its birth time can be interpreted in an inverse way; a cosmic string loop which has a length equal to $\ell$ at time $t$, has to have been born at a specific time $t_{\rm b}$ given by the solution of Eq.~\eqref{cslooplength}, and only at that time.

We will express the $n(\ell,t)$ function as a discrete two dimensional array, with cosmic time and cosmic string loop length as the axes. Each element $n(\ell_{\rm{i}},t_{\rm{j}})$ of this array we will be calculated in the following way. First, we calculate the corresponding birth time, $t_{\rm{b,j}}$, of these loops from Eq.~\eqref{cslooplength}. We define $\mathcal{N}(\ell ,t)d\ell$ to be the number of loops with length between $\ell$ and $\ell +d\ell$ present in our volume at time $t$. Of course, $\mathcal{N}(\ell_{\rm{i}} ,t_{\rm{j}})=\mathcal{N}(\ell_{\rm{b,j}},t_{\rm{b,j}})$. For the number of loops born at time $t_{\rm{b,j}}$ in our simulation volume we will have
\begin{equation}
\left.\mathcal{N}(\ell,t)d\ell\right\vert_{t=t_{\rm{b,j}},\ell=\ell_{\rm{b,j}}}=\left. dN_{\rm{loop}}\right\vert_{t=t_{\rm{b,j}}}\,.
\label{dNdt3}
\end{equation}
Since $d\ell=(f_{\rm r}\alpha\dot{d}_{\rm H}(t_b)-\Gamma G\mu/c)dt$, by substituting in Eq.~\eqref{dNdt3} and dividing by our simulation volume $V(t_{\rm j})$ we find that
\begin{equation}
n(\ell_{\rm{i}},t_{\rm{j}})=\frac{1}{V(t_{\rm{j}})\left[f_{\rm{r}}\alpha\dot{d}_{\rm{H}}(t_{\rm{b,j}})+\Gamma G\mu/c\right]} \left. \frac{dN_{\rm{loop}}}{dt}\right\vert_{t=t_{\rm{b,j}}}\,.
\label{nlt}
\end{equation}
This allows us to calculate all the elements of the $n(\ell,t)$ array and, by using multi-variate interpolation, we can construct the $n(\ell,t)$ function. Of course, we do not have to calculate the elements for which $\ell_{\rm{i}}>f_{\rm{r}}\alpha d_{\rm{H}}(t_{\rm{j}})$ which are equal to zero. The veracity of our results was tested against the analytic  formulas  for the loop number density in the radiation and matter dominated eras \cite{bb89b} for a range of values of $\alpha$ and $\Gamma G\mu$. The more recent and detailed analytic formulas found in \cite{lrs10} yield expressions which have to be fitted to evolution simulation results and do not have the flexibility of those by Bennett and Bouchet.

\end{subsection}


\begin{subsection}{Gravitational wave emission from string loops\label{sec:emissionmodel}}

A cosmic string loop oscillates relativistically under its tension and emits GWs in a series of harmonics with frequencies which depend only on the length of the loop, $\ell$, and the harmonic mode $n$. The period of the loop is $\ell/2c$ and the particular frequencies are harmonics of $\ell/2$, that is
\begin{equation}
f_n=\frac{2nc}{\ell}\,.
\label{emfreq}
\end{equation}
Given a particular string trajectory, in the Nambu-Goto approximation and ignoring the effects of radiation backreaction, one can compute the total radiated power from a loop, $P=\Gamma G\mu^2c$, which is independent of $\ell$. The value of $\Gamma$ depends on the specific trajectory, but as already noted it is $\approx 50$ for typical loop trajectories \cite{ca95}. The power emitted into each harmonic mode is given by
\begin{equation}
\frac{dE_{\rm{gw,loop}}}{dt}=P_nG\mu^2c\,,
\end{equation}
where
\begin{equation}
P_n=\Gamma n^{-q}/\sum_{m=1}^{\infty}m^{-q}\,,
\label{eq::powerco}
\end{equation}
is a coefficient for each mode which determines the amount of radiated energy that is emitted through the respective mode and $q$ is the spectral index. The value of $q$ can be computed for a specific trajectory: $q=4/3$ for the Kibble-Turok loops and $q=2$ for a square loop with kinks \cite{vs94}.  It can be argued that $q\approx 4/3$ for any string loop which has a cusp; something which is expected for string trajectories without a kink.

The effects of radiation backreaction are ignored in this calculation. The decay of the loop length can be described by the linear decay with time already discussed in the previous section. However, the precise details of the spectrum can have significant impact on the amplitude and slope of the SGWB in the nHz region that  is relevant to PTAs and this can also be affected by backreaction. It has been shown in full field theory simulations that the spectrum of Goldstone boson radiation from global strings is significantly softened by the effects of radiation backreaction  \cite{bs94}. In particular, it was shown that an initial $q=1$ spectrum was dominated by the fundamental mode after a small number of oscillations. The equivalent simulations are not possible in the context of gravitational radiation, but there are sufficient similarities between the radiative mechanisms to suggest that something similar may also take place in this case as well.

In order to model this effect we include an extra phenomenological parameter, $n_{*}$, as was done in \cite{cbs96}: $P_n=0$ for $n>n_{*}$ and the normalization factor is modified for $n\le n_{*}$ so that
\begin{equation}
P_n=\Gamma n^{-q}/\sum_{m=1}^{n_*}m^{-q}\,,
\label{eq::powerco1}
\end{equation}
and the total power emitted is unchanged. The value of $n_*$ is unknown, but it must be less than the ratio of the loop length to the string core width, $\delta$. For macroscopic strings, $\ell/\delta$ is typically very large, but the rounding of the cusps and kinks expected in a realistic network might mean that $n_* \approx R/\delta$ where $R$ is the local radius of curvature which could be much less, say in the range $\log_{10}n_*=3$ to 5. In what follows we will allow $n_*$ and $q$ to be free parameters which we will vary.

\subsection{Stochastic gravitational wave background}

The standard quantity used to quantify the amplitude of the SGWB is energy density in GWs per logarithmic frequency interval measured relative to the critical density, $\rho_{\rm crit}$, which is given by
\begin{equation}
\Omega_{\rm{gw}}(f)=\frac{1}{\rho_{\rm crit}}\frac{d\rho_{\rm{gw}}}{d\log f}\,.
\end{equation}
This can be related to the dimensionless strain of the GW by
\begin{equation}
h_{\rm gw}(f)=1.3\times 10^{-9}\sqrt{\Omega_{\rm gw}(f)h^2}\left({1\,{\rm nHz}\over f}\right)\,.
\end{equation}
Since $\rho_{\rm crit}=3H_0^2/8\pi G$ depends on the Hubble Constant $H_0=100h\,\rm{km\,s^{-1}\, Mpc^{-1}}$, it is conventional to plot the dimensionless quantity $\Omega_{\rm{gw}}(f)h^2$.

The starting point for our calculation is based on the formula that gives the spectral density of the emitted GWs derived in \cite{vs94},
\begin{align}
\frac{d\rho_{\rm{gw}}}{df}(t)&=2\pi \int_{t_{\rm f}}^t dt^{\prime}\left(\frac{a(t^{\prime})}{a(t)}\right)^3 \nonumber \\
&\times \int^{f_r\alpha d_{\rm H}(t^{\prime})}_{0}\ell d\ell n(\ell,t^{\prime})g\left(\frac{a(t_0)}{a(t^{\prime})}\frac{2\pi}{c} f\ell\right)\,,
\label{drhodt1}
\end{align}
where $f$ is the frequency of the GWs as we observe them today,  $t_0\approx 13.4\,{\rm Gyr}$ is the present time, $t_{\rm f}=t_{\rm pl}c^4/(G\mu)^2$ is the time of formation of the cosmic string network \cite{ca92} with $t_{\rm pl}$ the Planck time and $g(z)$ a function which describes the spectrum of radiation emitted by a loop and is normalized by $\int_{0}^{\infty}g(z)dz=\Gamma G\mu^2c$. We note that in \cite{vs94} the integral is written in terms of angular frequency $\omega =2\pi f$ and $g(z)$ is normalized by $\Gamma$ with the factor of $G\mu^2c$ included in the equivalent of Eq.~\eqref{drhodt1}.

We will model a discrete emission spectrum and therefore $g(z)$ will be a sum of $\delta$-functions given by
\begin{equation}
g(z)=G\mu^2 c\sum_{j=1}^{n_{*}}P_j\delta(z-4\pi j)\,.
\label{gz}
\end{equation}
If we set $z=(a(t_0)/a(t^{\prime}))(2\pi f\ell/c)$ then
\begin{equation}
zdz=4\pi^2\left(\frac{a(t_0)}{a(t^{\prime})}\right)^2\frac{f^2\ell}{c^2} d\ell\,,
\label{zdz}
\end{equation}
and hence substituting Eq.~\eqref{zdz} and Eq.~\eqref{gz}  into Eq.~\eqref{drhodt1} we find that
\begin{align}
\frac{d\rho_{\rm{gw}}}{df}&=\frac{2G\mu^2c^3}{f^2}\int_{t_{\rm f}}^{t_0}dt^{\prime}\nonumber\\
&\times\left(\frac{a(t^{\prime})}{a(t_0)}\right)^5\sum_{j=1}^{n_{*}}jP_jn\left(\frac{a(t^{\prime})}{a(t_0)}\frac{2jc}{f},t^{\prime}\right)\,.
\label{spectraldensity}
\end{align}

The integral of Eq.~\eqref{spectraldensity} requires the continuous calculation of the argument of $n(\ell,t)$ during its numerical evaluation. We can decouple these two calculations, something which also gives a better physical intuition into what Eq.~\eqref{spectraldensity} represents. A similar approach was also followed in DH07, but note that some quantities are expressed differently in our implementation. As we have already mentioned, the sum in Eq.~\eqref{spectraldensity} gives the contribution to the GW spectrum of each emission mode from each cosmic string loop. The quantity $a(t^{\prime})c(a(t_0)f)^{-1}$ in the number density is actually the length of the cosmic string loops which emit GWs at time $t^{\prime}$ which are observed at the present day with redshifted frequency $f$. This loop population is the only one we are interested in when we have to evaluate Eq.~\eqref{spectraldensity} in a specific frequency bin. It is reasonable then to express the integral in terms of a new function, say $n(f,t)$, which gives the number density of loops which at time $t$ emit GWs observed today with frequency $f$. In this case, $\Omega_{\rm{gw}}(f)$ can be written as
\begin{equation}
\Omega_{\rm{gw}}(f)=\frac{2G\mu^2c^3}{\rho_{\rm crit} a^5(t_0) f^2}\sum_{j=1}^{n_{*}}jP_j\int_{t_{\rm f}}^{t_0}a^5(t^{\prime})n_{j}(f,t^{\prime})dt^{\prime}\,.
\label{specint}
\end{equation}
Of course, the $n_j(f,t)$ function has to be constructed for every emission mode, $j$.

The $n_j(f,t)$ function can be constructed in a straightforward way once we have obtained $n(\ell,t)$. Using the same array we constructed for $n(\ell,t)$, we change the cosmic string loop length axis with the corresponding frequency axis by converting all lengths to frequencies using  Eq.~\eqref{emfreq}. Multi-variate interpolation with this new axis will give the $n_j(f_{\rm{in.}},t)$ function which gives the number density of loops which at time $t$ emit GWs at frequency $f_{\rm{in.}}$. We will use this $n_j(f_{\rm in}, t)$ to construct $n_j(f,t)$. We can calculate the elements $n_j(f_{\rm m},t_{\rm n})$ of the $n_j(f,t)$ data array in the following way. First, we calculate the frequency $f_{\rm m, in.}$ which the GW had when they were emitted, $f_{{\rm m, in.}}=(a(t_0)/a(t_{\rm n}))f_{\rm m}$. Once we have this information, we use the previously constructed $n(f_{\rm in.},t)$ function and calculate the respective loop number density. In this way we populate the $n(f,t)$ data array and we create the corresponding interpolating function.

A typical GW spectrum from cosmic strings is presented in Fig.~\ref{figspectrum}. It has two distinct features: a low frequency peak and a flat spectrum at higher frequencies. The flat part of the spectrum is created from GWs emitted by cosmic string loops which were created during the radiation era. Roughly, it is flat from $\gtrsim 1\,{\rm nHz}$ to around the $\rm{GHz}$ frequency range and then it drops rapidly in amplitude.  The low frequency peak, which dominates below $\sim 1\,{\rm nHz}$, is created from the more recent emission of cosmic string loops formed during the matter era. Interestingly the cross-over region between the two regimes is exactly that which is probed by the PTAs. As we will see, the point at which this cross-over actually takes place is sensitive to the parameters describing the string network and the radiation spectrum.
\begin{figure}[H]
\includegraphics[width=8.6cm]{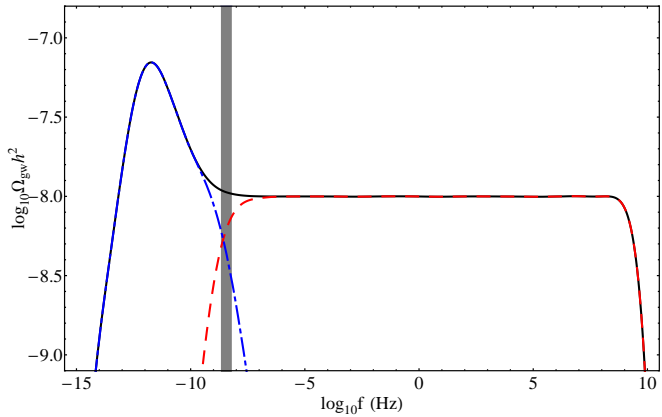}
\caption{The GW energy density per logarithmic frequency interval $\Omega_{\rm{gw}}(f)h^2$ of a cosmic string network with $G\mu/c^2=10^{-7}$, $\alpha=10^{-3}$ and $n_{*}=1$. The black (solid) line is the full spectrum from the network due to loops formed in both radiation and matter eras, whereas the red (dashed) line is that from the radiation-dominated era and the blue (dot-dashed) line is from the matter-dominated era. The grey shaded area shows the frequency window probed with the highest sensitivity by PTA experiments with duration between 5 and 10 years.\label{figspectrum}}
\end{figure}

\end{subsection}


\begin{subsection}{Intercommutation probability}

Whenever two field theory cosmic strings collide they exchange partners with an intercommutation probability $p=1$ \cite{she87}. This is not necessarily the case for cosmic superstrings however, which intercommute with a reduced intercommutation probability $p<1$. This can be attributed to the extra dimensions in which cosmic superstrings are moving, with a successful intercommutation requiring their collision in all dimensions and not just in the three spatial dimensions visible to us. If $p<1$ then the scaling density of long strings is increased in order to increase the number of intersections per unit time and hence allow the network to lose the requisite amount of energy necessary to maintain scaling. This will increase the number of loops and hence will increase the amplitude of the SGWB by a uniform scaling. There is, however, some controversy as to the exact dependence on $p$. Jones, Stoica and Tye \cite{jst03}, argued that the self-similar length scale, $L$, of the cosmic string network should scale as $L\propto pt$, which would mean that $\rho_{\infty}\propto L^{-2}\propto p^{-2}$. In that case, even a small decrease in $p$ would lead to a dramatic increase in the amplitude of the SGWB. However, in such a case the inter-string distance $d_{\rm s}$, due to the higher string density, is smaller than the length scale of the network $L$, whereas in the one-scale model $L\sim d_{\rm s}$, suggesting that this argument needs to be modified.

Sakellariadou \cite{sak05} has performed simulations of cosmic superstring networks in Minkowski spacetime which suggest that  $L\propto p^{1/2} t$, implying that $\rho_{\infty}\propto p^{-1}$. It was suggested the discrepancy with the results of Jones et al. stems from the small-scale structure of cosmic stings, which ensures more intersection points when two strings collide, and therefore there are more chances for successful loop production.

There are two techniques used to model the dynamics of strings in the Nambu-Goto approximation: one is the Minkowski spacetime approach used in \cite{sak05}; the other is to model the expansion of the Universe. The results of such simulations are reported by Avgoustidis and Shellard in \cite{as05,as06}. They find that when $p\leq0.1$ then $\rho_{\infty}\propto p^{-0.6}$, whereas for $0.1<p\leq 1.0$ they find $\rho_{\infty}\propto p^{-1}$. They also suggest that small-scale structure is responsible for the difference from the $\rho_{\infty}\propto p^{-2}$ scaling law and they propose a simple two-scale model which describes quite accurately their simulation results. The difference in the scaling laws of \cite{sak05} and \cite{as06} has to do with fitting model parameters to results of fundamentally different simulations, so the exact reasons for this discrepancy are not easy to trace.

In this work we will not make a judgement on the precise dependence of the scaling density of infinite strings as a function of $p$ except that it can be modeled by a power law
\begin{equation}
A(p)={A(1)\over p^{k}}\,,
\label{A(P)}
\end{equation}
where $k$ is the model parameter and  $A(1)=52$ and $A(1)=31$ in the radiation and matter eras respectively.  The results of \cite{sak05} suggest that $k=1$, whereas those of \cite{as05,as06} suggest $k=0.6$ for $p\leq 0.1$ and $k=1$ for $0.1<p\leq 1.0$. The consequence of this assumption is that the amplitude of the SWGB will scale as $\Omega_{\rm gw}(f)\propto p^{-k}$ independent of $f$.
\end{subsection}

\end{section}


\begin{section}{CHARACTERISTICS OF COSMIC STRING INDUCED SPECTRA}


\begin{subsection}{Low frequency cut-off due to newborn large loops.}
\label{sec:lowfreqcutoff}
\begin{figure}[ht]
\includegraphics[width=8.6cm]{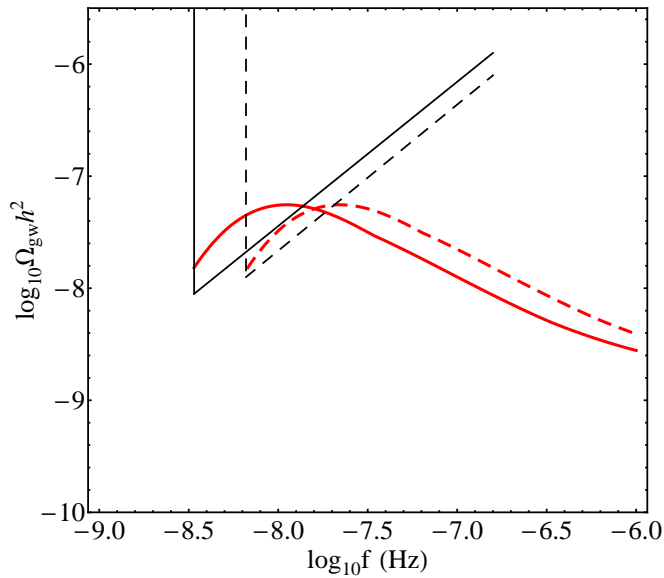}
\caption{The GW sensitivity curves for a 10-year (black thick line) and a 5-year (black dashed line) PTA experiment, with the 10-year experiment achieving slightly better maximum sensitivity. The frequencies where these experiments achieve maximum sensitivity are $3.2\,{\rm nHz}$ and $6.3\,{\rm nHz}$ respectively. The red thick line is the GW spectrum of a cosmic string network for $\alpha_1=5.7\times10^{-10}$ and the red dashed line is the spectrum for $\alpha_2=2.8\times10^{-10}$ network. While the 10-yr experiment has a greater overall sensitivity at its minimum frequency, it has a lower sensitivity at the frequencies to which the 5-year experiment is sensitive to (see text for details).\label{fig::ptasens}}
\end{figure}

As we mentioned in Sec.~\ref{sec:emissionmodel}, each cosmic string loop emits GWs into an ensemble of harmonics defined by $f_n=2nc/\ell$. This means that there is a low frequency cut-off on the GWs that a cosmic string network emits, defined by the first emission mode of the largest loops present. The largest loops are those created at the present time $t_0$ and have length $\ell_0=f_{\rm r}\alpha d_{\rm H}(t_0)$, with a corresponding low frequency cut-off $f_0\propto 1/\alpha t_0$. The redshifted frequencies of the GWs emitted by loops previously born will always be higher than $f_0$ in both the radiation- and matter-dominated eras. For example, in the radiation era the frequency of the first emission mode of a loop formed at time $t_1$ redshifted to the present is $f_{1}\propto t_{\rm eq}^{1/6}/\alpha(t_1^{1/2}t_0^{2/3})>f_0$, where $t_{\rm eq}\approx 25,000\,{\rm yrs}$ is the time of radiation-matter equality. The same calculation in the matter era gives $f_{1}\propto 1/\alpha(t_1^{1/3}t_0^{2/3})$, which is also greater than $f_0$. To demonstrate the strength of this inequality, in the matter era, the GWs of the first emission mode of a loop with $\alpha=0.1$ emitted at the time of its birth, say $t_1=10^{10}{\rm\, Gyrs}$, will be observed today at a frequency $\approx2.7\times10^{-17}{\rm\, Hz}$, whereas the corresponding GWs of a loop born at the present time will have a frequency $\approx 1.8\times10^{-17}{\rm\, Hz}$. Similarly, the GWs of a similar loop born in the radiation era at time, say $t_1=10^4{\rm \,yrs}$ will have at the present time a frequency $\approx4.3\times10^{-11}{\rm ,Hz}$.

This low frequency cut-off needs to be treated very carefully because of its strong dependence on $\alpha$, which is unknown. For small, but totally acceptable values of $\alpha$ (such as, $\alpha\sim 10^{-12}-10^{-16}$), the low frequency cut-off can be as high as a few microhertz. Although, ground-based observatories and LISA are not seriously affected by this low frequency limit since they are sensitive to frequencies $\sim100\,\rm Hz$ and $\sim10^{-3}\,{\rm Hz}$ respectively, PTAs can easily be rendered useless for detecting emission from such cosmic string networks. As we will show in the following paragraphs, PTAs can adequately probe the GW emission of cosmic string networks with $\alpha\lesssim10^{-9}$.

\begin{figure}[]
\includegraphics[width=7.5cm]{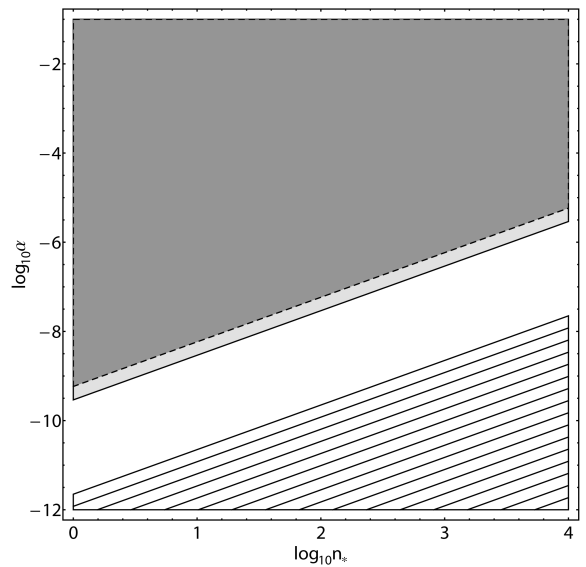}
\caption{Regions of the $\alpha -n$ parameter space which can be probed by PTA experiments. The dark gray region includes all the cosmic string network configurations which create a SGWB probed at maximum sensitivity by a 10-year PTA experiment. Additionally to this region, the light gray slice includes all the extra configurations which can be probed at maximum sensitivity by a 5-year PTA experiment. The white area includes all those configurations which are probed by the reduced sensitivity slope (see, Fig.~\ref{fig::ptasens}) for both 5- and 10-year experiments. The hatched area includes the configurations which are inaccessible to PTAs.\label{accessed_a-n}}
\end{figure}
A PTA sensitivity curve roughly has the shape of an inverted triangle, with its sensitivity peaking at wavelengths similar to the duration of the PTA experiment. In Fig.~\ref{fig::ptasens} we plot the sensitivity curves for two completely different PTA experiments, one with 5 and one with 10 years duration, which achieve maximum sensitivity at frequencies $\sim6.3\,\rm{nHz}$ and $\sim3.2\,\rm{nHz}$ respectively. For frequencies higher than these, the sensitivity decreases with a slope given by $R f$, where $R$ is the root-mean-square of the residuals in pulsar timing data \cite{lk05} and $f$ the frequency of the GWs. For the sensitivity curves of Fig.~\ref{fig::ptasens} we assumed that the maximum sensitivity of the 5-year experiment is at $\Omega_{\rm gw}h^2=1.2\times10^{-8}$ and that the 10-year experiment has a slightly better sensitivity, equal to $\Omega_{\rm gw}h^2=8.9\times10^{-9}$. This assumption actually implies that the RMS residuals in the 10-year PTA experiment are larger than those in the 5-year experiment. The RMS residuals of the time of arrival of pulses are expected to improve as $\propto N^{-1/2}$, where $N$ is the number of the time of arrivals (observations) used. Therefore, if the 10-year PTA experiment was performed at the same telescope/instrumentation with the 5-year experiment we would expect a much more improved sensitivity on $\Omega_{\rm gw}h^2$, due to the double amount of data. The value of the highest frequency in the sensitivity curve depends on the mean time between observations of the same MSP in the whole data span. As an example, for the EPTA where each MSP is typically observed once every two weeks, the maximum frequency is $\approx 830\,\rm{nHz}$. The PPTA \cite{man06} and NANOGrav \cite{jfl+09} follow a similar observing schedule.

Now, let us consider the sensitivity of PTAs to emission from string loops with size $\sim \alpha t$ emitting into the $n$th harmonic. In Fig.~\ref{accessed_a-n} we present the regions in the $\alpha -n$ parameter space which can, or cannot, be probed by present and future PTA experiments. The dark gray area includes all the $\alpha -n$ combinations which give a SGWB with a low frequency cut-off lower than the frequency at which a 10-year PTA experiment achieves its highest sensitivity. A shorter, 5-year experiment is sensitive to all these cosmic string networks plus the networks included in the light gray slice. Although counter-intuitive, by increasing the duration of a PTA experiment we reduce the number of different cosmic string network configurations which are observable at maximum sensitivity, even though we increase its overall sensitivity by collecting more data.

We can easily demonstrate this in Fig.~\ref{fig::ptasens}, where we present the GW spectra for two cosmic string networks with $\alpha_1=5.7\times10^{-10}$ (thick red curve) and $\alpha_2=2.8\times10^{-10}$ (dashed red curve) respectively. Both networks have $G\mu/c^2=10^{-7}$, $n_*=1$ and $q=4/3$. The specific values for $\alpha_1$, $\alpha_2$ were selected so that their low frequency cut-offs coincide with the frequencies which a 5-year and a 10-year PTA experiment achieves maximum sensitivity. In this example, the 5-year experiment probes part of both spectra, while the 10-year experiment probes the $\alpha_1$ network and just misses the $\alpha_2$ network. It is clear, that any network which has a low frequency cut-off between $3.2\,{\rm nHz}$ and $6.3\,{\rm nHz}$ will be probed with the reduced sensitivity slope from the 10-year experiment, something that doesn't happen with the 5-year experiment. It is interesting to see, that in the specific example, the 10-year experiment will not be able to detect the emission from the $\alpha_2$ network at all. Although in reality a 10-year experiment is expected to have much better RMS residuals of a 5-year one, such an event might be true when comparing PTA experiments of different observatories. The white area of Fig.~\ref{accessed_a-n} contains all the $\alpha -n$ combinations which can be probed by both experiments with reduced sensitivity, and therefore, higher $G\mu/c^2$ values are required to make a detection. The hatched area corresponds to the $\alpha -n$ combinations inaccessible by PTAs due to their high emission frequencies. From this plot we can see that networks which produce very small loops with $\alpha\leq10^{-12}$ cannot be detected by PTAs, regardless of their tension.

\end{subsection}


\begin{subsection}{Effects of cosmic string model parameters on the gravitational wave spectrum}
\label{sec:parameters}

Here, we present a detailed analysis of the effects of each cosmic string model parameter on the GW spectrum with the objective of building up a picture of how the various uncertainties can affect the spectrum. We will often make the distinction between whether loops at formation are smaller,  $\alpha<\Gamma G \mu/c^2$, or larger,  $\alpha>\Gamma G\mu/c^2$, than the gravitational backreaction scale; we will refer to these as small and large loops respectively. Since the time of death of the loop, $t_{\rm d}$, and its time of birth, $t_{\rm b}$, are related by $t_{\rm d}\approx[1+(\alpha c^2/\Gamma G\mu)]t_{\rm b}$, this distinction corresponds to the loops either dying within a Hubble time, or living for much longer, respectively. The fiducial set of parameters is $G\mu/c^2=10^{-7}$, $\alpha=10^{-7}$, $q=4/3$, $n_*=1$, $p=1$ and in the subsequent discussion we will vary each of them while keeping the others equal to their fiducial values. In our computations we used a scale factor suitable for a radiation-matter-$
\Lambda$ Universe. For the numerical values of the constants entering in the computation of the scale factor we used the WMAP 7-year results \cite{jbdg+11} and in particular the WMAP+BAO+$H_0$ parameter values, $h=0.704$ and $\Omega_{\rm \Lambda}=0.728$. We also used $\Omega_{\rm r}h^2=2.47\times 10^{-5}$ and $\Omega_{\rm m}=1-\Omega_{\Lambda}-\Omega_{\rm r}$ for a flat Universe.


\begin{subsubsection}{Varying $G\mu/c^2$}
\label{sec:tension}

One would expect heavier strings, with larger values of $G\mu/c^2$, to lead to larger SGWB amplitudes and indeed this is typically the case. However, when varying $G\mu/c^2$ there are other more subtle effects which can have an impact on the GW spectrum. We present plots of $\Omega_{\rm gw}h^2$ in Fig.~\ref{gmvarying} for different values of $G\mu/c^2$ keeping $\alpha$, $q$, $n_*$ and $p$ the same. In the small loops regime (thin red lines),  lowering the string tension reduces the amplitude of the spectrum with the shape remaining the same; one finds that  $\Omega_{\rm gw}h^2\propto G\mu/c^2$. However, in the large loops regime (thick blue lines), along with the expected decrease in amplitude, there are also changes in the shape of the spectrum. The peak frequency starts to shift towards higher frequencies, initially being $\propto (G\mu /c^2)^{-1/4}$ and rather quickly it settles to being $\propto (G\mu /c^2)^{-1}$. The SGWB amplitude also decreases $\propto (G\mu /c^2)^{1/2}$, slower than in the case where $\alpha<\Gamma G\mu/c^2$, which can clearly be seen in the flat part of the spectrum.
\begin{figure}[!ht]
\includegraphics[width=8.6cm]{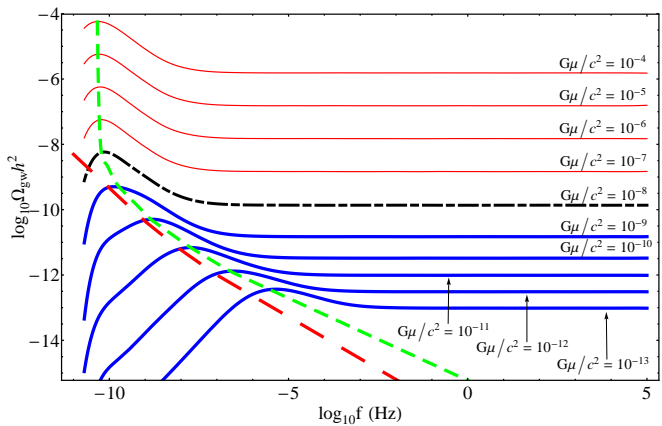}
\caption{Plots of normalized gravitational wave energy density per logarithmic frequency interval, $\Omega_{\rm gw}h^2$, due to cosmic string networks with different tensions but the same fiducial values of $\alpha$, $n_*$, $q$ and $p$. The thick blue lines are for networks in the large loop regime and the thin red lines are from networks in small loop regime. The dashed black line signifies the network for which $\alpha =\Gamma G\mu/c^2$. The analytic approximations of the peak frequency are also shown: the approximation found in CA92 (red long dashed curve) and our improved approximation (short dashed green curve).\label{gmvarying}}
\end{figure}

The frequency of the peak of the spectrum can be approximated analytically. The key question in making such an approximation is which loop population is responsible for the emission at the peak frequency. To answer this question, we need to define the birth time of loops $t_{\rm b}(t)$ as being the time of birth of a loop which dies at time $t$. The birth time of loops which die at the present time, assuming that we are in the matter era, can be calculated from Eq.~\eqref{cslooplength} setting $\ell =0$,
\begin{equation}
t_{\rm b}(t_0)=\left(1+\frac{3f_{\rm r}\alpha c^2}{\Gamma G\mu}\right)^{-1}t_0\,.
\label{eq::birthtime}
\end{equation}
In CA92, they suggested that the peak emission is created by the $n=1$ emission of the most recent, ``dominant population'' of loops. Since the birth rate of loops is continuously decreasing with time ($\propto t^{-2}$), they assumed that this ``dominant'' population was born at a time $\sim 2t_{\rm b}(t_0)$. This leads to a simple approximation of the peak frequency given by $f_{\rm peak}=2c^2/\Gamma G\mu t_0$. This approximation is presented in Fig.~\ref{gmvarying} with a long dashed red line. It is obvious that since this equation is independent of $\alpha$ it will not give a correct description on the different behavior in the large and small loops regimes and indeed this can be seen in Fig.~\ref{gmvarying}. Moreover, even in the large loop regime where it seems to be in reasonable agreement, the more we decrease the string tension the worse the approximation becomes.

We have managed to construct a better approximate formula for $f_{\rm peak}$, where we do not make any assumption about the birth time of the loop population responsible for the peak emission. Instead, we created a general, approximate formula and we determine when these loops were formed by comparing the analytic results with those of our computations.
\begin{figure}[!ht]
\includegraphics[width=8.6cm]{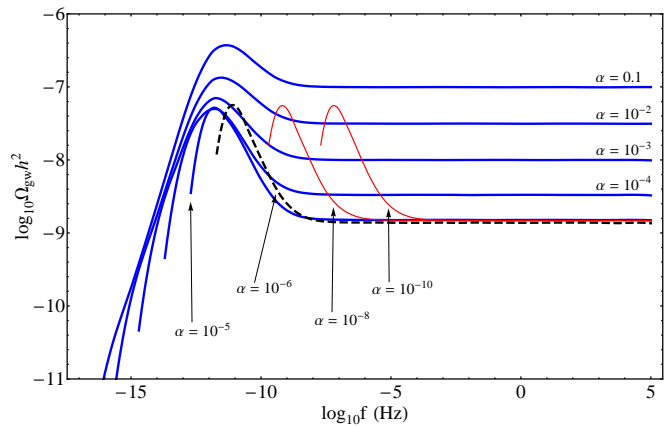}
\caption{$\Omega_{\rm gw}h^2$ for cosmic string networks with different values of $\alpha$ and the fiducial values of $G\mu/c^2$, $n_*$, $q$ and $p$. With thick blue lines we plot the networks in the regime of large loops and with thin red lines the networks in the regime of small loops. With dashed line we plot the network with $\alpha=\Gamma G\mu/c^2$ which signifies the critical point after which we have no amplitude decrease. \label{alphavarying}}
\end{figure}

\begin{figure*}[ht]
\includegraphics[width=17.5cm]{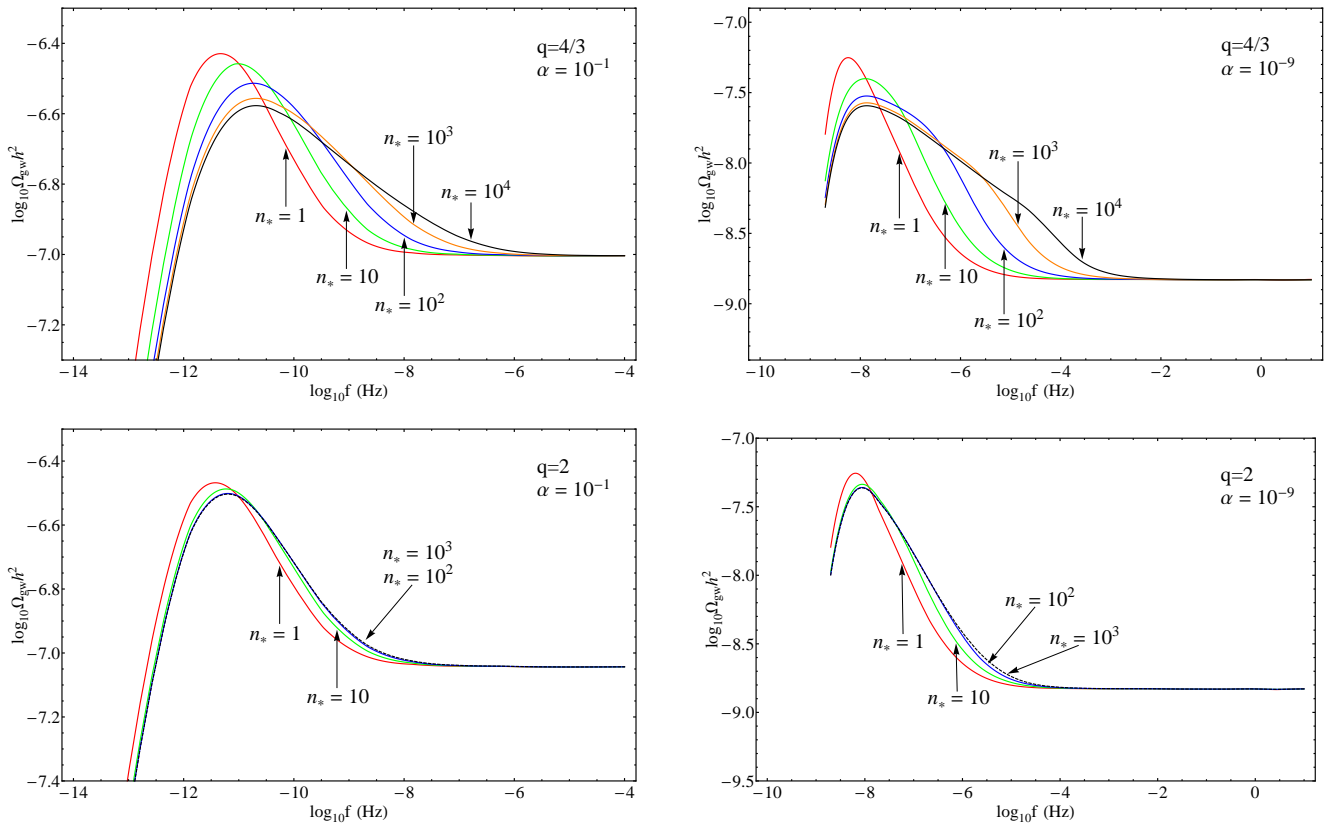}
\caption{$\Omega_{\rm gw}h^2$ for cosmic string networks with the same fiducial string tension value, but with different $n_{*}$ and $q$ values in the cases of large ($\alpha=0.1$) and small ($\alpha=10^{-9}$) loops. Different colors are used for different values of $n_*$, with the specific values denoted in each plot. In the $q=2$ case, the spectra for $n_*=10^3$ are plotted with a dashed line instead of a different color to better distinguish it from the $n_*=10^2$ case since the results are almost identical for extended frequency ranges.\label{modes}}
\end{figure*}
The peak frequency must originate from the redshifted emission in the $n=1$ mode of this population, the lowest frequency it ever emitted. Using Eq.~\eqref{eq::birthtime} for the birth time of loops we introduce the concept of loop generations, $g$. We will refer to loops which die right now, and therefore, were born at time $t_1=t_{\rm b}(t_0)$, as generation $g=1$ loops. The loops of generation $g=2$ are those which died when the loops of $g=1$ were born and have a birth time $t_2=t_{\rm b}(t_1)$. In the same way, the loops of generation $g$ are those which die when the loops of generation $g-1$ were born. From Eq.~\eqref{eq::birthtime} we find the birth time $t_g$ of generation $g$ loops to be
\begin{equation}
t_g=\left(1+\frac{3f_{\rm r}\alpha c^2}{\Gamma G\mu}\right)^{-g}t_0\,.
\end{equation}
 The lowest GW frequency ($n=1$) emitted by loops of generation $g$ in the matter era is
\begin{equation}
f_{g,\rm em}=\frac{2}{3f_{\rm r}\alpha t_g}=\frac{2}{3f_{\rm r}\alpha t_0}\left(1+\frac{3f_{\rm r}\alpha c^2}{\Gamma G\mu}\right)^g\,,
\label{eq::fgem}
\end{equation}
and when we redshift it to the present day, its observed frequency is
\begin{equation}
f_{g}=\frac{a(t_g)}{a(t_0)}\frac{2}{3f_{\rm r}\alpha t_g}=\frac{2}{3f_{\rm r}\alpha t_0}\left(1+\frac{3f_{\rm r}\alpha c^2}{\Gamma G\mu}\right)^{g/3}\,.
\label{eq::fgrs}
\end{equation}
Eq.~\eqref{eq::fgrs} is the general approximation for the peak frequency, without making any assumptions about which generation's loops created it. Using the results of our computations, we found out that the best approximation to the peak frequency is given by
\begin{equation}
f_{\rm peak}=\frac{2}{3f_{\rm r}\alpha t_0}\left(2+\frac{3f_{\rm r}\alpha c^2}{\Gamma G\mu}\right)^{10/9}\,,
\label{eq::fpeak}
\end{equation}
which is plotted with a short dashed green line in Fig.~\ref{gmvarying}. This means that the peak region is due to loops of generation $g\sim10/3$, i.e. of loops born just before the third generation loops. We have changed the numerical factor in the parenthesis of Eqs.~\eqref{eq::fgrs},~\eqref{eq::fpeak} from 1 to 2, so to achieve a perfect fit. In any case, this is a minor correction (less than $3\%$) which only affects networks with $\Gamma G\mu/c^2>\alpha$.

\end{subsubsection}


\begin{subsubsection}{Varying $\alpha$}\label{sec:length}

The effects of varying $\alpha$ in the large/small loop regions are the inverse of those seen when varying $G\mu/c^2$.  In Fig.~\ref{alphavarying} we present the GW spectra for cosmic string networks with the fiducial values of $G\mu/c^2$, $n_*$, $q$ and $p$ for various values of $\alpha$.

In the large loop regime (blue thick lines), as $\alpha$ decreases the most prominent feature is a decrease of the amplitude of the overall spectrum. This decrease is $\propto \alpha^{1/2}$ when $\alpha \gg \Gamma G\mu/c^2$, but the dependence becomes weaker, being $\propto \alpha^{1/4}$ when $\alpha$ gets close to the critical value, $\alpha=\Gamma G\mu/c^2$. The higher SGWB amplitude for large values of $\alpha$ is expected: large $\alpha$ means the loops persist for longer periods of time, and therefore have more time to emit their energy as GWs. Our results in this regime agree with those in DH07.

The situation is very different in the small loop regime (thin red lines). There is no decrease in the overall amplitude, nor a significant change in the amplitude difference between the peak and the flat part of the spectrum once $\alpha<\Gamma G\mu/c^2$. Instead of this we see a shifting of the spectrum to higher frequencies, something which agrees with the results presented in \cite{smc07}. The overall amplitude invariance to changes in $\alpha$ is a radiation era effect, where the small loops decay in less than a Hubble time. The independence of $\Omega_{\rm gw} h^2$ from the value of $\alpha$ can be clearly seen in the analytic approximations for $\Omega_{\rm gw}h^2$ in the radiation era (i.e., see equation 4 in \cite{bm10}) if we assume $\alpha\ll\Gamma G\mu$. The shifting of the peak frequency is $\propto \alpha^{-1}$, consistent with the minimum frequency emitted by a loop being $f_{\rm min}\approx2c/(\alpha t_{\rm b})$ and Eq.~\eqref{eq::fpeak}.

\end{subsubsection}


\begin{subsubsection}{Varying the emission spectrum parameters $q$ and $n_{*}$}\label{sec:emissionmodes}

\begin{figure}
\includegraphics[width=9 cm]{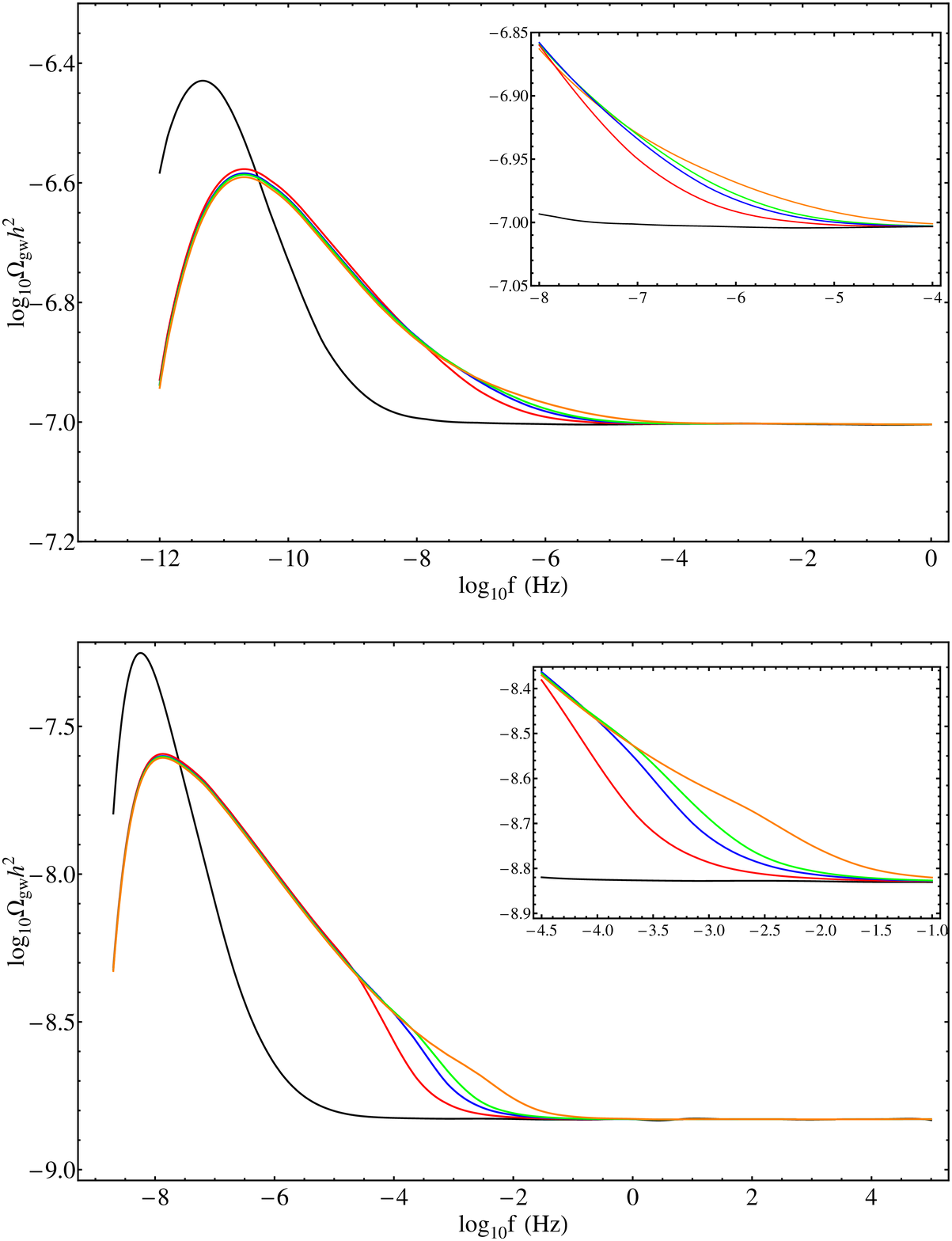}
\caption{Plots of $\Omega_{\rm gw}h^2$ for a cosmic string network with $G\mu/c^2=10^{-7}$, $\alpha=0.1$ (top), $\alpha=10^{-9}$ (bottom), and $q=4/3$ for $n_*=1$ (black), $n_*=10^4$ (red), $n_*=5\times 10^4$ (blue), $n_*=10^5$ (green) and $n_*=10^6$ (orange). In the upper right side of each plot a magnification of the area of interest is presented. \label{saturation}}
\end{figure}
\begin{table*}[ht]
\begin{tabular}{|l|c|c|c|c|c|c|c|c|}
\hline
 &\multicolumn{4}{|c|}{$$q=4/3}&\multicolumn{4}{|c|}{q=2}\\
\hline  \hline
 &\multicolumn{2}{|c|}{$\alpha >\Gamma G\mu/c^2$}&\multicolumn{2}{|c|}{$\alpha <\Gamma G\mu/c^2$}&\multicolumn{2}{|c|}{$\alpha >\Gamma G\mu/c^2$}&\multicolumn{2}{|c|}{$\alpha <\Gamma G\mu/c^2$}\\
\hline
$n_*$&Peak fr.&Peak amp.&Peak fr.&Peak amp.&Peak fr.&Peak amp.&Peak fr.&Peak amp.\\
\hline
$10$&$115\%$&$-7\%$&$122\%$&$-27\%$&$62\%$&$-5\%$&$60\%$&$-18\%$\\
\hline
$10^2$&$311\%$&$-18\%$&$129\%$&$-45\%$&$75\%$&$-8\%$&$60\%$&$-22\%$\\
\hline
$10^3$&$357\%$&$-26\%$&$130\%$&$-51\%$&$75\%$&$-8\%$&$60\%$&$-23\%$\\
\hline
$10^4$&$358\%$&$-29\%$&$130\%$&$-53\%$&$75\%$&$-8\%$&$60\%$&$-23\%$\\
\hline
\end{tabular}
\caption{The percentage increase in the peak frequency and decrease in the peak amplitude of $\Omega_{\rm gw} h^2$ for a particular value of $n_*$ relative to $n_*=1$ and whether we are in the large ($\alpha=10^{-1}$) or small loop ($\alpha=10^{-9}$) regimes. The results are categorized according to the value of the spectral index.\label{modestable} }
\end{table*}

The spectrum of gravitational radiation emitted by a cosmic string loop is still an open question. In the previous discussion we have introduced two parameters, $q$, the spectral index and, $n_*$, the cut-off in the radiation spectrum, to model the possible effects. In this section we investigate how these two parameters affect the observed spectrum. Previous works have used a range of values for $q$ and $n_*$ which have sometimes led them in imposing very strong constraints on the string tension. For example, DH07 typically used $n_*=1$, that is, they only considered emission in fundamental mode of the string loops. Although they investigated cases up to $n_*\approx5$, they found that this had only a small effect on the power spectrum of GWs and does not significantly effect the bounds on the string tension from PTAs. In contrast, CA92 used $n_{*}\rightarrow\infty$, which was done by replacing the summation in Eq.~\eqref{specint} with an integral in order to make the calculation tractable. As we will see, $n_*=1$ and $n_*=\infty$ give very different results. Damour and Vilenkin \cite{dv00,dv01} and Siemens et al. \cite{scm+06,smc07} followed a similar approach to CA92  making the strong assumption that $q=4/3$ and $n_*=\infty$ based on their study of cusp emission. A more conservative approach was taken in \cite{cbs96} who showed that constraints from PTAs are very sensitive to the choices of $q$ and $n_*$ and in particular that the predicted spectra for  $q=2$ and $n_*=\infty$ are very similar to $q=4/3$ and $n_*=10^3-10^4$ in the nHz frequency range.

\begin{figure*}[ht]
\includegraphics[width=17.5 cm]{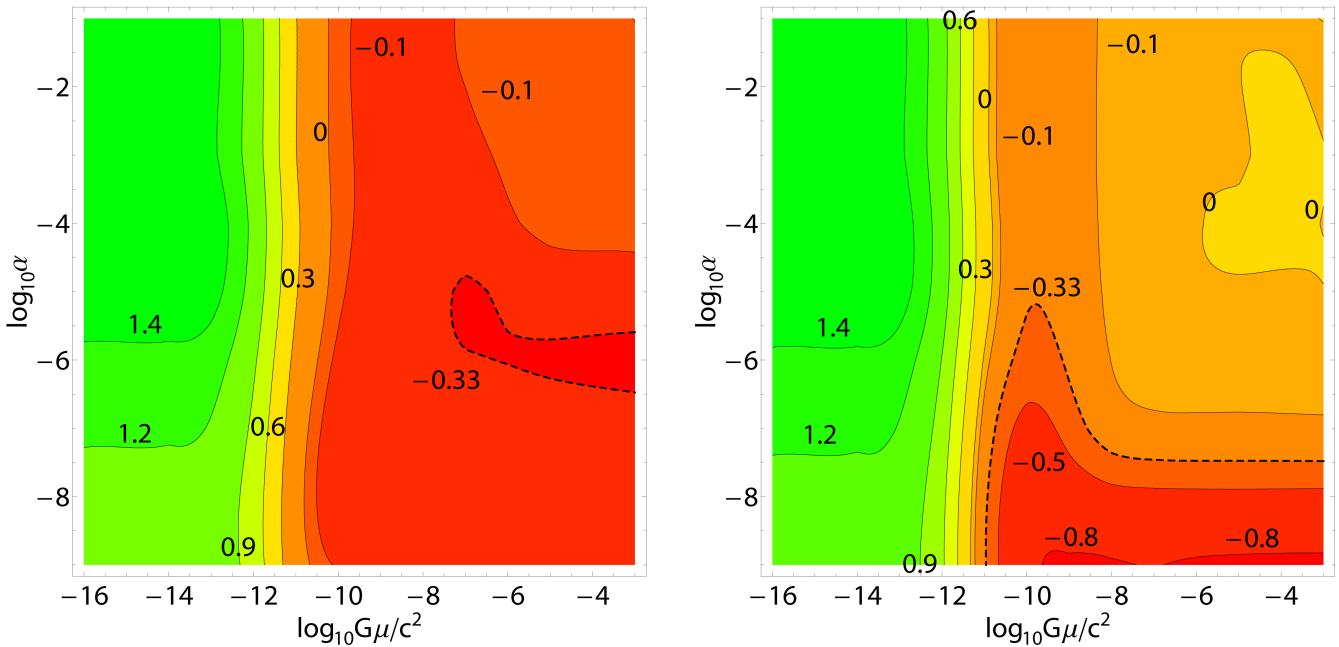}
\caption{The slope, $d_{\Omega}$, for cosmic string networks with $n_*=10^4$ (left) and $n_*=1$ (right) as a function of the string tension $G\mu/c^2$ and the birth scale $\alpha$ of the loops. For both plots, $q=4/3$. The dashed line corresponds to the spectral index $d=-7/6$ proposed in \cite{dv05}. We see that for a wide range of the parameter space $d$ is very different from $-7/6$.\label{slopes3d}}
\end{figure*}
As we will see, the values of $n_*$ and $q$ have a significant effect on the amplitude and the slope of the GW spectrum in the region of radiation to matter era transition, between the peak and the flat part of the spectrum. This is of critical importance for PTAs, since the frequency windows probed for the majority of $G\mu/c^2-\alpha$ combinations of interest are in this region. The PTA frequency window lies outside this region in three cases: (i) $G\mu/c^2\gtrsim 10^{-7}$ and $\alpha\gtrsim10^{-6}$, where it probes the flat part of the spectrum, (ii) in the case of very small tension networks, $G\mu/c^2\leq10^{-11}$ independent of $\alpha$, where it probes the region to the left of the peak (see, i.e., Fig.~\ref{gmvarying},~\ref{alphavarying}), and (iii) in the case $\alpha\lesssim 10^{-12}$ where the GW spectrum is always at higher frequencies irrespectively of the string tension.

In Fig.~\ref{modes} we present the GW spectra for two representative scenarios varying $q$ and $n_*$, a large loop case, $\alpha=0.1$ and a small loop case, $\alpha =10^{-9}$, both with $G\mu/c^2=10^{-7}$.  The first thing to note, as found in \cite{cbs96}, is that the spectrum is relatively independent of $n_*$ when $q=2$ both in the case of $\alpha=0.1$ and $\alpha=10^{-9}$. The modifications to the spectrum seen there are similar for the two values of $\alpha$, but slightly more pronounced for $\alpha=10^{-9}$.  Increasing $n_*$ from 1 appears to move the peak in the spectrum to slightly higher frequencies and its amplitude is also slightly reduced. For $n>n_{\rm sat}\approx 100$ the changes in the spectrum are almost negligible, where we define $n_{\rm sat}$ to be the saturation point for $n_*$, above which large increases of the value of $n_*$ result in negligible effects on the GW spectrum.

The case of $q=4/3$ is somewhat different. The shape of the spectra in the region of the peak is significantly affected by varying $n_*$. In the case of $\alpha=0.1$ there is a smooth broadening of the spectrum due to loops formed in the matter era, with the actual peak position moving to higher frequencies and the amplitude being reduced. The situation is similar for $\alpha=10^{-9}$; however, the spectrum appears to generate a hump as $n_*$ increases. There does not appear to be a convergence of the spectrum for the values of $n_*$ presented in Fig.~\ref{modes}. We have investigated the convergence of the spectrum in the case of large and small loops in Fig.~\ref{saturation}, where we present spectra for  $G\mu/c^2=10^{-7}$, $\alpha=0.1,\,10^{-9}$ and $q=4/3$ with  $n_{*}$ as high as $10^6$.  We see that in the case of large loops there are only minimal differences for $n_{*}>10^4$ suggesting that $n_{\rm sat}\approx 10^{4}$ for $q=4/3$. In the case of small loops a similar trend is observed, with the appearing hump moving along the radiation to matter era radiation tail of the spectrum towards higher frequencies until it reaches the flat part of the spectrum where it disappears (something that can be seen in Fig.~\ref{modes}). Whereas in general the spectrum has converged for $n_*\gtrsim 10^4$, the area near the tail of the spectrum still evolves until $n_*\gtrsim10^6$.

\begin{table}
\begin{tabular}{|l|c|c|c|c|}
\hline
&\multicolumn{2}{|c|}{$\alpha >\Gamma G\mu/c^2$}&\multicolumn{2}{|c|}{$\alpha <\Gamma G\mu/c^2$}\\
\hline
$n_*$&Peak fr.&Peak amp.&Peak fr.&Peak amp.\\
\hline
$10$&$32\%$&$-2\%$&$52\%$&$-8\%$\\
\hline
$10^2$&$135\%$&$-11\%$&$57\%$&$-27\%$\\
\hline
$10^3$&$158\%$&$-19\%$&$57\%$&$-36\%$\\
\hline
$10^4$&$161\%$&$-22\%$&$57\%$&$-36\%$\\
\hline
\end{tabular}
\caption{The percentage increase in the peak frequency and decrease in the peak amplitude of $\Omega_{gw}h^2$ for networks with $q=4/3$ relative to the values for $q=2$ as a function of $n_*$.\label{qtable}}
\end{table}

In order to re-enforce the results of Fig.~\ref{modes}, in Table~\ref{modestable} we present the percentage differences in the peak frequency and amplitude between models with the same values of $G\mu/c^2=10^{-7}$, $\alpha=0.1,\,10^{-9}$ and $q=4/3$ but different values of $n_*$ relative to that for $n_*=1$. In Table~\ref{qtable} we present similar information for a change in $q$, keeping all parameters fixed and changing $q$ from $2$ to $4/3$. The results vary numerically for other combinations of $G\mu/c^2-\alpha$, but exhibit exactly the same trend in both the large and the small loop regimes.

When we discuss the constraints on $G\mu/c^2$ due to observations we will be interested in the amplitude of the spectrum at the appropriate frequency and also the slope of the spectrum. If we define $d$ and $d_{\Omega}$ such that $h_{\rm gw}(f)\propto f^d $ and $\Omega_{\rm gw}\propto f^{d_{\Omega}}$ then $d_\Omega=2(d+1)$. It is often suggested \cite{dv05} that $d=-7/6$ for cosmic strings, but from a cursory examination of the spectra presented in Fig.~\ref{gmvarying}, \ref{alphavarying}, \ref{modes}  it is clear that this is not the case if $G\mu/c^2$, $\alpha$, $q$ and $n_*$ are allowed to vary. We measured $d_{\Omega}$ in the region between $31\,{\rm nHz}$ and $32\,{\rm nHz}$ as a function of $G\mu/c^2$ and $\alpha$ for $q=4/3$ and two values for $n_*$, $n_*=1$ and $n_*=10^4$. In general, realistic PTA experiments are sensitive to GWs with frequencies of a few nHz. However, later in this work we will calculate the string tension constraints based on limits placed for a frequency $(1{\rm yr})^{-1}$ and therefore we have chosen this range.

In Fig.~\ref{slopes3d} we present a plot of $d_{\Omega}(G\mu/c^2,\alpha)$ at a frequency $f=(1{\rm yr})^{-1}$ for cosmic string networks with $n_{*}=1,\,10^4$ and $q=4/3$. The results for other values of $n_*$ and $q$ are of course numerically different, but they follow a very similar trend. We see that describing the cosmic string GW spectrum with a simple power law in the frequencies probed by PTAs is far from reality. In the tension range $G\mu/c^2>10^{-11}$ the slope is generally negative, and PTAs probe the whole area of the spectrum which lies between the matter era peak and the radiation era flat part of the spectrum. However, when we go to small tensions, $G\mu/c^2<10^{-11}$, the PTA frequency window falls to the left of the matter era peak and the spectrum slope becomes positive.

\end{subsubsection}

\begin{subsubsection}{Varying $p$}

In order to calculate the GW spectra of cosmic string networks with $p< 1$ we used the same one-scale model based code, only changing the value of $A$ as in Eq.~\eqref{A(P)}. In Fig.~\ref{icp_plot} we present the results for different values of the intercommutation probability ($p=0.1,10^{-2},10^{-3}$) and for both scaling laws ($k=1$ and $k=0.6$). In all computations we used $G\mu/c^2=10^{-7}$, $\alpha=0.1$, $n_*=1$ and $q=4/3$.

 The reduced intercommutation probability leads to an increased number density of cosmic string loops, and therefore, an increased number of GW sources which give higher SGWB. The uniform scaling across the frequency band make the effects of decreasing $p$ similar to those of increasing $G\mu/c^2$ in the small loop regime, see Fig.~\ref{gmvarying}.

\begin{figure}[h]
\includegraphics[width=8.6 cm]{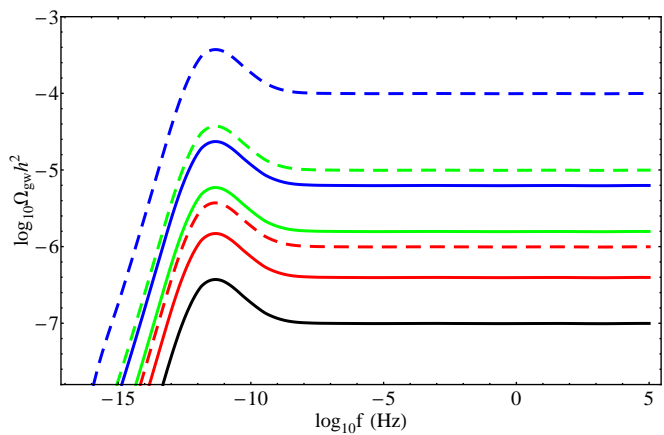}
\caption{The effects of varying the intercommutation probability $p$ for $G\mu/c^2=10^{-7}$ and $\alpha=0.1$. The red, green and blue lines are for $p=0.1,10^{-2},10^{-3}$ respectively for $k=0.6$. With the same color scheme but with dashed lines we show the equivalent results for $k=1$. The black solid line is for $p=1$.\label{icp_plot}}
\end{figure}

\end{subsubsection}

\end{subsection}


\begin{subsection}{Improved modeling of loop production}

\begin{figure*}[ht]
\includegraphics[width=17cm]{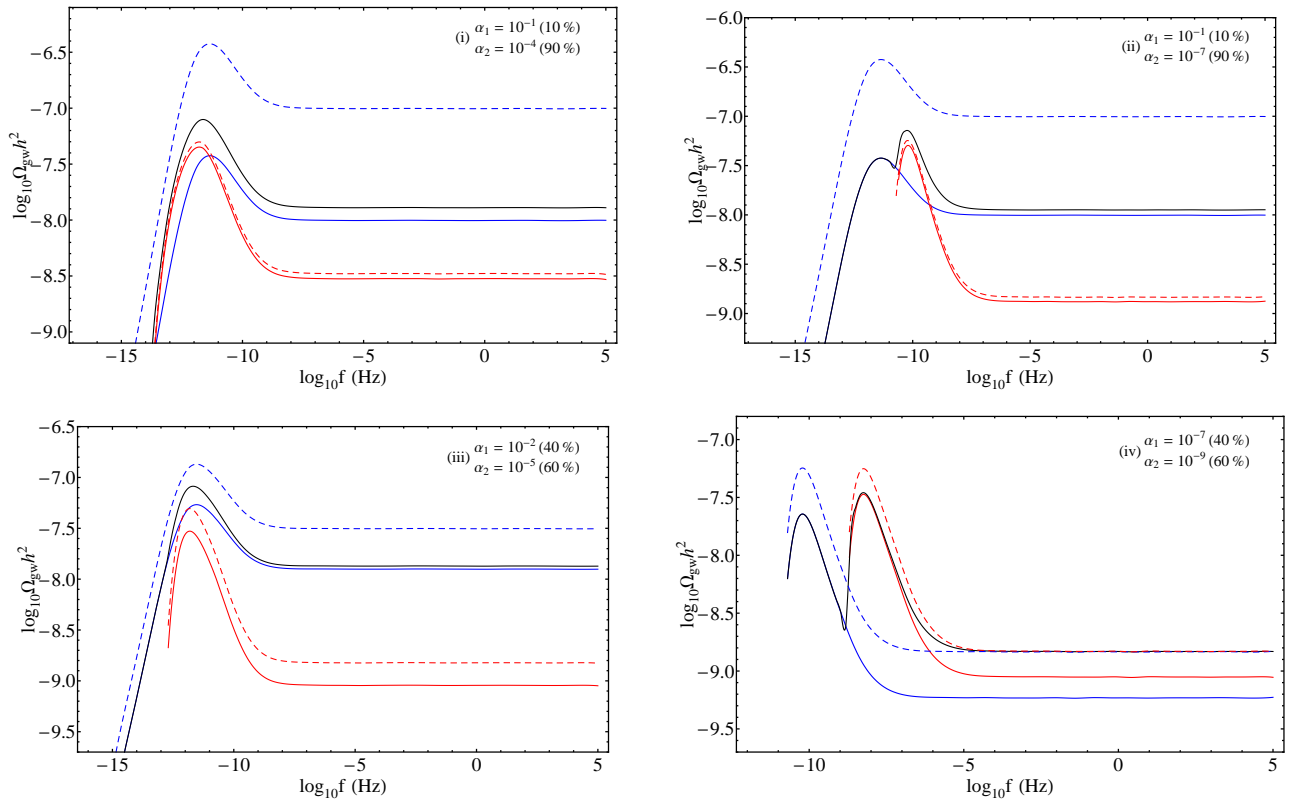}
\caption{$\Omega_{\rm gw}h^2$ for networks which produce two different sizes of loops corresponding to scales $\alpha_1$ and $\alpha_2$ relative to the horizon. In all cases we used $G\mu/c^2=10^{-7}$, $q=4/3$, $n_*=1$ and $p=1$. The $\alpha_1$, $\alpha_2$ and the relative percentages used are shown in the individual figures. The black lines are the total spectra for the overall 2-scale network. The solid blue and red lines are the individual contributions to the total spectrum from loops with initial sizes relative to the horizon, $\alpha_1$ and $\alpha_2$ respectively. For reference we have also included  spectra for the networks with $\alpha_1$ and $\alpha_2$ (blue and red dashed lines respectively) assuming that all of the energy was channeled into each of the individual loop sizes.\label{2scalefig}}
\end{figure*}
It is extremely difficult to model the distribution of loops produced by a cosmic string network. The calculations presented so far are based on the assumption that loops are produced with a single size relative to the horizon, $\alpha$. In the initial work on cosmic strings it was believed that large loops are born, with $\alpha \approx 0.1$ \cite{kib85,at89}. The work that followed \cite{bb89a,as90} argued against this, supporting the idea of a smaller length scale close to the scale of the gravitational backreaction $\alpha\approx \Gamma G\mu/c^2$. Subsequent work has led to a somewhat confusing situation. Some appear to  suggest large loops \cite{ms06,rsb07,vov05,vov06,ov07} ($\alpha\approx10^{-1}-10^{-3}$) while others \cite{so01,sov02,pr06,pr07a,ck09} support the view that the loops are small ($\alpha <\Gamma G\mu/c^2$).  Some even suggest microscopic loops with $l_{\rm b}\approx\delta$ \cite{vhs97,vah97,bhku08}, where $\delta$ is the string core width. For this reason we have allowed $\alpha$ to vary as an unknown parameter.

There is, of course, nothing to prevent loops being born over a range of different scales, both small and large. Recent work \cite{pol08,dpr08} has presented arguments supporting the idea that loops are created at two different scales, with $90\%$ of the loops created at the gravitational backreaction scale ($\alpha\approx \Gamma G\mu/c^2$) and $10\%$ at large scales ($\alpha\approx 0.1$). In this section we will discuss the effects of relaxing the assumption of a single loop production scale. We will consider two possibilities: the first is a two-scale model for loop production motivated by \cite{pol08,dpr08}, and the second is for the initial loop distribution to have a log-normal distribution with the mean being $\propto t_{\rm b}$.


\begin{subsubsection}{2-scale networks}
\begin{figure}[ht]
\includegraphics[width=8.6cm]{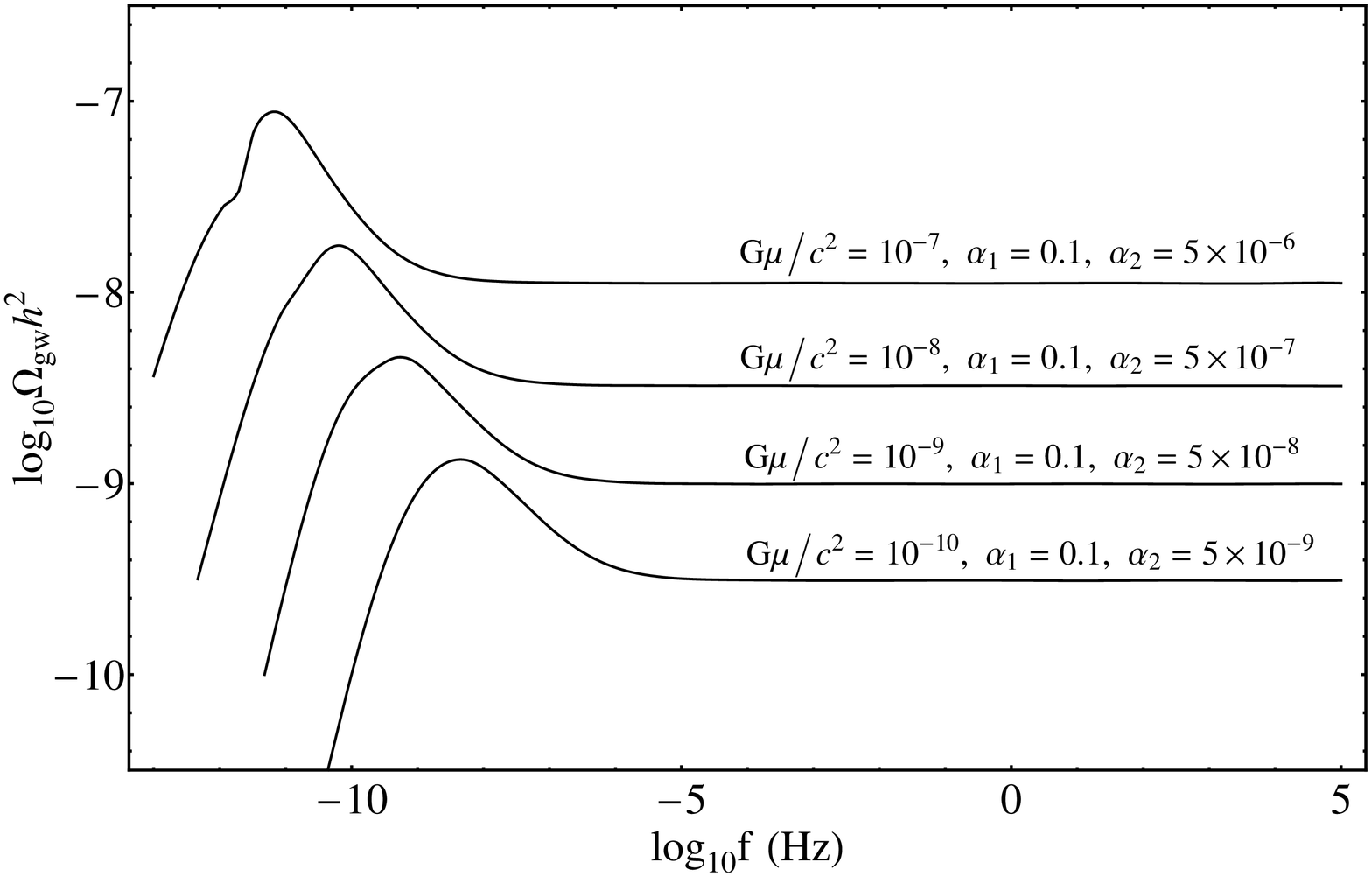}
\caption{$\Omega_{\rm gw}h^2$ plots for 2-scale networks with $10\%$ of the energy lost into the production of large loops ($\alpha_1=0.1$) and $90\%$ to loops at the gravitational backreaction scale ($\alpha_2=\Gamma G\mu/c^2$). GW spectra for various $G\mu/c^2$ values are presented. As we go to smaller tensions, the sharp peak created by the small loops starts to weaken until we reach the value $G\mu/c^2=10^{-10}$, below which the total GW spectrum of the 2-scale network is actually created only by the large loops. \label{prfig}}
\end{figure}

The one-scale model can be easily adapted in order to describe networks with more than one scale for the newborn loops. A network that produces loops of two scales, will create the same shape of SGWB as the one created by two distinct one-scale networks, but with the relative amplitude of the two carefully normalized. In order to impose this normalization we need to enforce the condition that a specific amount of energy has to be ``lost'' from the network in the form of loops per unit time; a requirement necessary for scaling. If we want to describe the loop distribution with two separate networks of $\alpha_1$ and $\alpha_2$ respectively, the amount of energy transferred to loops per unit time will be
\begin{equation}
\frac{dE_{2\rm{-scale}}}{dt}=c_1^{\rm E}\frac{dE_1}{dt}+c_2^{\rm E}\frac{dE_2}{dt}\,,
\end{equation}
where $dE_1/dt$, $dE_2/dt$ are the energies transferred to the two different types of loops and $c_1^{\rm E}, c_2^{\rm E}$ are the appropriate weighting factors. The values of $c_1^{\rm E}, c_2^{\rm E}$ are the relative amounts of energy we want to be channeled in the loops of each scale.

From Eq.~\eqref{dE-dN} we can see that the energy channeled in loops of a specific scale is proportional to the number of loops created
\begin{equation}
\frac{dE_{1,2}}{dt}\propto\frac{dN_{1,2}}{dt}\,,
\end{equation}
and therefore the normalizing constants $c_{1,2}^{\rm E}$ can also be used as normalizing constants for the loop number density
\begin{equation}
c_{1,2}^{\rm E}=c_{1,2}^{\rm N}\,.
\end{equation}
 In this way we can calculate the SGWB for each network individually, using the method described before, but this time we will normalize the number density of loops with $c_1^{\rm N}$ and $c_2^{\rm N}$ respectively. Once we have calculated the individual SGWBs, we add them and we get the SGWB produced by the 2-scale network.

In Fig.~\ref{2scalefig} we present results for representative models for which newborn loops are created with two distinct different scales, $\alpha_1$ and $\alpha_2$ and all other parameters given by their fiducial values. The solid blue and red lines are the individual contributions due to loops with $\alpha_1$ and $\alpha_2$ respectively, normalized to their relative contributions to the total emission.  For reference, the dashed lines are the results we would expect if all the energy was channeled into loops with either $\alpha_1$ or $\alpha_2$. In Fig.~\ref{2scalefig}(i) the spectrum corresponds to a case with $\alpha_1 ,\alpha_2 >\Gamma G\mu/c^2$. In this case, even for a very small amount of energy ($10\%$) channeled into the creation of the largest scale loops ($\alpha_1=0.1$), the GWs emitted by them dominate the overall result, with the smaller scale loops ($\alpha_2=10^{-4},90\%$) only dominating the overall spectrum at very low frequencies, $f<10^{-12}\,{\rm Hz}$. This effect is even more obvious in Fig.~\ref{2scalefig}(iii), where there is  a larger percentage of the energy channeled into the creation of the large scale loops ($\alpha_1=10^{-2}, 40\%$). There we see that the contribution from the large loops contributes most of the overall result. From these results, and the other cases we have investigated but not included in the figure, we conclude that the spectrum of the 2-scale model has a similar shape to ones created in cases when there is only one loop production size when $\alpha_1,\alpha_2>\Gamma G\mu/c^2$.

The situation is very different if  $\alpha_1 >\Gamma G\mu/c^2 >\alpha_2$, as illustrated in Fig.~\ref{2scalefig}(ii). In that case, we find that the spectrum of the SGWB has two peaks, one due to each loop population. This is consistent with the discussion of Sec.~\ref{sec:length}, where we showed that the peak of the spectrum for  $\alpha <\Gamma G\mu/c^2$ is shifted towards higher frequencies. This means that its peak will be higher than that due to the emission from the loops with $\alpha >\Gamma G\mu/c^2$ in a frequency higher than the $f_{\rm peak}$ of the $\alpha >\Gamma G\mu/c^2$ loops. In Fig.~\ref{2scalefig}(ii), $90\%$ of the energy goes to the creation of small scale loops ($\alpha_2$) and therefore the peak corresponding to the $\alpha_2$ (higher frequency peak) will be more prominent. In general, which of the two peaks dominates over the other depends on the amount of energy that the network allocates to the respective loop sizes. The same behavior is exhibited in Fig.~\ref{2scalefig}(iv), where $\alpha_1 ,\alpha_2 <\Gamma G\mu/c^2$. We see the two peaks are even more clear in this case, with the second being slightly higher since a larger amount of energy goes into the  creation of loops at that scale ($\alpha_2$). Based on a more general investigation of the parameter space, the basic feature of two peaks is something which appears to be generic in the case when one, or both, of the loop production scales is lower than the gravitational backreaction scale.

\begin{figure*}
\includegraphics[width=17 cm]{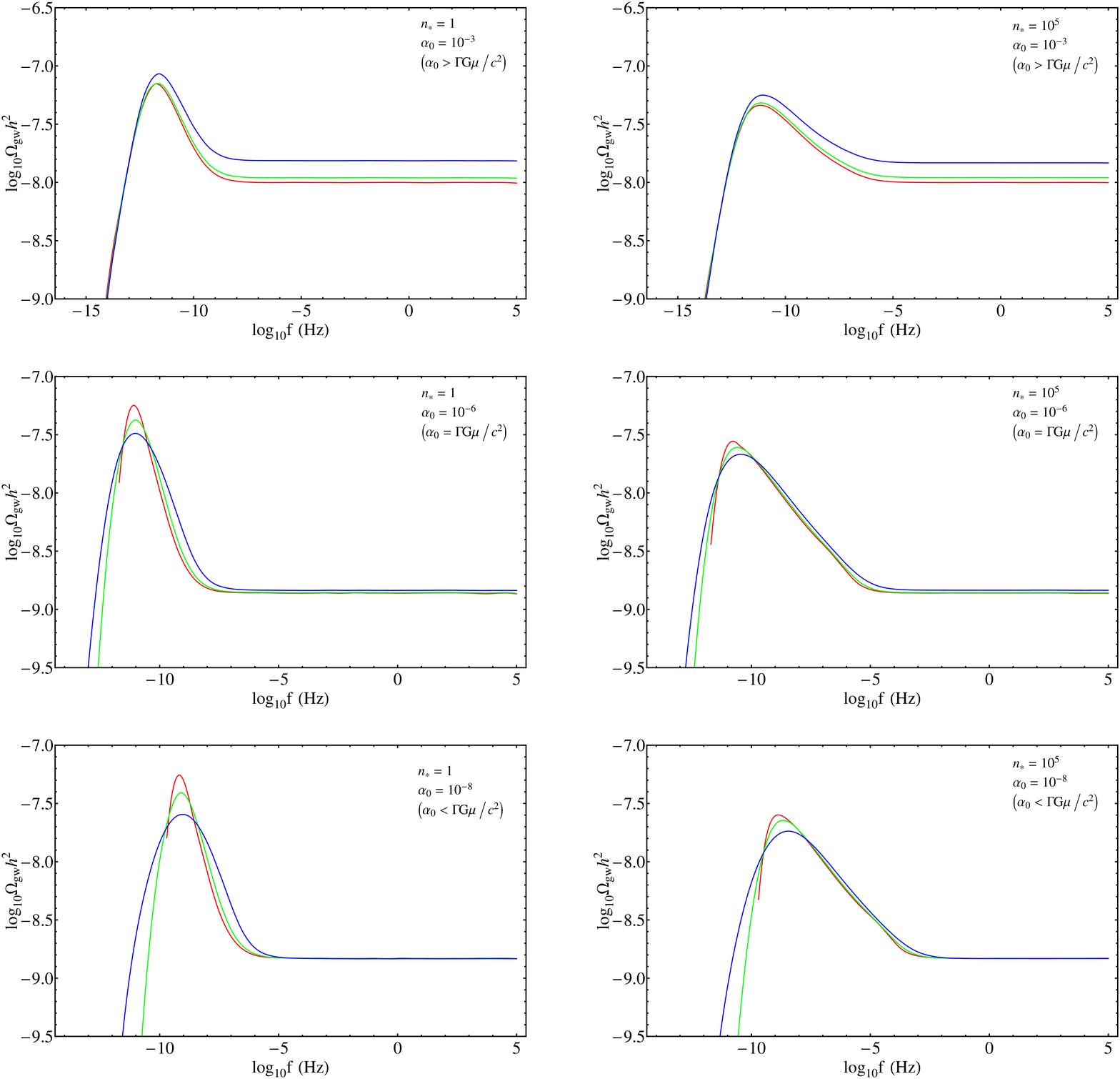}
\caption{$\Omega_{\rm gw}h^2$ plots for networks in which cosmic string loops are born with $\alpha$  given by a log-normal distribution. All the computations were performed for a network with $G\mu/c^2=10^{-7}$, $q=4/3$, $n_*=1$ and $p=1$. In the left column are the plots with $n_{*}=1$ and in the right column are results with $n_{*}=10^5$. The green and blue lines are created using the log-normal loop production distribution with a mean value of $\alpha_0$ and with $\sigma=0.4$ and $\sigma=0.8$ respectively. The red lines are the results for a single loop production size $\alpha=\alpha_0$.\label{plotgaussianlengths}}
\end{figure*}
In Fig.~\ref{prfig} we present results for networks which specifically follow the predictions of \cite{pol08,dpr08}, in which $10\%$ of the energy lost by the network goes into large loops of scale $\alpha_1\approx 0.1$ and the remaining $90\%$ goes into loops at the gravitational backreaction scale $(\alpha_2\approx \Gamma G\mu/c^2)$. We present results for various values of $G\mu/c^2$. For $G\mu/c^2>10^{-10}$, we find a two-peak spectrum, as in Fig.~\ref{2scalefig}(ii), but much less prominent. The position of the peak created by the small loops is almost the same as the one created by the large loops, creating a spectrum similar to that of a pure one-scale network but with a sharper peak. However, the peak amplitude continuously decreases as we decrease $G\mu/c^2$, and when $G\mu/c^2\lesssim 10^{-10}$ the peak disappears and the large scale loops' contribution totally dominates the spectrum.

\end{subsubsection}


\begin{subsubsection}{Log-normal distribution for loop production}

Another possibility to consider is that the loops are born with size $\ell(t_{\rm b})=f_{\rm r}\alpha d_{\rm H}(t_{\rm b})$ but with $\alpha$ having a distribution, ${\cal P}(\alpha)$, an idea qualitatively justified by some recent simulations \cite{vov06,rsb07,dpr08,bos11}.

One can model any distribution using an adaptation of the methods used in the earlier sections by splitting the distribution into $N$ populations with loop production size relative to the horizon $\alpha_i$ for $i=1,..,N$ and a fraction of loops in each bin of size $\Delta\alpha$ given by  ${\cal P}(\alpha_i)\Delta\alpha$, assuming that
\begin{equation}
\int_0^{\infty}d\alpha\,{\cal P}(\alpha)=1\,.
\end{equation}
If $\Omega_{\rm gw}^{(i)}(f)$ is the spectrum computed from loops with $\alpha=\alpha_i$ then the overall spectrum is
\begin{equation}
\Omega_{\rm gw}(f)=\sum_{i=1}^{N}c_i\Omega_{\rm gw}^{(i)}(f)\,,
\end{equation}
where the $c_i={\cal P}(\alpha_i)\Delta\alpha$ are computed in order to enforce the overall energy loss required to maintain scaling.

We have chosen to use the log-normal distribution to model the loop distribution
\begin{equation}
{\cal P}(\alpha)={1\over \sigma\sqrt{2\pi}\alpha}\exp\left[-{1\over 2\sigma^2}\left[\log\left({\alpha\over{\alpha_0}}\right)\right]^2\right]\,,
\end{equation}
which corresponds to a Gaussian distribution in $\log\alpha$ with mean $\log\alpha_0$ and variance $\sigma$. Typically, we will discretise the distribution in the range  $-9\leq\log_{10}\alpha\leq -1$) with $21$ bins with $\Delta(\log_{10}\alpha)=8/21$. We also experimented with smaller bin sizes, but we didn't see any change in the final results, making our choice the most computationally effective.

The results we present here use the fiducial set of parameters, except that we consider the two cases $n_*=1$ and $n_*=10^{5}$, and we have varied $\alpha_0$ and $\sigma$. To exhibit the important effects, we chose to use three different mean values, $\alpha_0=10^{-3}$ $(\alpha >\Gamma G\mu/c^2)$, $\alpha_0=10^{-8}$ ($\alpha <\Gamma G \mu/c^2$) and $\alpha_0 =5\times 10^{-6}$ ($\alpha =\Gamma G\mu/c^2$), and two values of the variance $\sigma=0.4$ and $\sigma=0.8$ to demonstrate the behavior of multiple scale networks in the case of large loops, small loops and those produced at the gravitational backreaction scale respectively.

In Fig.~\ref{plotgaussianlengths} we present the results of our computations. In the upper right of each plot is the mean value of $\alpha$ used. The red lines are for the power spectra of a one-scale model whose value of $\alpha$ equal to the mean used for the multiscale networks. The green and blue lines are the spectra for the multiscale networks with $\sigma=0.4$ and $\sigma=0.8$ respectively. The spectra have different behavior in the case of large and small loops. In the case of large loop creation ($\alpha_0 >\Gamma G\mu/c^2$), the multiscale networks give higher SGWB amplitudes than the corresponding one-scale networks and moreover, the increase is higher for larger values of $\sigma$ since the spectrum is dominated by the few very large loops. This amplitude increase is seen over the whole frequency region for $f>f_{\rm{peak}}$. This behavior reverses in the case of networks with small loops or loops near the gravitational backreaction scale ($\alpha_0\lesssim\Gamma G\mu/c^2$). The flat part of the spectrum, as expected, is not affected when all the loops are born small ($\alpha<\Gamma G\mu/c^2$). However, when $\alpha_0=\Gamma G\mu/c^2$ a small increase in the amplitude of the flat part is seen due to the loops at the tail of the distribution, born with $\alpha >\Gamma G\mu/c^2$. Although there are very few of them, they can have a noticeable effect on the spectrum.

On the other hand, the peak region created from GW emission during the matter era exhibits a richer behavior. A common feature is the decrease of the peak amplitude combined with a broadening of the whole peak region, when compared to the corresponding results from the one-scale model. This is even more prominent in the case where $\alpha_0\lesssim \Gamma G\mu/c^2$, because of the shift in the peak frequency from the loops born with different values of $\alpha$. Interestingly, we see that the low frequency cut-off of the one-scale network is no longer present, with the SGWB spectrum extending to lower frequencies. This extra emission into these low frequencies is created by the few large loops at the tail of the distribution which individually, have lower cut-offs. Finally, as seen from Fig.~\ref{plotgaussianlengths}, the number of emission modes does not have any significant effect on the spectra of the multiscale networks but in general suppresses the effects of amplitude decrease of the flat portion of the spectrum, peak amplitude decrease and peak region broadening already discussed.

\end{subsubsection}

\end{subsection}
\end{section}


\begin{section}{Pulsar Timing constraints on the cosmic string tension}

In this section we will use the constraints on the SGWB from pulsars to impose a constraint on the dimensionless cosmic string tension parameter $G\mu/c^2$ using the EPTA SGWB limit \cite{vlj+11} and discuss the possibilities of using the future constraints from the LEAP \cite{fvb+10} project which should be available in the near future. There is a long literature on this subject \cite{td96,jhv+06,vlj+11,ca92,cbs96,dv05,scm+06,smc07,bm10,oms10} based on many different assumptions for the string network evolution, radiation emission, and pulsar timing data. Often there are contradictory constraints based on the same data since some authors make very strong assumptions about the expected SGWB spectrum whereas others are more conservative. This is at best confusing to those uninitiated in the details of string evolution, something which this paper is aimed at clarifying. Therefore, we will take the view that none of the parameters  ($\alpha$, $q$ and $n_{*}$) which we have described in the earlier sections are known, extending the work done in \cite{cbs96}. The headline constraints which we will quote will be the highest upper bound possible for any parameter, this being the absolute constraint on $G\mu/c^2$ - one that is conservative and which we can never go back on! In addition we will present constraints for various specific models which have been discussed in the literature.

\begin{subsection}{EPTA constraint on $\Omega_{\rm gw}h^2$}

\begin{table}
\begin{tabular}{|l|c|}
\hline

Authors&$\Omega_{\rm{gw}}h^2$ bound\\
\hline
\hline
Bertotti, Carr, Rees \cite{bcr83}&$1.0\times10^{-3}\rm{@} 20\rm{nHz}$\\
\hline
Kaspi, Taylor, Ryba \cite{ktr94}&$6.0\times10^{-8}\rm{@}4.5\rm{nHz}$\\
\hline
Thorsett, Dewey \cite{td96}&$1.0\times 10^{-8}\rm{@}4.5\rm{nHz}$\\
\hline
McHugh, Zalamansky, Vertotte \cite{mzvl96}&$9.3\times 10^{-8}\rm{@}4.5\rm{nHz}$\\
\hline
Lommen \cite{lom02}&$2.0\times 10^{-9}\rm{@}1.9\rm{nHz}$\\
\hline
Jenet et al. \cite{jhv+06} (PPTA)&$1.9\times 10^{-8}\rm{
@}4.0\rm{nHz}$\\
\hline
van Haasteren et al. \cite{vlj+11}&$5.6\times 10^{-9}\rm{
@}4.0\rm{nHz}$\\
\hline
\end{tabular}
\caption{The $\Omega_{\rm gw}(f)h^2$ limits published in the literature so far. The Jenet et al. limit is derived for supermassive black holes ($d_{\Omega}=2/3$) and cosmic strings ($d_{\Omega}=-1/3$) as sources of the SGWB. The van Haasteren et al. limit quoted here, is for supermassive black holes at a frequency of $4.0\rm nHz$, for direct comparison with the Jenet et al. limit. The Thorsett and Dewey limit was criticized by McHugh et al. because of the statistical approach used. The Lommen limit, although the most stringent so far, was also criticized for similar reasons. \label{omegatable}}
\end{table}

The SGWB limit we will use is that which comes from the EPTA who have computed a 95\% exclusion curve for the SGWB amplitude $h_{\rm gw}$ as a function of the local slope, $d$, at $f=(1{\rm yr})^{-1}\approx32{\rm nHz}$. Previous work on the subject has typically quoted an upper limit on $\Omega_{\rm gw}h^2$ at a particular frequency. Over the last 30 years various data from PTA experiments have set constraints, as presented in Table \ref{omegatable}. The limits which we will discuss in the subsequent sections will use the full exclusion curve from \cite{vlj+11}. However, we shall follow a different approach in the limit predictions for the LEAP project since such an exclusion curve is obviously not available yet. In order to have a conservative projection for LEAP, we will use a spectral index $d_{\Omega}=0$ to make a projected bound. By making this choice we guarantee that the constraints on $\Omega_{\rm{gw}}(f)$ will be the most conservative applicable to cosmic strings for the majority of $G\mu/c^2 - \alpha$ combinations. Note that $d_{\Omega}<0$ in the radiation to matter transition epoch frequency range for any cosmic string model. We will come again to the  applicability of this idea when we will calculate the projected LEAP constraints. Moreover, the LEAP constraints are evaluated at a frequency of $(5{\rm yr})^{-1}$, which is calculated from the duration of a typical PTA experiment.

In \cite{vlj+11} the strain of the SGWB is described by a power law of the form $h_{\rm gw}=A(f/{\rm yr}^{-1})^d$ and an upper bound $A=6\times 10^{-15}$ was established in the case of supermassive black holes ($d=-2/3$) at $f=(1{\rm yr})^{-1}$. It is interesting to estimate the corresponding constraint on $\Omega_{\rm gw}h^2$ in order to confirm that the EPTA SGWB limit is the strongest possible one with which to perform our analysis. The previously published limit in \cite{jhv+06} (see Table~\ref{omegatable}), was quoted for $d=-2/3$ at $f=(8{\rm yr})^{-1}$. For the same parameter values, the projection of the EPTA limit gives
\begin{equation}
\Omega_{\rm{gw,EPTA}}(f)h^2\lesssim 5.6\times10^{-9},
\label{eptalimit}
\end{equation}
 at $95\%$ confidence level, a significant improvement in comparison to the Jenet et al. limit \cite{jhv+06}.

We note that there is a constraint on $G\mu/c^2$ presented in \cite{vlj+11}
 \begin{equation}
  G\mu/c^2<4.0\times10^{-9},
  \end{equation}
 that claims to  improve the previous one by Jenet et al. \cite{jhv+06} which was
    \begin{equation}
    G\mu/c^2<1.5\times10^{-8}.
    \nonumber
    \end{equation}
However,  both of these limits are based on the approach of Damour and Vilenkin \cite{dv05} who make strong assumptions about the amplitude and slope of the SGWB from strings. In particular, they describe the amplitude of the cosmic string loop generated SGWB using a model based on the emission from cusps on cosmic string loops with a slope $\Omega_{\rm gw}h^2\propto f^{-7/6}$. As we have already discussed, loop decay solely through cusp emission is a rather strong assumption, and the possible behavior of the cosmic string GW spectrum is much richer than this. In addition, the analytic results of \cite{dv05} are based on a simplified cosmic string evolution model (i.e., the loop number density $n(\ell,t)$ calculation). Damour and Vilenkin assume $n(t)\sim(\Gamma G\mu t^3/c^2)^{-1}$ independent of $\alpha$, whereas the analytic model of \cite{ben86a,ben86b} which is compatible with our computations yields $n(t)\sim(1+\Gamma G\mu\alpha^{-1}c^{-2})(\alpha t^3)^{-1}$. This agrees with that of Damour and Vilenkin when $\alpha =\Gamma G\mu/c^2$, that is, the loop production is at the gravitational backreaction scale.
\end{subsection}

\begin{subsection}{Conservative constraint on $G\mu/c^2$}

\begin{figure*}
\includegraphics[width=14 cm]{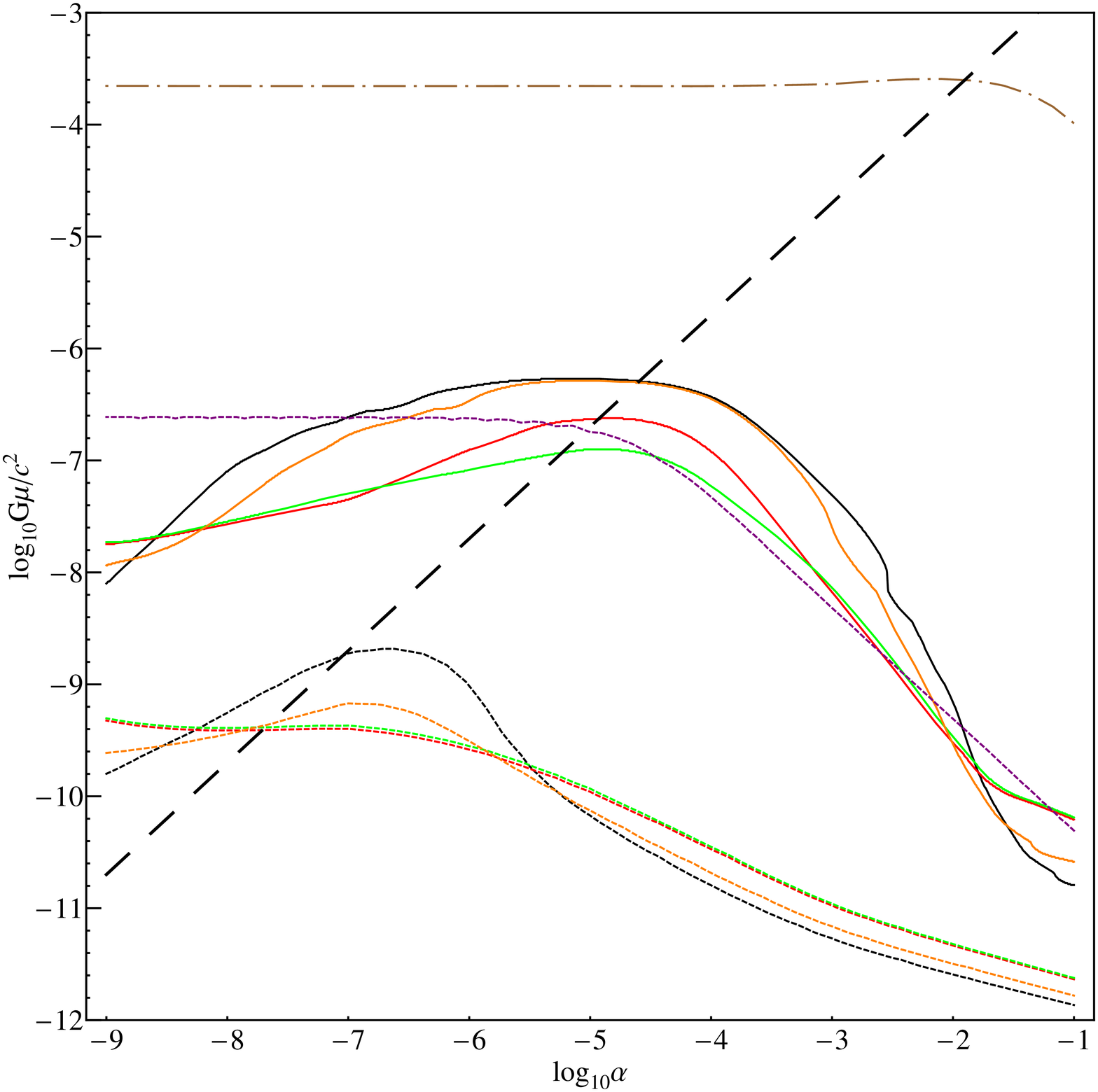}
\caption{Exclusion limits for different cosmic string network configurations with the same $\Omega_{\rm gw}h^2$ value at a frequency $f=(1{\rm yr})^{-1}$ in the $G\mu/c^2 - \alpha$ parameter space. The solid lines are for the EPTA $(1{\rm yr})^{-1}$ limit with $q=4/3$ and $n_{*}=1$ (black), $n_{*}=10^3$ (red), $n_{*}=10^4$ (green) and for $q=2$, $n_{*}=10^2$ (orange). The purple dashed curve is the analytic approximation of the flat part of the spectrum as described in \cite{bm10}, again for the EPTA limit on $\Omega_{\rm gw}h^2$. The dot-dashed brown line is the present LIGO limit. The constraint set by LIGO is almost independent of $n_{*}$ and $q$ and is also only very weakly dependent on $\alpha$. The dashed lines are the corresponding curves the planned LEAP sensitivity at a frequency $f=(5{\rm yr})^{-1}$. The thick dashed black line shows $\alpha=\Gamma G\mu/c^2$. \label{eptabounds}}
\end{figure*}

In order to calculate the constraints on the cosmic string tension, we have to find all the cosmic string network parameter combinations which lead to a SGWB amplitude which is in agreement with the EPTA SGWB limit at a frequency of $(1\,\rm yr)^{-1}$. For this, we computed the SGWB ($\Omega_{\rm gw}h^2(f)$) for more than 3000 parameter sets covering almost all the range of the theoretically expected values for $G\mu/c^2,\, \alpha,\, n_*, q$ and $p$. For each set of fixed $n_*,\, q$ and $p$, we deduced from our computations the quantity $\Omega_{\rm gw}(G\mu/c^2,\alpha,d_{\Omega})h^2$, which gives the amplitude of the SGWB as a function of $G\mu/c^2, \alpha$ and the local slope $d_{\Omega}$ at a frequency of $(1\,\rm yr)^{-1}$. The EPTA limit is given in the form of a $95\%$ exclusion curve of the form $h_{\rm{gw,EPTA}}(d)$. From this, we calculated the corresponding $\Omega_{\rm{gw,EPTA}}(d_{\Omega})h^2$ $95\%$ exclusion curve which would be applied to our results. The $G\mu/c^2-\alpha$ combinations which provide the constraint curve for each set of $n_*,\, q$ and $p$ in the $G\mu/c^2-\alpha$ parameter space was calculated by requiring $\Omega_{\rm gw}(G\mu/c^2,\alpha,d_{\Omega})h^2=\Omega_{\rm{gw,EPTA}}(d_{\Omega})h^2$.

In Fig.~\ref{eptabounds} we set $p=1$ and present constraints for various values of $q$ and $n_*$ which satisfy the EPTA limit. Specifically, we have plotted the cosmic string network families for $n_*=1$ (blue); $n_{*}=10^2$ and $q=4/3$ (red), $n_{*}=10^3$ and $q=4/3$ (green); $n_{*}=10^4$ and $q=4/3$ (black); $n_*=10^2$ and $q=2$ (orange). The dashed black line is $\alpha=\Gamma G\mu/c^2$ separating the large and the small loop production regions. The constraints for high-$n_{*}$ models are stronger than those with low-$n_{*}$ for most of the $\alpha-G\mu/c^2$ combinations except for the case of very small $G\mu/c^2$ or $\alpha$, where the opposite takes place. In the same figure we also present (purple dashed line) the constraints of the analytic approximation presented in \cite{bm10} that is just from the radiation era contribution. This is a good approximation to our results in the case of large loops, however it is only a conservative upper limit for lower values of $\alpha$.

The most conservative and generic constraint on the cosmic string tension can be set by the curves presented in Fig.~\ref{eptabounds}. This is provided by the cosmic string networks with $\alpha\approx 10^{-5}$ and $n_*=1$, and is
\begin{equation}
G\mu/c^2<5.3\times10^{-7}\,,
\end{equation}
which is a $95\%$ upper bound for this specific set of parameters.

 For comparison, we have also used the bound on the SGWB set by LIGO to obtain a constraint on the string tension. We expect this to be worse than the one set by PTAs not only because the limit on $\Omega_{\rm gw}h^2$ itself is smaller, but also because the LIGO frequency band ($\sim 1 \rm{kHz}$) is probing the radiation era part of the spectrum which has a smaller amplitude than the matter era equivalent. The LIGO limit is \cite{aaa+09}
\begin{equation}
\Omega_{\rm{gw,LIGO}}h^2<3.6\times 10^{-6}\,,
\end{equation}
at $95\%$ confidence level. In Fig.~\ref{eptabounds} with the dot-dashed brown line we show all the cosmic string configurations compatible with this bound at $1\rm{kHz}$. For $p=1$ the most conservative constraint is
\begin{equation}
G\mu/c^2\lesssim 2.6\times 10^{-4}\,.
\end{equation}
 Advanced LIGO is expected to make sufficient improvement, although it is unlikely to compete with PTAs in the short term. Conversely, one of its biggest advantages is that at these frequencies the spectrum is independent of the modeling of the string radiation spectrum, that is, $q$ and $n_*$.

The prospects of improving the PTA limit in the near future are very promising. As part of the EPTA, the LEAP \cite{fvb+10} project, is a large collaboration of the major telescopes of the EPTA members (Effelsberg Telescope, Lovell Telescope, Nancay Telescope, Westerbork Synthesis Radio Telescope, Sardinia Radio Telescope) which will use long-baseline array techniques to provide pulsar timing data equivalent to that of a 200-meter telescope. It is estimated that LEAP data will push the sensitivity in GWs down to $\Omega_{\rm{gw, LEAP}}(f)h^2<10^{-10}$ \cite{ks10} in a few years. In Fig.~\ref{eptabounds}, with dashed lines and the same color scheme as for the EPTA limits, we present the constraints one might expect from such a limit on the SGWB. If we assume that LEAP does not make a detection, then we would expect to set a limit of $G\mu/c^2<2.0\times 10^{-9}$; an improvement of more than two orders of magnitude. The careful reader will notice that for the LEAP constraint the order of the lines is reversed for high-$n_*$ values and the high-$n_*$ models are above the low-$n_*$ cases. This is expected, since we are in the low $\alpha-G\mu/c^2$ region of the parameter space and the PTA frequency window probes the part of the spectrum which is on the left side of the peak, where we can see from Fig.~\ref{modes} that configurations with high-$n_{*}$ give lower amplitudes.

\begin{figure*}[ht]
\includegraphics[width=14 cm]{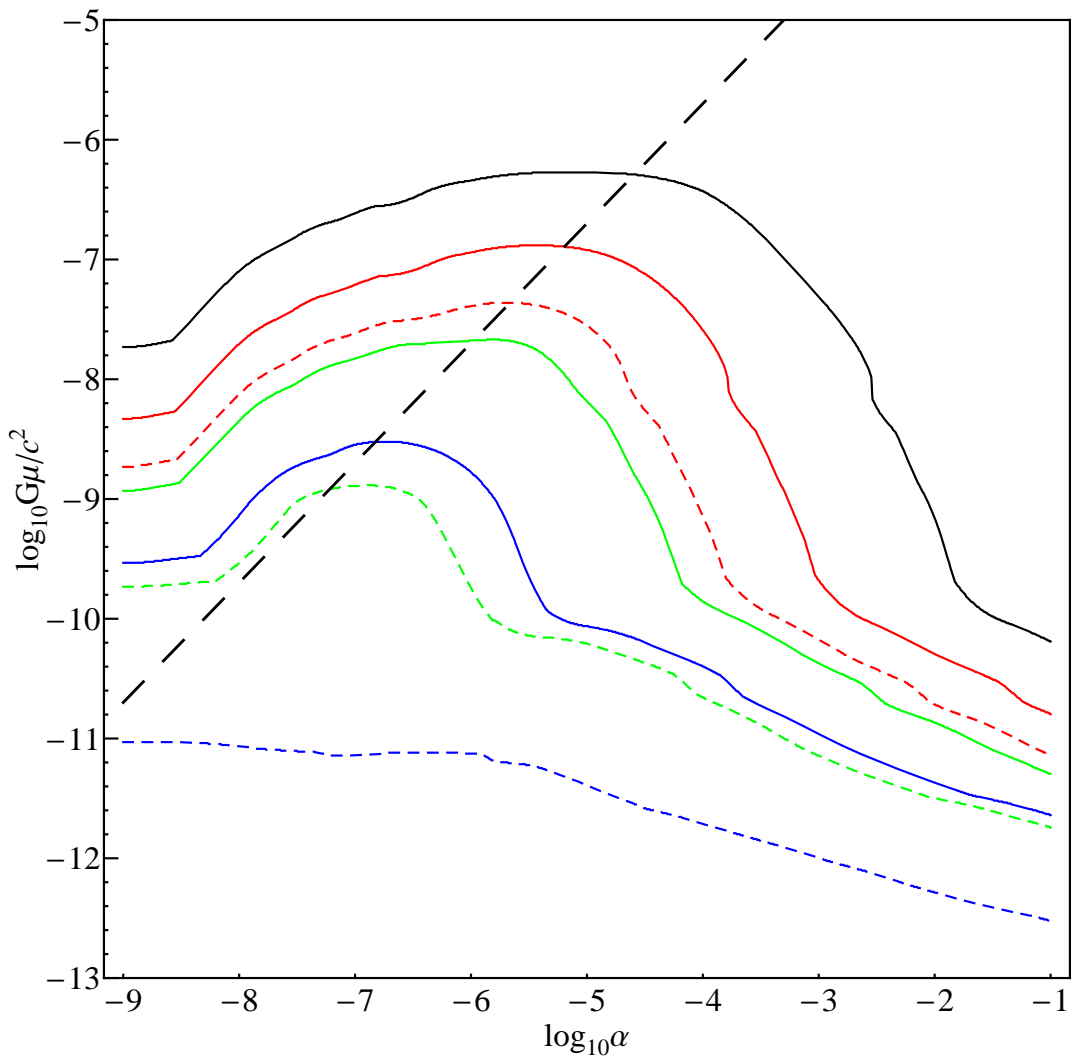}
\caption{The exclusion limits for cosmic string networks with different intercommutation probabilities and different scaling predictions ($\rho_{\infty}\propto p^{-1}$ and $\rho_{\infty}\propto p^{-0.6}$). These curves where created by combining the equivalent lines for $n_{*}=1$ (for most of the range of $\alpha$) and $n_{*}=10^4$ (for large/small $\alpha$). All of these network configurations give an amplitude equal to $\Omega_{\rm gw, EPTA}h^2$ at the frequency of $(1\,\rm yr)^{-1}$. The solid black line is for networks with $p=1$ and the solid red, blue and green are for networks with $p=0.1,10^{-2},10^{-3}$ respectively for $\rho_{\infty}\propto p^{-0.6}$. The dashed lines are the networks with $\rho_{\infty}\propto p^{-1}$ using the same color scheme. The thick dashed black line shows $\alpha=\Gamma G\mu/c^2$.
\label{icpbounds}}
\end{figure*}

We will conclude this section with some comments on our choice of a spectral index $d_{\Omega}=0$ for the calculation of the projected LEAP constraints. As we mentioned in Sec.~\ref{sec:emissionmodes}, the PTA window probes the area on the left of the peak (positive slope) for cosmic string networks with $G\mu/c^2<10^{-11}$. The constraint lines from the projected LEAP limit correspond to models with $G\mu/c^2>10^{-11}$ in the range $\alpha\in [10^{-9},10^{-3}]$. For all these models, indeed our assumption behaves well and moreover, it gives slightly overestimated values for $G\mu/c^2$ since the actual slope in this region is negative. The projected tension constraint is given by the cosmic string networks with $n_*=1$ (blue dashed line in Fig.~\ref{eptabounds}) which lie in this region, so our results are robust. On the other hand, in the region $\alpha\in[10^{-3},0.1]$ the cosmic string models have $G\mu/c^2<10^{-11}$ and their slope is positive. For these networks, our assumption is invalid and the tensions are slightly underestimated. This under-estimation, however, is not sufficient to overcome the much stronger constraints provided by the $n_*=1$ networks.

\end{subsection}

\begin{subsection}{Conservative tension constraints for $p\ne 1$}

We have used the EPTA $h_{\rm gw}$ $95\%$ exclusion curve to calculate conservative constraints on $G\mu/c^2$ for the cases of $p< 1$ for both scaling laws, $\rho_{\infty}\propto p^{-1}$ and $\rho_{\infty}\propto p^{-0.6}$ and the results are presented in Fig.~\ref{icpbounds}. In this plot, the curves correspond to the highest tension values only, meaning that a significant part of each curve is provided by the $n_*=1$ case (in approximately the mid-$\alpha$ range) and the rest by $n_*=10^4$ (in the low and high $\alpha$ ranges). The constraints for the $\rho_{\infty}=p^{-0.6}$ scaling law are plotted with solid red, blue and green lines for the cases of $p=0.1,10^{-2}$ and $10^{-3}$ respectively. The dashed lines are the corresponding results for the $\rho_{\infty}=p^{-1}$ scaling law. The conservative $G\mu/c^2$ constraints are presented in Table~\ref{icpcontable}. The results are indicative of how sensitive the tension constraints are, not only for the value of the intercommutation probability, but also to the exact scaling law as well.
\begin{table}[!hb]
\begin{tabular}{|l|c|c|}
\hline
\,\,\,  $p$&$G\mu/c^2$&$G\mu/c^2$\\
&($\rho_{\infty}=p^{-0.6}$)&($\rho_{\infty}=p^{-1}$)\\
\hline
\hline
$0.1$&$1.2\times 10^{-7}$&$4.4\times 10^{-8}$\\
\hline
$0.01$&$2.0\times 10^{-8}$&$1.3\times 10^{-9}$\\
\hline
$0.001$&$2.8\times 10^{-9}$&$9.3\times 10^{-12}$\\
\hline
\end{tabular}
\caption{Cosmic string tension constraints for models with $p< 1$ and for the two scaling laws discussed in the literature.\label{icpcontable}}
\end{table}

\end{subsection}

\begin{subsection}{Constraints on specific scenarios}

In addition to the previous, generic constraint on the string tension, we have also computed constraints for some specific cosmic string network scenarios popular in the literature. These models are based either on evolution specific simulations of cosmic string networks or theoretical arguments which estimate the birth scale of the loops, $\alpha$. Unfortunately, as we mentioned before no definite conclusion has been drawn yet. Depending on the value of $\alpha$ we can separate these models into four categories
\begin{enumerate}[i.]
\item Large loops: it has been suggested that most of the cosmic string loops are born with a size comparable to the horizon radius, with $\alpha \approx 0.1$ \cite{vov06,ov07}. Such simulations have been performed for both flat and expanding spacetimes. The production of small loops along with the large ones has also been observed in these simulations but it was suggested that they are a transient effect which disappears when the dynamic range of the simulations is increased. The most up-to-date simulations in this category are those in \cite{bos11} where a slightly different value of $\alpha \approx 0.05$ was suggested.
\item Intermediate loops: in some simulations \cite{ms06,rsb07} it has been argued against the creation of such large scale loops supporting the idea that the smaller loops are not a transient phenomenon but instead dominate the loop production. Both of these simulations were performed for expanding spacetimes and they concluded that $\alpha \approx 10^{-2}-10^{-3}$ without observing any significant production of large scale loops.
\item Loop production is governed by the gravitational backreaction scale : in this scenario cosmic string loops are created at the scale of gravitational backreaction $\alpha \approx \Gamma G \mu/c^2$ \cite{bb89a} as a direct result of the small-scale structure on the cosmic string network. In more recent simulations \cite{sov02} it has been proposed that the scale is even smaller and is $\alpha \approx (\Gamma G\mu/c^2)^k$ with $k=3/2$ in the radiation era and $k=5/2$ in the matter era.
\item Two scale loops: Some simulations \cite{dpr08,pol08} suggest that cosmic string loops are born with two different scales. Specifically, $10\%$ of the string energy converted to loops per unit time is channeled into loops with $\alpha \approx 0.1$ and the rest $90\%$ into loops with $\alpha \approx \Gamma G\mu/c^2$ scale loops.
\end{enumerate}
There are also simulations which suggest that cosmic string loops are born at a fixed size that does not scale with cosmic time and it is equal to the string width, that is $\alpha \approx 0$. Such tiny loops are expected to decay mainly through particle emission \cite{vah98} and are of no interest for this work. However, even if their GW emission is significant, it will take place at frequencies much higher than those probed by PTAs.

We have computed the GW spectra of cosmic string models in each of these scenarios in the tension range $G\mu/c^2\in[10^{-5},10^{-16}]$ and applied the EPTA $2\sigma$-limit on the SGWB at the frequency of $f_{\rm EPTA}=(1{\rm yr})^{-1}$ to set constraints on the string tension. The results for all categories are presented in Table~\ref{tensionstable}. Constraints for models with $\alpha \approx (\Gamma G\mu/c^2)^k$ have not been computed since their low frequency cut-off is at frequencies higher than those probed by PTAs, rendering them unobservable. In the categories (ii), (iii) and (iv), the most conservative constraint is given by $n_*=1$, whereas in case (i) it is given by $n_*=10^4$ (see in Fig.~\ref{eptabounds}, that these networks give higher tensions near $\alpha=0.1$).

In the case of large loop production which is favored by the most recent simulations, a very stringent limit of $G\mu/c^2<6.5\times10^{-11}$ is obtained. This limit is slightly weaker than the corresponding one found by the authors in \cite{bm10}, $G\mu/c^2<5.0\times 10^{-11}$ which is based on the analytic approximation of the amplitude of the radiation era part of the spectrum. In the regime $\alpha\ll \Gamma G\mu/c^2$, the smallest loops that PTAs are sensitive to have $\alpha\simeq 10^{-9}$. From Fig.~\ref{eptabounds} we get a constraint of $G\mu/c^2<1.9\times 10^{-8}$ for such loops.

\begin{table}[H]
\begin{center}
\begin{tabular}{|l|c|c|}
\hline
Loop scale ($\alpha$)&$G\mu/c^2$ bound&Ref.\\
\hline
\hline
$0.1$&$6.5\times10^{-11}$&\cite{vov06,ov07}\\
\hline
$0.05$&$8.8\times10^{-11}$&\cite{bos11}\\
\hline
$10^{-2}$
&$7.0\times 10^{-10}$&\cite{ms06,rsb07}\\
\hline
$\Gamma G\mu/c^2$&$5.3\times 10^{-7}$&\cite{bb89a}\\
\hline
$10\%\hspace{1mm}\alpha=0.1\hspace{2mm}+\hspace{2mm}90\%\hspace{1mm}\alpha=\Gamma G\mu/c^2$&$4.1\times 10^{-8}$&\cite{pol08,dpr08}\\
\hline
$\simeq10^{-9}$&$1.9\times 10^{-8}$& \\
\hline
\end{tabular}
\end{center}
\caption{Cosmic string tension limits for individual cosmic string models with specific $\alpha$ values predicted by particular simulations.\label{tensionstable}}
\end{table}

\end{subsection}

\end{section}


\begin{section}{CONCLUSIONS}

The constraints on the cosmic string tension from PTAs suffer from the many uncertainties concerning the parameters describing the loop production size, number density and GW emission properties. Particular treatments of the emission mechanism presented in the literature has led to different conflicting constraints \cite{scm+06,oms10,dh07} with the most stringent of them \cite{jhv+06,vlj+11} being based on rather strong assumptions \cite{dv00,dv01,dv05}. In this paper, we have expand on the work performed in \cite{cbs96}, performing a detailed investigation of the constraints for a wide range of scenarios. Using the recent limit on the SGWB at a frequency $f=(1{\rm yr})^{-1}$ set by the EPTA \cite{vlj+11}, we have managed to set an absolute, model independent upper limit on the string tension without making any assumptions for the emission properties. This limit is
\begin{equation}
G\mu/c^2<5.3\times 10^{-7}\,,
\nonumber
\end{equation}
at $95\%$ confidence level. Such an approach is particularly necessary in the case of PTAs, since the frequency range they probe is potentially sensitive to all the network and emission parameters. Additionally, we have calculated constraints for cosmic string networks with $p<1$ and for networks with specific loop birth scales which have been proposed in the literature.

In order to achieve this, we investigated the effects on the SGWB for each of the cosmic string model parameters in the range of values which are interesting for PTAs. This allowed us to delineate the fundamentally different SGWB behavior of cosmic string networks in the large and small loop production regimes, defined by the gravitational backreaction scale $\Gamma G\mu/c^2$. Of special interest to the PTAs is the low frequency cut-off, which can render the GW signatures of small loop size cosmic string networks undetectable even for high tension values. Additionally, we considered extensions to the standard one-scale model in order to investigate the differences in the SGWB created by more realistic cosmic string networks with a 2-scale and a log-normal loop production function.

It is worth comparing our limits to those placed on $G\mu/c^2$ from other observational approaches.
\begin{itemize}
\item CMB observations: cosmic strings could be responsible for a fraction of the CMB anisotropies due to the Kaiser-Stebbins effect \cite{ks84}. Data from various CMB experiments have been used to place constraints on the string tension, the most stringent of which are those set by combining 7-year Wilkinson Microwave Anisotropy Probe (WMAP) data and Atacama Cosmology Telescope (ACT) observations \cite{dhs+11}
    \begin{equation}
    G\mu/c^2<1.6\times10^{-7}\,,
    \nonumber
    \end{equation}
    at $95\%$ confidence level. This uses the methods described in \cite{bm10} designed to model the simulations of \cite{as90}. Weaker constraints are claimed in the case of Abelian-Higgs simulations \cite{ubhk11}, the most constraining of which is $G\mu<4.2\times10^{-7}$, derived using all available CMB data. See \cite{bm10} for a discussion of the veracity of these different approaches.
\item Gravitational lensing: cosmic strings can create gravitational lensing events due to the conical shape that spacetime acquires globally around them \cite{vil81c,vs94}. Such events have special features which can distinguish them from ``normal'' lenses (that is, no magnification, identical images, odd number of images) and can be observed in searches for gravitational lenses. The most recent constraint set from such searches was in \cite{cajg+08,caf+11} where the authors used archival data from the HST (GOODS and COSMOS surveys respectively) to search for lensing events that could have been caused by straight cosmic strings, managing to set a bound slightly weaker than the CMB one
\begin{equation}
G\mu/c^2 <3.0\times 10^{-7}\,,
\nonumber
\end{equation}
at $95\%$ confidence level. We note, that these constraints are also sensitive to some of the modeling issues which are important in the CMB case.
\end{itemize}

So far, these constraints were considered more reliable than those from PTAs, due to the many uncertainties concerning the cosmic string loop distribution and the GW emission assumptions of previous implementations. In this paper we believe that we have overcome all these difficulties and provide equally reliable, PTA constraints. In addition, the prospects for PTAs are much better than those of other experimental approaches. In the very near future, LEAP is expected to improve on the present upper limit on $G\mu/c^2$ by more than two orders of magnitude to $G\mu/c^2<2.0\times 10^{-9}$, if no detection of the SGWB takes place. This is a stronger constraint than any other expected from present or near future CMB experiments. The projected limit for {\it Planck} satellite \cite{tmp+10} is $G\mu/c^2<6.5\times 10^{-8}$ \cite{bgms08}, approximately one order of magnitude better than the present one. In \cite{fms11,dwh11}, the authors come to similar pessimistic conclusions about the prospects of present and future planned CMB experiments to detect CMB polarization.

Future PTA experiments could provide an unprecedented insight on the physics of cosmic strings. It is clear that projects like LEAP and the growing International Pulsar Timing Array \cite{haa+10} will likely make a detection of the SGWB from supermassive black hole binaries on the timescale of the next five years. However, the sensitivity that will be achieved with the next generation radio telescope, the SKA \citep{dhsl09}, will give us the opportunity to study the GW spectrum in detail. As it has been shown in this paper for cosmic strings and for supermassive black holes in \cite{sv10}, measuring the shape of the spectrum is essential to be able to both distinguish between the source of the SGWB and also to extract more information about the source. In the case of supermassive black hole binaries, information about the supermassive black hole population and the evolution of binary systems can be extracted. In the case of cosmic strings, we could potentially determine all the fundamental model parameters (loop birth scale $\alpha$, intercommutation probability $p$) and get a definitive answer about the exact GW emission mechanism and small scale structure of cosmic string loops (i.e. cusps or kinks).

\end{section}

\begin{acknowledgments}
The authors would like to thank all the EPTA and LEAP members for useful discussions and suggestions during the duration of this work and especially Rutger van Haasteren for providing the data file for the EPTA $(1\,{\rm yr)^{-1}}$ limit on the amplitude of the SGWB as a function of the spectral index. We would also like to thank Paul Shellard, Xavier Siemens and Alkistis Pourtsidou for helpful discussions.

\end{acknowledgments}


\begin{thebibliography}{123}
\expandafter\ifx\csname natexlab\endcsname\relax\def\natexlab#1{#1}\fi
\expandafter\ifx\csname bibnamefont\endcsname\relax
  \def\bibnamefont#1{#1}\fi
\expandafter\ifx\csname bibfnamefont\endcsname\relax
  \def\bibfnamefont#1{#1}\fi
\expandafter\ifx\csname citenamefont\endcsname\relax
  \def\citenamefont#1{#1}\fi
\expandafter\ifx\csname url\endcsname\relax
  \def\url#1{\texttt{#1}}\fi
\expandafter\ifx\csname urlprefix\endcsname\relax\def\urlprefix{URL }\fi
\providecommand{\bibinfo}[2]{#2}
\providecommand{\eprint}[2][]{\url{#2}}

\bibitem[{\citenamefont{Vilenkin and Shellard}(1994)}]{vs94}
\bibinfo{author}{\bibfnamefont{A.}~\bibnamefont{Vilenkin}} \bibnamefont{and}
  \bibinfo{author}{\bibfnamefont{E.~P.~S.} \bibnamefont{Shellard}},
  \emph{\bibinfo{title}{Cosmic Strings and Other Topological Defects}}
  (\bibinfo{publisher}{Cambridge University Press},
  \bibinfo{address}{Cambridge}, \bibinfo{year}{1994}).

\bibitem[{\citenamefont{Kibble}(1976)}]{kib76}
\bibinfo{author}{\bibfnamefont{T.~W.~B.} \bibnamefont{Kibble}},
  \bibinfo{journal}{J. Phys. A} \textbf{\bibinfo{volume}{9}},
  \bibinfo{pages}{1387} (\bibinfo{year}{1976}).

\bibitem[{\citenamefont{Vilenkin}(1981{\natexlab{a}})}]{vil81a}
\bibinfo{author}{\bibfnamefont{A.}~\bibnamefont{Vilenkin}},
  \bibinfo{journal}{Phys. Rev. Lett.} \textbf{\bibinfo{volume}{46, 17}},
  \bibinfo{pages}{1169} (\bibinfo{year}{1981}{\natexlab{a}}).

\bibitem[{\citenamefont{Silk and Vilenkin}(1984)}]{sv84}
\bibinfo{author}{\bibfnamefont{J.}~\bibnamefont{Silk}} \bibnamefont{and}
  \bibinfo{author}{\bibfnamefont{A.}~\bibnamefont{Vilenkin}},
  \bibinfo{journal}{Phys. Rev. Lett.} \textbf{\bibinfo{volume}{53, 17}},
  \bibinfo{pages}{1700} (\bibinfo{year}{1984}).

\bibitem[{\citenamefont{Turok}(1984)}]{tur84}
\bibinfo{author}{\bibfnamefont{N.}~\bibnamefont{Turok}}, \bibinfo{journal}{Nuc.
  Phys. B} \textbf{\bibinfo{volume}{242}}, \bibinfo{pages}{520}
  (\bibinfo{year}{1984}).

\bibitem[{\citenamefont{Bouchet and Bennett}(1990{\natexlab{a}})}]{bb90b}
\bibinfo{author}{\bibfnamefont{F.~R.} \bibnamefont{Bouchet}} \bibnamefont{and}
  \bibinfo{author}{\bibfnamefont{D.}~\bibnamefont{Bennett}},
  \bibinfo{journal}{ApJ} \textbf{\bibinfo{volume}{354}}, \bibinfo{pages}{L41}
  (\bibinfo{year}{1990}{\natexlab{a}}).

\bibitem[{\citenamefont{Albrecht et~al.}(1997)\citenamefont{Albrecht, Battye,
  and Robinson}}]{abr97}
\bibinfo{author}{\bibfnamefont{A.}~\bibnamefont{Albrecht}},
  \bibinfo{author}{\bibfnamefont{R.}~\bibnamefont{Battye}}, \bibnamefont{and}
  \bibinfo{author}{\bibfnamefont{J.}~\bibnamefont{Robinson}},
  \bibinfo{journal}{Phys. Rev. Lett.} \textbf{\bibinfo{volume}{79, 24}},
  \bibinfo{pages}{4736} (\bibinfo{year}{1997}).

\bibitem[{\citenamefont{Albrecht et~al.}(1999)\citenamefont{Albrecht, Battye,
  and Robinson}}]{abr99}
\bibinfo{author}{\bibfnamefont{A.}~\bibnamefont{Albrecht}},
  \bibinfo{author}{\bibfnamefont{R.}~\bibnamefont{Battye}}, \bibnamefont{and}
  \bibinfo{author}{\bibfnamefont{J.}~\bibnamefont{Robinson}},
  \bibinfo{journal}{Phys. Rev. D} \textbf{\bibinfo{volume}{69, 023508}}
  (\bibinfo{year}{1999}).

\bibitem[{\citenamefont{Pogosian}(2001)}]{pog01}
\bibinfo{author}{\bibfnamefont{L.}~\bibnamefont{Pogosian}},
  \bibinfo{journal}{Int. J. Mod. Phys. A} \textbf{\bibinfo{volume}{16}},
  \bibinfo{pages}{1043} (\bibinfo{year}{2001}).

\bibitem[{\citenamefont{{Battye} et~al.}(2006)\citenamefont{{Battye},
  {Garbrecht}, and {Moss}}}]{bgm06}
\bibinfo{author}{\bibfnamefont{R.~A.} \bibnamefont{{Battye}}},
  \bibinfo{author}{\bibfnamefont{B.}~\bibnamefont{{Garbrecht}}},
  \bibnamefont{and} \bibinfo{author}{\bibfnamefont{A.}~\bibnamefont{{Moss}}},
  \bibinfo{journal}{JCAP} \textbf{\bibinfo{volume}{9}}, \bibinfo{pages}{7}
  (\bibinfo{year}{2006}).

\bibitem[{\citenamefont{{Bevis} et~al.}(2008)\citenamefont{{Bevis},
  {Hindmarsh}, {Kunz}, and {Urrestilla}}}]{bhku08}
\bibinfo{author}{\bibfnamefont{N.}~\bibnamefont{{Bevis}}},
  \bibinfo{author}{\bibfnamefont{M.}~\bibnamefont{{Hindmarsh}}},
  \bibinfo{author}{\bibfnamefont{M.}~\bibnamefont{{Kunz}}}, \bibnamefont{and}
  \bibinfo{author}{\bibfnamefont{J.}~\bibnamefont{{Urrestilla}}},
  \bibinfo{journal}{Phys. Rev. Lett.} \textbf{\bibinfo{volume}{100}},
  \bibinfo{pages}{021301} (\bibinfo{year}{2008}).

\bibitem[{\citenamefont{Battye and Moss}(2010)}]{bm10}
\bibinfo{author}{\bibfnamefont{R.}~\bibnamefont{Battye}} \bibnamefont{and}
  \bibinfo{author}{\bibfnamefont{A.}~\bibnamefont{Moss}},
  \bibinfo{journal}{Phys. Rev. D} \textbf{\bibinfo{volume}{82, 023521}}
  (\bibinfo{year}{2010}).

\bibitem[{\citenamefont{Jeannerot}(1996)}]{jea96}
\bibinfo{author}{\bibfnamefont{R.}~\bibnamefont{Jeannerot}},
  \bibinfo{journal}{Phys. Rev. D} \textbf{\bibinfo{volume}{53, 10}},
  \bibinfo{pages}{5426} (\bibinfo{year}{1996}).

\bibitem[{\citenamefont{Jeannerot et~al.}(2003)\citenamefont{Jeannerot, Rocher,
  and Sakellariadou}}]{jrs03}
\bibinfo{author}{\bibfnamefont{R.}~\bibnamefont{Jeannerot}},
  \bibinfo{author}{\bibfnamefont{J.}~\bibnamefont{Rocher}}, \bibnamefont{and}
  \bibinfo{author}{\bibfnamefont{M.}~\bibnamefont{Sakellariadou}},
  \bibinfo{journal}{Phys. Rev. D} \textbf{\bibinfo{volume}{68, 103514}},
  \bibinfo{pages}{1} (\bibinfo{year}{2003}).

\bibitem[{\citenamefont{{Kachru} et~al.}(2003)\citenamefont{{Kachru},
  {Kallosh}, {Linde}, {Maldacena}, {McAllister}, and {Trivedi}}}]{kkl+03}
\bibinfo{author}{\bibfnamefont{S.}~\bibnamefont{{Kachru}}},
  \bibinfo{author}{\bibfnamefont{R.}~\bibnamefont{{Kallosh}}},
  \bibinfo{author}{\bibfnamefont{A.}~\bibnamefont{{Linde}}},
  \bibinfo{author}{\bibfnamefont{J.}~\bibnamefont{{Maldacena}}},
  \bibinfo{author}{\bibfnamefont{L.}~\bibnamefont{{McAllister}}},
  \bibnamefont{and} \bibinfo{author}{\bibfnamefont{S.~P.}
  \bibnamefont{{Trivedi}}}, \bibinfo{journal}{JCAP}
  \textbf{\bibinfo{volume}{10}}, \bibinfo{pages}{13} (\bibinfo{year}{2003}).

\bibitem[{\citenamefont{{Sarangi} and {Tye}}(2002)}]{st02}
\bibinfo{author}{\bibfnamefont{S.}~\bibnamefont{{Sarangi}}} \bibnamefont{and}
  \bibinfo{author}{\bibfnamefont{S.-H.~H.} \bibnamefont{{Tye}}},
  \bibinfo{journal}{Phys. Lett. B} \textbf{\bibinfo{volume}{536}},
  \bibinfo{pages}{185} (\bibinfo{year}{2002}).

\bibitem[{\citenamefont{{Dvali} and {Vilenkin}}(2004)}]{dv04a}
\bibinfo{author}{\bibfnamefont{G.}~\bibnamefont{{Dvali}}} \bibnamefont{and}
  \bibinfo{author}{\bibfnamefont{A.}~\bibnamefont{{Vilenkin}}},
  \bibinfo{journal}{JCAP} \textbf{\bibinfo{volume}{3}}, \bibinfo{pages}{10}
  (\bibinfo{year}{2004}).

\bibitem[{\citenamefont{{Copeland} et~al.}(2004)\citenamefont{{Copeland},
  {Myers}, and {Polchinski}}}]{cmp04}
\bibinfo{author}{\bibfnamefont{E.~J.} \bibnamefont{{Copeland}}},
  \bibinfo{author}{\bibfnamefont{R.~C.} \bibnamefont{{Myers}}},
  \bibnamefont{and}
  \bibinfo{author}{\bibfnamefont{J.}~\bibnamefont{{Polchinski}}},
  \bibinfo{journal}{J. High Energy Phys.} \textbf{\bibinfo{volume}{6}},
  \bibinfo{pages}{13} (\bibinfo{year}{2004}).

\bibitem[{\citenamefont{Jones et~al.}(2003)\citenamefont{Jones, Stoica, and
  Tye}}]{jst03}
\bibinfo{author}{\bibfnamefont{N.~T.} \bibnamefont{Jones}},
  \bibinfo{author}{\bibfnamefont{H.}~\bibnamefont{Stoica}}, \bibnamefont{and}
  \bibinfo{author}{\bibfnamefont{S.~H.~H.} \bibnamefont{Tye}},
  \bibinfo{journal}{Phys. Lett. B} \textbf{\bibinfo{volume}{563}},
  \bibinfo{pages}{6} (\bibinfo{year}{2003}).

\bibitem[{\citenamefont{Bhattacharjee and Sigl}(2000)}]{bs00}
\bibinfo{author}{\bibfnamefont{P.}~\bibnamefont{Bhattacharjee}}
  \bibnamefont{and} \bibinfo{author}{\bibfnamefont{G.}~\bibnamefont{Sigl}},
  \bibinfo{journal}{Phys. Rep.} \textbf{\bibinfo{volume}{327}},
  \bibinfo{pages}{109} (\bibinfo{year}{2000}).

\bibitem[{\citenamefont{Vachaspati}(2010)}]{vac10}
\bibinfo{author}{\bibfnamefont{T.}~\bibnamefont{Vachaspati}},
  \bibinfo{journal}{Phys. Rev. D} \textbf{\bibinfo{volume}{81, 043531}},
  \bibinfo{pages}{1} (\bibinfo{year}{2010}).

\bibitem[{\citenamefont{{Babul} et~al.}(1987)\citenamefont{{Babul},
  {Paczynski}, and {Spergel}}}]{bps87}
\bibinfo{author}{\bibfnamefont{A.}~\bibnamefont{{Babul}}},
  \bibinfo{author}{\bibfnamefont{B.}~\bibnamefont{{Paczynski}}},
  \bibnamefont{and}
  \bibinfo{author}{\bibfnamefont{D.}~\bibnamefont{{Spergel}}},
  \bibinfo{journal}{ApJ} \textbf{\bibinfo{volume}{316}}, \bibinfo{pages}{49}
  (\bibinfo{year}{1987}).

\bibitem[{\citenamefont{Berezinsky et~al.}(2001)\citenamefont{Berezinsky,
  Hnatyk, and Vilenkin}}]{bhv01}
\bibinfo{author}{\bibfnamefont{V.}~\bibnamefont{Berezinsky}},
  \bibinfo{author}{\bibfnamefont{B.}~\bibnamefont{Hnatyk}}, \bibnamefont{and}
  \bibinfo{author}{\bibfnamefont{A.}~\bibnamefont{Vilenkin}},
  \bibinfo{journal}{Phys. Rev. D} \textbf{\bibinfo{volume}{64}}
  (\bibinfo{year}{2001}).

\bibitem[{\citenamefont{Vachaspati}(2008)}]{vac08}
\bibinfo{author}{\bibfnamefont{T.}~\bibnamefont{Vachaspati}},
  \bibinfo{journal}{Phys. Rev. Lett.} \textbf{\bibinfo{volume}{101, 141301}}
  (\bibinfo{year}{2008}).

\bibitem[{\citenamefont{Chudnovsky et~al.}(1986)\citenamefont{Chudnovsky,
  Field, Spergel, and Vilenkin}}]{cfsv86}
\bibinfo{author}{\bibfnamefont{E.~M.} \bibnamefont{Chudnovsky}},
  \bibinfo{author}{\bibfnamefont{G.~B.} \bibnamefont{Field}},
  \bibinfo{author}{\bibfnamefont{D.~N.} \bibnamefont{Spergel}},
  \bibnamefont{and} \bibinfo{author}{\bibfnamefont{A.}~\bibnamefont{Vilenkin}},
  \bibinfo{journal}{Phys. Rev. D} \textbf{\bibinfo{volume}{34, 4}},
  \bibinfo{pages}{944} (\bibinfo{year}{1986}).

\bibitem[{\citenamefont{Chu et~al.}(2010)\citenamefont{Chu, Mathur, and
  Vachaspati}}]{cmv10}
\bibinfo{author}{\bibfnamefont{Y.}~\bibnamefont{Chu}},
  \bibinfo{author}{\bibfnamefont{H.}~\bibnamefont{Mathur}}, \bibnamefont{and}
  \bibinfo{author}{\bibfnamefont{T.}~\bibnamefont{Vachaspati}},
  \bibinfo{journal}{Phys. Rev. D} \textbf{\bibinfo{volume}{82, 063515}},
  \bibinfo{pages}{1} (\bibinfo{year}{2010}).

\bibitem[{\citenamefont{Jones-Smith et~al.}(2010)\citenamefont{Jones-Smith,
  Mathur, and Vachaspati}}]{jmv10}
\bibinfo{author}{\bibfnamefont{K.}~\bibnamefont{Jones-Smith}},
  \bibinfo{author}{\bibfnamefont{H.}~\bibnamefont{Mathur}}, \bibnamefont{and}
  \bibinfo{author}{\bibfnamefont{T.}~\bibnamefont{Vachaspati}},
  \bibinfo{journal}{Phys. Rev. D} \textbf{\bibinfo{volume}{81, 043501}},
  \bibinfo{pages}{1} (\bibinfo{year}{2010}).

\bibitem[{\citenamefont{Vilenkin}(1981{\natexlab{b}})}]{vil81c}
\bibinfo{author}{\bibfnamefont{A.}~\bibnamefont{Vilenkin}},
  \bibinfo{journal}{Phys. Rev. D} \textbf{\bibinfo{volume}{23, 4}},
  \bibinfo{pages}{852} (\bibinfo{year}{1981}{\natexlab{b}}).

\bibitem[{\citenamefont{Gasparini et~al.}(2008)\citenamefont{Gasparini,
  Marshall, Treu, Morganson, and Dubath}}]{gmtm+08}
\bibinfo{author}{\bibfnamefont{M.~A.} \bibnamefont{Gasparini}},
  \bibinfo{author}{\bibfnamefont{P.}~\bibnamefont{Marshall}},
  \bibinfo{author}{\bibfnamefont{T.}~\bibnamefont{Treu}},
  \bibinfo{author}{\bibfnamefont{E.}~\bibnamefont{Morganson}},
  \bibnamefont{and} \bibinfo{author}{\bibfnamefont{F.}~\bibnamefont{Dubath}},
  \bibinfo{journal}{MNRAS} \textbf{\bibinfo{volume}{389}},
  \bibinfo{pages}{1959} (\bibinfo{year}{2008}).

\bibitem[{\citenamefont{Kuijken et~al.}(2008)\citenamefont{Kuijken, Siemens,
  and Vachaspati}}]{ksv08}
\bibinfo{author}{\bibfnamefont{K.}~\bibnamefont{Kuijken}},
  \bibinfo{author}{\bibfnamefont{X.}~\bibnamefont{Siemens}}, \bibnamefont{and}
  \bibinfo{author}{\bibfnamefont{T.}~\bibnamefont{Vachaspati}},
  \bibinfo{journal}{MNRAS} \textbf{\bibinfo{volume}{384}}, \bibinfo{pages}{161}
  (\bibinfo{year}{2008}).

\bibitem[{\citenamefont{Kaiser and Stebbins}(1984)}]{ks84}
\bibinfo{author}{\bibfnamefont{N.}~\bibnamefont{Kaiser}} \bibnamefont{and}
  \bibinfo{author}{\bibfnamefont{A.}~\bibnamefont{Stebbins}},
  \bibinfo{journal}{Nature} \textbf{\bibinfo{volume}{310}},
  \bibinfo{pages}{391} (\bibinfo{year}{1984}).

\bibitem[{\citenamefont{{Turok} et~al.}(1998)\citenamefont{{Turok}, {Pen}, and
  {Seljak}}}]{tps98}
\bibinfo{author}{\bibfnamefont{N.}~\bibnamefont{{Turok}}},
  \bibinfo{author}{\bibfnamefont{U.-L.} \bibnamefont{{Pen}}}, \bibnamefont{and}
  \bibinfo{author}{\bibfnamefont{U.}~\bibnamefont{{Seljak}}},
  \bibinfo{journal}{Phys. Rev. D} \textbf{\bibinfo{volume}{58}},
  \bibinfo{pages}{023506} (\bibinfo{year}{1998}).

\bibitem[{\citenamefont{{Battye}}(1998)}]{bat98}
\bibinfo{author}{\bibfnamefont{R.~A.} \bibnamefont{{Battye}}}, in
  \emph{\bibinfo{booktitle}{Fundamental Parameters in Cosmology, Recontres de
  Moriond}} (\bibinfo{year}{1998}), \eprint{arXiv:astro-ph/9806115}.

\bibitem[{\citenamefont{{Bevis} et~al.}(2007)\citenamefont{{Bevis},
  {Hindmarsh}, {Kunz}, and {Urrestilla}}}]{bhku07}
\bibinfo{author}{\bibfnamefont{N.}~\bibnamefont{{Bevis}}},
  \bibinfo{author}{\bibfnamefont{M.}~\bibnamefont{{Hindmarsh}}},
  \bibinfo{author}{\bibfnamefont{M.}~\bibnamefont{{Kunz}}}, \bibnamefont{and}
  \bibinfo{author}{\bibfnamefont{J.}~\bibnamefont{{Urrestilla}}},
  \bibinfo{journal}{Phys. Rev. D} \textbf{\bibinfo{volume}{76}},
  \bibinfo{pages}{043005} (\bibinfo{year}{2007}).

\bibitem[{\citenamefont{Fraisse et~al.}(2008)\citenamefont{Fraisse, Ringeval,
  Spergel, and Bouchet}}]{frsb08}
\bibinfo{author}{\bibfnamefont{A.~A.} \bibnamefont{Fraisse}},
  \bibinfo{author}{\bibfnamefont{C.}~\bibnamefont{Ringeval}},
  \bibinfo{author}{\bibfnamefont{D.}~\bibnamefont{Spergel}}, \bibnamefont{and}
  \bibinfo{author}{\bibfnamefont{F.}~\bibnamefont{Bouchet}},
  \bibinfo{journal}{Phys. Rev. D} \textbf{\bibinfo{volume}{78, 043535}},
  \bibinfo{pages}{1} (\bibinfo{year}{2008}).

\bibitem[{\citenamefont{{Pogosian} and {Wyman}}(2008)}]{pw08}
\bibinfo{author}{\bibfnamefont{L.}~\bibnamefont{{Pogosian}}} \bibnamefont{and}
  \bibinfo{author}{\bibfnamefont{M.}~\bibnamefont{{Wyman}}},
  \bibinfo{journal}{Phys. Rev. D} \textbf{\bibinfo{volume}{77}},
  \bibinfo{pages}{083509} (\bibinfo{year}{2008}).

\bibitem[{\citenamefont{Takahashi et~al.}(2009)\citenamefont{Takahashi, Nakuro,
  Sendouda, Yamauchi, Yoo, and Sasaki}}]{tnsy+09}
\bibinfo{author}{\bibfnamefont{K.}~\bibnamefont{Takahashi}},
  \bibinfo{author}{\bibfnamefont{A.}~\bibnamefont{Nakuro}},
  \bibinfo{author}{\bibfnamefont{Y.}~\bibnamefont{Sendouda}},
  \bibinfo{author}{\bibfnamefont{D.}~\bibnamefont{Yamauchi}},
  \bibinfo{author}{\bibfnamefont{C.}~\bibnamefont{Yoo}}, \bibnamefont{and}
  \bibinfo{author}{\bibfnamefont{M.}~\bibnamefont{Sasaki}},
  \bibinfo{journal}{JCAP} \textbf{\bibinfo{volume}{10, 003}},
  \bibinfo{pages}{1} (\bibinfo{year}{2009}).

\bibitem[{\citenamefont{{Garc{\'{\i}}a-Bellido}
  et~al.}(2011)\citenamefont{{Garc{\'{\i}}a-Bellido}, {Durrer}, {Fenu},
  {Figueroa}, and {Kunz}}}]{gdf+11}
\bibinfo{author}{\bibfnamefont{J.}~\bibnamefont{{Garc{\'{\i}}a-Bellido}}},
  \bibinfo{author}{\bibfnamefont{R.}~\bibnamefont{{Durrer}}},
  \bibinfo{author}{\bibfnamefont{E.}~\bibnamefont{{Fenu}}},
  \bibinfo{author}{\bibfnamefont{D.~G.} \bibnamefont{{Figueroa}}},
  \bibnamefont{and} \bibinfo{author}{\bibfnamefont{M.}~\bibnamefont{{Kunz}}},
  \bibinfo{journal}{Phys. Lett. B} \textbf{\bibinfo{volume}{695}},
  \bibinfo{pages}{26} (\bibinfo{year}{2011}).

\bibitem[{\citenamefont{{Brandenberger}
  et~al.}(2010)\citenamefont{{Brandenberger}, {Danos}, {Hern{\'a}ndez}, and
  {Holder}}}]{bdhh10}
\bibinfo{author}{\bibfnamefont{R.~H.} \bibnamefont{{Brandenberger}}},
  \bibinfo{author}{\bibfnamefont{R.~J.} \bibnamefont{{Danos}}},
  \bibinfo{author}{\bibfnamefont{O.~F.} \bibnamefont{{Hern{\'a}ndez}}},
  \bibnamefont{and} \bibinfo{author}{\bibfnamefont{G.~P.}
  \bibnamefont{{Holder}}}, \bibinfo{journal}{JCAP}
  \textbf{\bibinfo{volume}{12}}, \bibinfo{pages}{28} (\bibinfo{year}{2010}).

\bibitem[{\citenamefont{Khatri and Wandelt}(2008)}]{kw08}
\bibinfo{author}{\bibfnamefont{R.}~\bibnamefont{Khatri}} \bibnamefont{and}
  \bibinfo{author}{\bibfnamefont{B.}~\bibnamefont{Wandelt}},
  \bibinfo{journal}{Phys. Rev. Lett.} \textbf{\bibinfo{volume}{100, 091302}}
  (\bibinfo{year}{2008}).

\bibitem[{\citenamefont{Vilenkin}(1981{\natexlab{c}})}]{vil81d}
\bibinfo{author}{\bibfnamefont{A.}~\bibnamefont{Vilenkin}},
  \bibinfo{journal}{Phys. Lett. B} \textbf{\bibinfo{volume}{107}},
  \bibinfo{pages}{47} (\bibinfo{year}{1981}{\natexlab{c}}).

\bibitem[{\citenamefont{Hogan and Rees}(1984)}]{hr84}
\bibinfo{author}{\bibfnamefont{C.~J.} \bibnamefont{Hogan}} \bibnamefont{and}
  \bibinfo{author}{\bibfnamefont{M.~J.} \bibnamefont{Rees}},
  \bibinfo{journal}{Nature} \textbf{\bibinfo{volume}{311}},
  \bibinfo{pages}{109} (\bibinfo{year}{1984}).

\bibitem[{\citenamefont{Vachaspati and Vilenkin}(1985)}]{vv85}
\bibinfo{author}{\bibfnamefont{T.}~\bibnamefont{Vachaspati}} \bibnamefont{and}
  \bibinfo{author}{\bibfnamefont{A.}~\bibnamefont{Vilenkin}},
  \bibinfo{journal}{Phys. Rev. B} \textbf{\bibinfo{volume}{31}},
  \bibinfo{pages}{3052} (\bibinfo{year}{1985}).

\bibitem[{\citenamefont{{Brandenberger}
  et~al.}(1986)\citenamefont{{Brandenberger}, {Albrecht}, and {Turok}}}]{bat86}
\bibinfo{author}{\bibfnamefont{R.~H.} \bibnamefont{{Brandenberger}}},
  \bibinfo{author}{\bibfnamefont{A.}~\bibnamefont{{Albrecht}}},
  \bibnamefont{and} \bibinfo{author}{\bibfnamefont{N.}~\bibnamefont{{Turok}}},
  \bibinfo{journal}{Nucl. Phys. B} \textbf{\bibinfo{volume}{277}},
  \bibinfo{pages}{605} (\bibinfo{year}{1986}).

\bibitem[{\citenamefont{{Accetta} and {Krauss}}(1989)}]{ac89}
\bibinfo{author}{\bibfnamefont{F.~S.} \bibnamefont{{Accetta}}}
  \bibnamefont{and} \bibinfo{author}{\bibfnamefont{L.~M.}
  \bibnamefont{{Krauss}}}, \bibinfo{journal}{Nucl. Phys. B}
  \textbf{\bibinfo{volume}{319}}, \bibinfo{pages}{747} (\bibinfo{year}{1989}).

\bibitem[{\citenamefont{Bouchet and Bennett}(1990{\natexlab{b}})}]{bb90a}
\bibinfo{author}{\bibfnamefont{F.~R.} \bibnamefont{Bouchet}} \bibnamefont{and}
  \bibinfo{author}{\bibfnamefont{D.~P.} \bibnamefont{Bennett}},
  \bibinfo{journal}{Phys. Rev. D} \textbf{\bibinfo{volume}{41}},
  \bibinfo{pages}{720} (\bibinfo{year}{1990}{\natexlab{b}}).

\bibitem[{\citenamefont{Caldwell and Allen}(1992)}]{ca92}
\bibinfo{author}{\bibfnamefont{R.~R.} \bibnamefont{Caldwell}} \bibnamefont{and}
  \bibinfo{author}{\bibfnamefont{B.}~\bibnamefont{Allen}},
  \bibinfo{journal}{Phys. Rev. D} \textbf{\bibinfo{volume}{45}},
  \bibinfo{pages}{3447} (\bibinfo{year}{1992}).

\bibitem[{\citenamefont{Caldwell et~al.}(1996)\citenamefont{Caldwell, Battye,
  and Shellard}}]{cbs96}
\bibinfo{author}{\bibfnamefont{R.~R.} \bibnamefont{Caldwell}},
  \bibinfo{author}{\bibfnamefont{R.~A.} \bibnamefont{Battye}},
  \bibnamefont{and} \bibinfo{author}{\bibfnamefont{E.~P.~S.}
  \bibnamefont{Shellard}}, \bibinfo{journal}{Phys. Rev. D}
  \textbf{\bibinfo{volume}{54}}, \bibinfo{pages}{7146} (\bibinfo{year}{1996}).

\bibitem[{\citenamefont{{Allen}}(1988)}]{all88}
\bibinfo{author}{\bibfnamefont{B.}~\bibnamefont{{Allen}}},
  \bibinfo{journal}{Phys. Rev. D} \textbf{\bibinfo{volume}{37}},
  \bibinfo{pages}{2078} (\bibinfo{year}{1988}).

\bibitem[{\citenamefont{{Smith} et~al.}(2006)\citenamefont{{Smith},
  {Kamionkowski}, and {Cooray}}}]{skc06}
\bibinfo{author}{\bibfnamefont{T.~L.} \bibnamefont{{Smith}}},
  \bibinfo{author}{\bibfnamefont{M.}~\bibnamefont{{Kamionkowski}}},
  \bibnamefont{and} \bibinfo{author}{\bibfnamefont{A.}~\bibnamefont{{Cooray}}},
  \bibinfo{journal}{Phys. Rev. D} \textbf{\bibinfo{volume}{73}},
  \bibinfo{pages}{023504} (\bibinfo{year}{2006}).

\bibitem[{\citenamefont{{Jones-Smith} et~al.}(2008)\citenamefont{{Jones-Smith},
  {Krauss}, and {Mathur}}}]{jkm08}
\bibinfo{author}{\bibfnamefont{K.}~\bibnamefont{{Jones-Smith}}},
  \bibinfo{author}{\bibfnamefont{L.~M.} \bibnamefont{{Krauss}}},
  \bibnamefont{and} \bibinfo{author}{\bibfnamefont{H.}~\bibnamefont{{Mathur}}},
  \bibinfo{journal}{Phys. Rev. Lett.} \textbf{\bibinfo{volume}{100}},
  \bibinfo{eid}{131302} (\bibinfo{year}{2008}).

\bibitem[{\citenamefont{{Fenu} et~al.}(2009)\citenamefont{{Fenu}, {Figueroa},
  {Durrer}, and {Garc{\'{\i}}a-Bellido}}}]{ffdg09}
\bibinfo{author}{\bibfnamefont{E.}~\bibnamefont{{Fenu}}},
  \bibinfo{author}{\bibfnamefont{D.~G.} \bibnamefont{{Figueroa}}},
  \bibinfo{author}{\bibfnamefont{R.}~\bibnamefont{{Durrer}}}, \bibnamefont{and}
  \bibinfo{author}{\bibfnamefont{J.}~\bibnamefont{{Garc{\'{\i}}a-Bellido}}},
  \bibinfo{journal}{JCAP} \textbf{\bibinfo{volume}{10}}, \bibinfo{pages}{5}
  (\bibinfo{year}{2009}).

\bibitem[{\citenamefont{Detweiler}(1979)}]{det79}
\bibinfo{author}{\bibfnamefont{S.}~\bibnamefont{Detweiler}},
  \bibinfo{journal}{ApJ} \textbf{\bibinfo{volume}{234}}, \bibinfo{pages}{1100}
  (\bibinfo{year}{1979}).

\bibitem[{\citenamefont{Foster and Backer}(1990)}]{fb90}
\bibinfo{author}{\bibfnamefont{R.~S.} \bibnamefont{Foster}} \bibnamefont{and}
  \bibinfo{author}{\bibfnamefont{D.~C.} \bibnamefont{Backer}},
  \bibinfo{journal}{ApJ} \textbf{\bibinfo{volume}{361}}, \bibinfo{pages}{300}
  (\bibinfo{year}{1990}).

\bibitem[{\citenamefont{Hellings and Downs}(1983)}]{hd83}
\bibinfo{author}{\bibfnamefont{R.~W.} \bibnamefont{Hellings}} \bibnamefont{and}
  \bibinfo{author}{\bibfnamefont{G.~S.} \bibnamefont{Downs}},
  \bibinfo{journal}{ApJ} \textbf{\bibinfo{volume}{265}}, \bibinfo{pages}{L39}
  (\bibinfo{year}{1983}).

\bibitem[{\citenamefont{Damour and Vilenkin}(2000)}]{dv00}
\bibinfo{author}{\bibfnamefont{T.}~\bibnamefont{Damour}} \bibnamefont{and}
  \bibinfo{author}{\bibfnamefont{A.}~\bibnamefont{Vilenkin}},
  \bibinfo{journal}{Phys. Rev. Lett.} \textbf{\bibinfo{volume}{85, 18}},
  \bibinfo{pages}{3761} (\bibinfo{year}{2000}).

\bibitem[{\citenamefont{Damour and Vilenkin}(2001)}]{dv01}
\bibinfo{author}{\bibfnamefont{T.}~\bibnamefont{Damour}} \bibnamefont{and}
  \bibinfo{author}{\bibfnamefont{A.}~\bibnamefont{Vilenkin}},
  \bibinfo{journal}{Phys. Rev. D} \textbf{\bibinfo{volume}{64}},
  \bibinfo{pages}{064008} (\bibinfo{year}{2001}).

\bibitem[{\citenamefont{{Damour} and {Vilenkin}}(2005)}]{dv05}
\bibinfo{author}{\bibfnamefont{T.}~\bibnamefont{{Damour}}} \bibnamefont{and}
  \bibinfo{author}{\bibfnamefont{A.}~\bibnamefont{{Vilenkin}}},
  \bibinfo{journal}{Phys. Rev. D} \textbf{\bibinfo{volume}{71}},
  \bibinfo{pages}{063510} (\bibinfo{year}{2005}).

\bibitem[{\citenamefont{{Hogan}}(2006)}]{hog06}
\bibinfo{author}{\bibfnamefont{C.~J.} \bibnamefont{{Hogan}}},
  \bibinfo{journal}{Phys. Rev. D} \textbf{\bibinfo{volume}{74}}
  (\bibinfo{year}{2006}).

\bibitem[{\citenamefont{{Siemens} et~al.}(2006)\citenamefont{{Siemens},
  {Creighton}, {Maor}, {Majumder}, {Cannon}, and {Read}}}]{scm+06}
\bibinfo{author}{\bibfnamefont{X.}~\bibnamefont{{Siemens}}},
  \bibinfo{author}{\bibfnamefont{J.}~\bibnamefont{{Creighton}}},
  \bibinfo{author}{\bibfnamefont{I.}~\bibnamefont{{Maor}}},
  \bibinfo{author}{\bibfnamefont{S.~R.} \bibnamefont{{Majumder}}},
  \bibinfo{author}{\bibfnamefont{K.}~\bibnamefont{{Cannon}}}, \bibnamefont{and}
  \bibinfo{author}{\bibfnamefont{J.}~\bibnamefont{{Read}}},
  \bibinfo{journal}{Phys. Rev. D} \textbf{\bibinfo{volume}{73}},
  \bibinfo{pages}{105001} (\bibinfo{year}{2006}).

\bibitem[{\citenamefont{{Jenet} et~al.}(2006)\citenamefont{{Jenet}, {Hobbs},
  {van Straten}, {Manchester}, {Bailes}, {Verbiest}, {Edwards}, {Hotan},
  {Sarkissian}, and {Ord}}}]{jhv+06}
\bibinfo{author}{\bibfnamefont{F.~A.} \bibnamefont{{Jenet}}},
  \bibinfo{author}{\bibfnamefont{G.~B.} \bibnamefont{{Hobbs}}},
  \bibinfo{author}{\bibfnamefont{W.}~\bibnamefont{{van Straten}}},
  \bibinfo{author}{\bibfnamefont{R.~N.} \bibnamefont{{Manchester}}},
  \bibinfo{author}{\bibfnamefont{M.}~\bibnamefont{{Bailes}}},
  \bibinfo{author}{\bibfnamefont{J.~P.~W.} \bibnamefont{{Verbiest}}},
  \bibinfo{author}{\bibfnamefont{R.~T.} \bibnamefont{{Edwards}}},
  \bibinfo{author}{\bibfnamefont{A.~W.} \bibnamefont{{Hotan}}},
  \bibinfo{author}{\bibfnamefont{J.~M.} \bibnamefont{{Sarkissian}}},
  \bibnamefont{and} \bibinfo{author}{\bibfnamefont{S.~M.} \bibnamefont{{Ord}}},
  \bibinfo{journal}{ApJ} \textbf{\bibinfo{volume}{653}}, \bibinfo{pages}{1571}
  (\bibinfo{year}{2006}).

\bibitem[{\citenamefont{{Siemens} et~al.}(2007)\citenamefont{{Siemens},
  {Mandic}, and {Creighton}}}]{smc07}
\bibinfo{author}{\bibfnamefont{X.}~\bibnamefont{{Siemens}}},
  \bibinfo{author}{\bibfnamefont{V.}~\bibnamefont{{Mandic}}}, \bibnamefont{and}
  \bibinfo{author}{\bibfnamefont{J.}~\bibnamefont{{Creighton}}},
  \bibinfo{journal}{Phys. Rev. Lett.} \textbf{\bibinfo{volume}{98}},
  \bibinfo{pages}{111101} (\bibinfo{year}{2007}).

\bibitem[{\citenamefont{DePies and Hogan}(2007)}]{dh07}
\bibinfo{author}{\bibfnamefont{M.~R.} \bibnamefont{DePies}} \bibnamefont{and}
  \bibinfo{author}{\bibfnamefont{C.}~\bibnamefont{Hogan}},
  \bibinfo{journal}{Phys. Rev. D} \textbf{\bibinfo{volume}{75, 12}}
  (\bibinfo{year}{2007}).

\bibitem[{\citenamefont{{{\"O}lmez} et~al.}(2010)\citenamefont{{{\"O}lmez},
  {Mandic}, and {Siemens}}}]{oms10}
\bibinfo{author}{\bibfnamefont{S.}~\bibnamefont{{{\"O}lmez}}},
  \bibinfo{author}{\bibfnamefont{V.}~\bibnamefont{{Mandic}}}, \bibnamefont{and}
  \bibinfo{author}{\bibfnamefont{X.}~\bibnamefont{{Siemens}}},
  \bibinfo{journal}{Phys. Rev. D} \textbf{\bibinfo{volume}{81}},
  \bibinfo{pages}{104028} (\bibinfo{year}{2010}).

\bibitem[{\citenamefont{{van Haasteren} et~al.}(2011)\citenamefont{{van
  Haasteren}, {Levin}, {Janssen}, {Lazaridis}, {Kramer}, {Stappers},
  {Desvignes}, {Purver}, {Lyne}, {Ferdman} et~al.}}]{vlj+11}
\bibinfo{author}{\bibfnamefont{R.}~\bibnamefont{{van Haasteren}}},
  \bibinfo{author}{\bibfnamefont{Y.}~\bibnamefont{{Levin}}},
  \bibinfo{author}{\bibfnamefont{G.~H.} \bibnamefont{{Janssen}}},
  \bibinfo{author}{\bibfnamefont{K.}~\bibnamefont{{Lazaridis}}},
  \bibinfo{author}{\bibfnamefont{M.}~\bibnamefont{{Kramer}}},
  \bibinfo{author}{\bibfnamefont{B.~W.} \bibnamefont{{Stappers}}},
  \bibinfo{author}{\bibfnamefont{G.}~\bibnamefont{{Desvignes}}},
  \bibinfo{author}{\bibfnamefont{M.~B.} \bibnamefont{{Purver}}},
  \bibinfo{author}{\bibfnamefont{A.~G.} \bibnamefont{{Lyne}}},
  \bibinfo{author}{\bibfnamefont{R.~D.} \bibnamefont{{Ferdman}}},
  \bibnamefont{et~al.}, \bibinfo{journal}{MNRAS} p. \bibinfo{pages}{568}
  (\bibinfo{year}{2011}).

\bibitem[{\citenamefont{Kibble}(1985)}]{kib85}
\bibinfo{author}{\bibfnamefont{T.~W.~B.} \bibnamefont{Kibble}},
  \bibinfo{journal}{Nucl. Phys. B} \textbf{\bibinfo{volume}{252}},
  \bibinfo{pages}{227} (\bibinfo{year}{1985}).

\bibitem[{\citenamefont{Bennett}(1986{\natexlab{a}})}]{ben86a}
\bibinfo{author}{\bibfnamefont{D.~P.} \bibnamefont{Bennett}},
  \bibinfo{journal}{Phys. Rev. D} \textbf{\bibinfo{volume}{33, 4}},
  \bibinfo{pages}{872} (\bibinfo{year}{1986}{\natexlab{a}}).

\bibitem[{\citenamefont{Bennett}(1986{\natexlab{b}})}]{ben86b}
\bibinfo{author}{\bibfnamefont{D.~P.} \bibnamefont{Bennett}},
  \bibinfo{journal}{Phys. Rev. D} \textbf{\bibinfo{volume}{34, 12}},
  \bibinfo{pages}{3592} (\bibinfo{year}{1986}{\natexlab{b}}).

\bibitem[{\citenamefont{{Bin{\'e}truy}
  et~al.}(2012)\citenamefont{{Bin{\'e}truy}, {Boh{\'e}}, {Caprini}, and
  {Dufaux}}}]{bbcd12}
\bibinfo{author}{\bibfnamefont{P.}~\bibnamefont{{Bin{\'e}truy}}},
  \bibinfo{author}{\bibfnamefont{A.}~\bibnamefont{{Boh{\'e}}}},
  \bibinfo{author}{\bibfnamefont{C.}~\bibnamefont{{Caprini}}},
  \bibnamefont{and} \bibinfo{author}{\bibfnamefont{J.-F.}
  \bibnamefont{{Dufaux}}}, \bibinfo{journal}{ArXiv e-prints}
  (\bibinfo{year}{2012}), \eprint{1201.0983}.

\bibitem[{\citenamefont{Shellard}(1987)}]{she87}
\bibinfo{author}{\bibfnamefont{E.~P.~S.} \bibnamefont{Shellard}},
  \bibinfo{journal}{Nucl. Phys. B} \textbf{\bibinfo{volume}{283, 624}}
  (\bibinfo{year}{1987}).

\bibitem[{\citenamefont{{Vincent} et~al.}(1997)\citenamefont{{Vincent},
  {Hindmarsh}, and {Sakellariadou}}}]{vhs97}
\bibinfo{author}{\bibfnamefont{G.~R.} \bibnamefont{{Vincent}}},
  \bibinfo{author}{\bibfnamefont{M.}~\bibnamefont{{Hindmarsh}}},
  \bibnamefont{and}
  \bibinfo{author}{\bibfnamefont{M.}~\bibnamefont{{Sakellariadou}}},
  \bibinfo{journal}{Phys. Rev. D} \textbf{\bibinfo{volume}{56}},
  \bibinfo{pages}{637} (\bibinfo{year}{1997}).

\bibitem[{\citenamefont{{Vincent} et~al.}(1998)\citenamefont{{Vincent},
  {Antunes}, and {Hindmarsh}}}]{vah98}
\bibinfo{author}{\bibfnamefont{G.}~\bibnamefont{{Vincent}}},
  \bibinfo{author}{\bibfnamefont{N.~D.} \bibnamefont{{Antunes}}},
  \bibnamefont{and}
  \bibinfo{author}{\bibfnamefont{M.}~\bibnamefont{{Hindmarsh}}},
  \bibinfo{journal}{Phys. Rev. Lett.} \textbf{\bibinfo{volume}{80}},
  \bibinfo{pages}{2277} (\bibinfo{year}{1998}).

\bibitem[{\citenamefont{{Hindmarsh}}(2011)}]{hin11}
\bibinfo{author}{\bibfnamefont{M.}~\bibnamefont{{Hindmarsh}}},
  \bibinfo{journal}{Prog.Theor.Phys.Suppl.} \textbf{\bibinfo{volume}{190}},
  \bibinfo{pages}{197} (\bibinfo{year}{2011}).

\bibitem[{\citenamefont{Bennett and Bouchet}(1989{\natexlab{a}})}]{bb89a}
\bibinfo{author}{\bibfnamefont{D.~P.} \bibnamefont{Bennett}} \bibnamefont{and}
  \bibinfo{author}{\bibfnamefont{F.~R.} \bibnamefont{Bouchet}},
  \bibinfo{journal}{Phys. Rev. Lett.} \textbf{\bibinfo{volume}{63, 26}},
  \bibinfo{pages}{2776} (\bibinfo{year}{1989}{\natexlab{a}}).

\bibitem[{\citenamefont{{Blanco-Pillado}
  et~al.}(2011)\citenamefont{{Blanco-Pillado}, {Olum}, and {Shlaer}}}]{bos11}
\bibinfo{author}{\bibfnamefont{J.~J.} \bibnamefont{{Blanco-Pillado}}},
  \bibinfo{author}{\bibfnamefont{K.~D.} \bibnamefont{{Olum}}},
  \bibnamefont{and} \bibinfo{author}{\bibfnamefont{B.}~\bibnamefont{{Shlaer}}},
  \bibinfo{journal}{Phys. Rev. D} \textbf{\bibinfo{volume}{83}},
  \bibinfo{pages}{083514} (\bibinfo{year}{2011}).

\bibitem[{\citenamefont{Bennett and Bouchet}(1989{\natexlab{b}})}]{bb89b}
\bibinfo{author}{\bibfnamefont{D.~P.} \bibnamefont{Bennett}} \bibnamefont{and}
  \bibinfo{author}{\bibfnamefont{F.~R.} \bibnamefont{Bouchet}},
  \bibinfo{journal}{Phys. Rev. D} \textbf{\bibinfo{volume}{41, 2}},
  \bibinfo{pages}{720} (\bibinfo{year}{1989}{\natexlab{b}}).

\bibitem[{\citenamefont{Allen and Caldwell}(1991)}]{ac91a}
\bibinfo{author}{\bibfnamefont{B.}~\bibnamefont{Allen}} \bibnamefont{and}
  \bibinfo{author}{\bibfnamefont{R.~R.} \bibnamefont{Caldwell}},
  \bibinfo{journal}{Phys. Rev. D} \textbf{\bibinfo{volume}{43, 10}},
  \bibinfo{pages}{3173} (\bibinfo{year}{1991}).

\bibitem[{\citenamefont{Casper and Allen}(1995)}]{ca95}
\bibinfo{author}{\bibfnamefont{P.}~\bibnamefont{Casper}} \bibnamefont{and}
  \bibinfo{author}{\bibfnamefont{B.}~\bibnamefont{Allen}},
  \bibinfo{journal}{Phys. Rev. D} \textbf{\bibinfo{volume}{52, 8}}
  (\bibinfo{year}{1995}).

\bibitem[{\citenamefont{Allen and Casper}(1994)}]{ac94}
\bibinfo{author}{\bibfnamefont{B.}~\bibnamefont{Allen}} \bibnamefont{and}
  \bibinfo{author}{\bibfnamefont{P.}~\bibnamefont{Casper}},
  \bibinfo{journal}{Phys. Rev. D} \textbf{\bibinfo{volume}{50, 4}}
  (\bibinfo{year}{1994}).

\bibitem[{\citenamefont{Allen et~al.}(1994)\citenamefont{Allen, Casper, and
  Ottewill}}]{aco94}
\bibinfo{author}{\bibfnamefont{B.}~\bibnamefont{Allen}},
  \bibinfo{author}{\bibfnamefont{P.}~\bibnamefont{Casper}}, \bibnamefont{and}
  \bibinfo{author}{\bibfnamefont{A.}~\bibnamefont{Ottewill}},
  \bibinfo{journal}{Phys. Rev. D} \textbf{\bibinfo{volume}{50, 6}},
  \bibinfo{pages}{3703} (\bibinfo{year}{1994}).

\bibitem[{\citenamefont{{Lorenz} et~al.}(2010)\citenamefont{{Lorenz},
  {Ringeval}, and {Sakellariadou}}}]{lrs10}
\bibinfo{author}{\bibfnamefont{L.}~\bibnamefont{{Lorenz}}},
  \bibinfo{author}{\bibfnamefont{C.}~\bibnamefont{{Ringeval}}},
  \bibnamefont{and}
  \bibinfo{author}{\bibfnamefont{M.}~\bibnamefont{{Sakellariadou}}},
  \bibinfo{journal}{JCAP} \textbf{\bibinfo{volume}{10}}, \bibinfo{pages}{3}
  (\bibinfo{year}{2010}).

\bibitem[{\citenamefont{{Battye} and {Shellard}}(1994)}]{bs94}
\bibinfo{author}{\bibfnamefont{R.~A.} \bibnamefont{{Battye}}} \bibnamefont{and}
  \bibinfo{author}{\bibfnamefont{E.~P.~S.} \bibnamefont{{Shellard}}},
  \bibinfo{journal}{Nucl. Phys. B} \textbf{\bibinfo{volume}{423}},
  \bibinfo{pages}{260} (\bibinfo{year}{1994}).

\bibitem[{\citenamefont{Sakellariadou}(2005)}]{sak05}
\bibinfo{author}{\bibfnamefont{M.}~\bibnamefont{Sakellariadou}},
  \bibinfo{journal}{JCAP} \textbf{\bibinfo{volume}{04}}, \bibinfo{pages}{003}
  (\bibinfo{year}{2005}).

\bibitem[{\citenamefont{Avgoustidis and Shellard}(2005)}]{as05}
\bibinfo{author}{\bibfnamefont{A.}~\bibnamefont{Avgoustidis}} \bibnamefont{and}
  \bibinfo{author}{\bibfnamefont{E.~P.~S.} \bibnamefont{Shellard}},
  \bibinfo{journal}{Phys. Rev. D} \textbf{\bibinfo{volume}{71,12}}
  (\bibinfo{year}{2005}).

\bibitem[{\citenamefont{Avgoustidis and Shellard}(2006)}]{as06}
\bibinfo{author}{\bibfnamefont{A.}~\bibnamefont{Avgoustidis}} \bibnamefont{and}
  \bibinfo{author}{\bibfnamefont{E.~P.~S.} \bibnamefont{Shellard}},
  \bibinfo{journal}{Phys. Rev. D} \textbf{\bibinfo{volume}{73,4}}
  (\bibinfo{year}{2006}).

\bibitem[{\citenamefont{Lorimer and Kramer}(2005)}]{lk05}
\bibinfo{author}{\bibfnamefont{D.~R.} \bibnamefont{Lorimer}} \bibnamefont{and}
  \bibinfo{author}{\bibfnamefont{M.}~\bibnamefont{Kramer}},
  \emph{\bibinfo{title}{Handbook of Pulsar Astronomy}}
  (\bibinfo{publisher}{Cambridge University Press}, \bibinfo{year}{2005}).

\bibitem[{\citenamefont{{Manchester}}(2006)}]{man06}
\bibinfo{author}{\bibfnamefont{R.~N.} \bibnamefont{{Manchester}}},
  \bibinfo{journal}{Chin. J. Astron. Astrophys., Suppl. 2}
  \textbf{\bibinfo{volume}{6}}, \bibinfo{pages}{139} (\bibinfo{year}{2006}),
  \eprint{arXiv:astro-ph/0604288}.

\bibitem[{\citenamefont{{Jenet} et~al.}(2009)\citenamefont{{Jenet}, {Finn},
  {Lazio}, {Lommen}, {McLaughlin}, {Stairs}, {Stinebring}, {Verbiest},
  {Archibald}, {Arzoumanian} et~al.}}]{jfl+09}
\bibinfo{author}{\bibfnamefont{F.}~\bibnamefont{{Jenet}}},
  \bibinfo{author}{\bibfnamefont{L.~S.} \bibnamefont{{Finn}}},
  \bibinfo{author}{\bibfnamefont{J.}~\bibnamefont{{Lazio}}},
  \bibinfo{author}{\bibfnamefont{A.}~\bibnamefont{{Lommen}}},
  \bibinfo{author}{\bibfnamefont{M.}~\bibnamefont{{McLaughlin}}},
  \bibinfo{author}{\bibfnamefont{I.}~\bibnamefont{{Stairs}}},
  \bibinfo{author}{\bibfnamefont{D.}~\bibnamefont{{Stinebring}}},
  \bibinfo{author}{\bibfnamefont{J.}~\bibnamefont{{Verbiest}}},
  \bibinfo{author}{\bibfnamefont{A.}~\bibnamefont{{Archibald}}},
  \bibinfo{author}{\bibfnamefont{Z.}~\bibnamefont{{Arzoumanian}}},
  \bibnamefont{et~al.}, \bibinfo{journal}{ArXiv e-prints}
  (\bibinfo{year}{2009}), \eprint{0909.1058}.

\bibitem[{\citenamefont{Jarosik~et al.}(2011)}]{jbdg+11}
\bibinfo{author}{\bibfnamefont{N.}~\bibnamefont{Jarosik~et al.}},
  \bibinfo{journal}{Astrophys. J. Supp. Series}
  \textbf{\bibinfo{volume}{192,14}} (\bibinfo{year}{2011}).

\bibitem[{\citenamefont{Albrecht and Turok}(1989)}]{at89}
\bibinfo{author}{\bibfnamefont{A.}~\bibnamefont{Albrecht}} \bibnamefont{and}
  \bibinfo{author}{\bibfnamefont{N.}~\bibnamefont{Turok}},
  \bibinfo{journal}{Phys. Rev. D} \textbf{\bibinfo{volume}{40}},
  \bibinfo{pages}{973} (\bibinfo{year}{1989}).

\bibitem[{\citenamefont{Allen and Shellard}(1990)}]{as90}
\bibinfo{author}{\bibfnamefont{B.}~\bibnamefont{Allen}} \bibnamefont{and}
  \bibinfo{author}{\bibfnamefont{E.~P.~S.} \bibnamefont{Shellard}},
  \bibinfo{journal}{Phys. Rev. Lett.} \textbf{\bibinfo{volume}{64, 2}},
  \bibinfo{pages}{119} (\bibinfo{year}{1990}).

\bibitem[{\citenamefont{Martins and Shellard}(2006)}]{ms06}
\bibinfo{author}{\bibfnamefont{C.~J. A.~P.} \bibnamefont{Martins}}
  \bibnamefont{and} \bibinfo{author}{\bibfnamefont{E.~P.~S.}
  \bibnamefont{Shellard}}, \bibinfo{journal}{Phys. Rev. D}
  \textbf{\bibinfo{volume}{73}}, \bibinfo{pages}{043515}
  (\bibinfo{year}{2006}).

\bibitem[{\citenamefont{Ringeval et~al.}(2007)\citenamefont{Ringeval,
  Sakellariadou, and Bouchet}}]{rsb07}
\bibinfo{author}{\bibfnamefont{C.}~\bibnamefont{Ringeval}},
  \bibinfo{author}{\bibfnamefont{M.}~\bibnamefont{Sakellariadou}},
  \bibnamefont{and} \bibinfo{author}{\bibfnamefont{F.~R.}
  \bibnamefont{Bouchet}}, \bibinfo{journal}{JCAP} \textbf{\bibinfo{volume}{2}},
  \bibinfo{pages}{023} (\bibinfo{year}{2007}).

\bibitem[{\citenamefont{Vanchurin et~al.}(2005)\citenamefont{Vanchurin, Olum,
  and Vilenkin}}]{vov05}
\bibinfo{author}{\bibfnamefont{V.}~\bibnamefont{Vanchurin}},
  \bibinfo{author}{\bibfnamefont{K.}~\bibnamefont{Olum}}, \bibnamefont{and}
  \bibinfo{author}{\bibfnamefont{A.}~\bibnamefont{Vilenkin}},
  \bibinfo{journal}{Phys. Rev. D} \textbf{\bibinfo{volume}{72}},
  \bibinfo{pages}{063514} (\bibinfo{year}{2005}).

\bibitem[{\citenamefont{Vanchurin et~al.}(2006)\citenamefont{Vanchurin, Olum,
  and Vilenkin}}]{vov06}
\bibinfo{author}{\bibfnamefont{V.}~\bibnamefont{Vanchurin}},
  \bibinfo{author}{\bibfnamefont{K.}~\bibnamefont{Olum}}, \bibnamefont{and}
  \bibinfo{author}{\bibfnamefont{A.}~\bibnamefont{Vilenkin}},
  \bibinfo{journal}{Phys. Rev. D} \textbf{\bibinfo{volume}{74}},
  \bibinfo{pages}{063527} (\bibinfo{year}{2006}).

\bibitem[{\citenamefont{Olum and Vanchurin}(2007)}]{ov07}
\bibinfo{author}{\bibfnamefont{K.}~\bibnamefont{Olum}} \bibnamefont{and}
  \bibinfo{author}{\bibfnamefont{V.}~\bibnamefont{Vanchurin}},
  \bibinfo{journal}{Phys. Rev. D} \textbf{\bibinfo{volume}{75}},
  \bibinfo{pages}{063521} (\bibinfo{year}{2007}).

\bibitem[{\citenamefont{Siemens and Olum}(2001)}]{so01}
\bibinfo{author}{\bibfnamefont{X.}~\bibnamefont{Siemens}} \bibnamefont{and}
  \bibinfo{author}{\bibfnamefont{K.}~\bibnamefont{Olum}},
  \bibinfo{journal}{Nucl. Phys. B} \textbf{\bibinfo{volume}{611}},
  \bibinfo{pages}{125} (\bibinfo{year}{2001}).

\bibitem[{\citenamefont{Siemens et~al.}(2002)\citenamefont{Siemens, Olum, and
  Vilenkin}}]{sov02}
\bibinfo{author}{\bibfnamefont{X.}~\bibnamefont{Siemens}},
  \bibinfo{author}{\bibfnamefont{K.}~\bibnamefont{Olum}}, \bibnamefont{and}
  \bibinfo{author}{\bibfnamefont{A.}~\bibnamefont{Vilenkin}},
  \bibinfo{journal}{Phys. Rev. D} \textbf{\bibinfo{volume}{66}},
  \bibinfo{pages}{043501} (\bibinfo{year}{2002}).

\bibitem[{\citenamefont{Polchinski and Rocha}(2006)}]{pr06}
\bibinfo{author}{\bibfnamefont{J.}~\bibnamefont{Polchinski}} \bibnamefont{and}
  \bibinfo{author}{\bibfnamefont{J.~V.} \bibnamefont{Rocha}},
  \bibinfo{journal}{Phys. Rev. D} \textbf{\bibinfo{volume}{74}},
  \bibinfo{pages}{083504} (\bibinfo{year}{2006}).

\bibitem[{\citenamefont{Polchinski and Rocha}(2007)}]{pr07a}
\bibinfo{author}{\bibfnamefont{J.}~\bibnamefont{Polchinski}} \bibnamefont{and}
  \bibinfo{author}{\bibfnamefont{J.~V.} \bibnamefont{Rocha}},
  \bibinfo{journal}{Phys. Rev. D} \textbf{\bibinfo{volume}{75}},
  \bibinfo{pages}{123503} (\bibinfo{year}{2007}).

\bibitem[{\citenamefont{Copeland and Kibble}(2009)}]{ck09}
\bibinfo{author}{\bibfnamefont{E.~J.} \bibnamefont{Copeland}} \bibnamefont{and}
  \bibinfo{author}{\bibfnamefont{T.~W.~B.} \bibnamefont{Kibble}},
  \bibinfo{journal}{Phys. Rev. D} \textbf{\bibinfo{volume}{77}},
  \bibinfo{pages}{123523} (\bibinfo{year}{2009}).

\bibitem[{\citenamefont{Vincent et~al.}(1997)\citenamefont{Vincent, Antunes,
  and Hindmarsch}}]{vah97}
\bibinfo{author}{\bibfnamefont{G.}~\bibnamefont{Vincent}},
  \bibinfo{author}{\bibfnamefont{N.~D.} \bibnamefont{Antunes}},
  \bibnamefont{and}
  \bibinfo{author}{\bibfnamefont{M.}~\bibnamefont{Hindmarsch}},
  \bibinfo{journal}{Phys. Rev. Lett.} \textbf{\bibinfo{volume}{80, 11}},
  \bibinfo{pages}{2277} (\bibinfo{year}{1997}).

\bibitem[{\citenamefont{{Polchinski}}(2008)}]{pol08}
\bibinfo{author}{\bibfnamefont{J.}~\bibnamefont{{Polchinski}}}, in
  \emph{\bibinfo{booktitle}{The Eleventh Marcel Grossmann Meeting On Recent
  Developments in Theoretical and Experimental General Relativity, Gravitation
  and Relativistic Field Theories}}, edited by
  \bibinfo{editor}{\bibnamefont{{H.~Kleinert, R.~T.~Jantzen, \& R.~Ruffini}}}
  (\bibinfo{year}{2008}), pp. \bibinfo{pages}{105--125}.

\bibitem[{\citenamefont{Dubath et~al.}(2008)\citenamefont{Dubath, Polchinski,
  and Rocha}}]{dpr08}
\bibinfo{author}{\bibfnamefont{F.}~\bibnamefont{Dubath}},
  \bibinfo{author}{\bibfnamefont{J.}~\bibnamefont{Polchinski}},
  \bibnamefont{and} \bibinfo{author}{\bibfnamefont{J.~V.} \bibnamefont{Rocha}},
  \bibinfo{journal}{Phys. Rev. D} \textbf{\bibinfo{volume}{77}},
  \bibinfo{pages}{123528} (\bibinfo{year}{2008}).

\bibitem[{\citenamefont{Ferdman and van Haasteren~et al.}(2010)}]{fvb+10}
\bibinfo{author}{\bibfnamefont{R.~D.} \bibnamefont{Ferdman}} \bibnamefont{and}
  \bibinfo{author}{\bibfnamefont{R.}~\bibnamefont{van Haasteren~et al.}},
  \bibinfo{journal}{Class. Quant Grav.} \textbf{\bibinfo{volume}{27, 8}}
  (\bibinfo{year}{2010}).

\bibitem[{\citenamefont{Thorsett and Dewey}(1996)}]{td96}
\bibinfo{author}{\bibfnamefont{S.~E.} \bibnamefont{Thorsett}} \bibnamefont{and}
  \bibinfo{author}{\bibfnamefont{R.~J.} \bibnamefont{Dewey}},
  \bibinfo{journal}{Phys. Rev. D} \textbf{\bibinfo{volume}{53}},
  \bibinfo{pages}{3468} (\bibinfo{year}{1996}).

\bibitem[{\citenamefont{Bertotti et~al.}(1983)\citenamefont{Bertotti, Carr, and
  Rees}}]{bcr83}
\bibinfo{author}{\bibfnamefont{B.}~\bibnamefont{Bertotti}},
  \bibinfo{author}{\bibfnamefont{B.~J.} \bibnamefont{Carr}}, \bibnamefont{and}
  \bibinfo{author}{\bibfnamefont{M.~J.} \bibnamefont{Rees}},
  \bibinfo{journal}{MNRAS} \textbf{\bibinfo{volume}{203}}, \bibinfo{pages}{945}
  (\bibinfo{year}{1983}).

\bibitem[{\citenamefont{Kaspi et~al.}(1994)\citenamefont{Kaspi, Taylor, and
  Ryba}}]{ktr94}
\bibinfo{author}{\bibfnamefont{V.~M.} \bibnamefont{Kaspi}},
  \bibinfo{author}{\bibfnamefont{J.~H.} \bibnamefont{Taylor}},
  \bibnamefont{and} \bibinfo{author}{\bibfnamefont{M.}~\bibnamefont{Ryba}},
  \bibinfo{journal}{ApJ} \textbf{\bibinfo{volume}{428}}, \bibinfo{pages}{713}
  (\bibinfo{year}{1994}).

\bibitem[{\citenamefont{{McHugh} et~al.}(1996)\citenamefont{{McHugh},
  {Zalamansky}, {Vernotte}, and {Lantz}}}]{mzvl96}
\bibinfo{author}{\bibfnamefont{M.~P.} \bibnamefont{{McHugh}}},
  \bibinfo{author}{\bibfnamefont{G.}~\bibnamefont{{Zalamansky}}},
  \bibinfo{author}{\bibfnamefont{F.}~\bibnamefont{{Vernotte}}},
  \bibnamefont{and} \bibinfo{author}{\bibfnamefont{E.}~\bibnamefont{{Lantz}}},
  \bibinfo{journal}{Phys. Rev. D} \textbf{\bibinfo{volume}{54}},
  \bibinfo{pages}{5993} (\bibinfo{year}{1996}).

\bibitem[{\citenamefont{{Lommen}}(2002)}]{lom02}
\bibinfo{author}{\bibfnamefont{A.~N.} \bibnamefont{{Lommen}}}, in
  \emph{\bibinfo{booktitle}{WE-Heraeus Seminar on Neutron Stars, Pulsars, and
  Supernova Remnants}}, edited by
  \bibinfo{editor}{\bibfnamefont{W.}~\bibnamefont{Becker}},
  \bibinfo{editor}{\bibfnamefont{H.}~\bibnamefont{Lesch}}, \bibnamefont{and}
  \bibinfo{editor}{\bibfnamefont{J.}~\bibnamefont{Tr\"umper}}
  (\bibinfo{publisher}{Max-Plank-Institut f\"ur Extraterrestrische Physik},
  \bibinfo{address}{Garching}, \bibinfo{year}{2002}), pp.
  \bibinfo{pages}{114--125}.

\bibitem[{\citenamefont{{Abbott} et~al.}(2009)\citenamefont{{Abbott}, {Abbott},
  {Acernese}, {Adhikari}, {Ajith}, {Allen}, {Allen}, {Alshourbagy}, {Amin},
  {Anderson} et~al.}}]{aaa+09}
\bibinfo{author}{\bibfnamefont{B.~P.} \bibnamefont{{Abbott}}},
  \bibinfo{author}{\bibfnamefont{R.}~\bibnamefont{{Abbott}}},
  \bibinfo{author}{\bibfnamefont{F.}~\bibnamefont{{Acernese}}},
  \bibinfo{author}{\bibfnamefont{R.}~\bibnamefont{{Adhikari}}},
  \bibinfo{author}{\bibfnamefont{P.}~\bibnamefont{{Ajith}}},
  \bibinfo{author}{\bibfnamefont{B.}~\bibnamefont{{Allen}}},
  \bibinfo{author}{\bibfnamefont{G.}~\bibnamefont{{Allen}}},
  \bibinfo{author}{\bibfnamefont{M.}~\bibnamefont{{Alshourbagy}}},
  \bibinfo{author}{\bibfnamefont{R.~S.} \bibnamefont{{Amin}}},
  \bibinfo{author}{\bibfnamefont{S.~B.} \bibnamefont{{Anderson}}},
  \bibnamefont{et~al.}, \bibinfo{journal}{Nature}
  \textbf{\bibinfo{volume}{460}}, \bibinfo{pages}{990} (\bibinfo{year}{2009}),
  \eprint{0910.5772}.

\bibitem[{\citenamefont{Kramer and Stappers}(2010)}]{ks10}
\bibinfo{author}{\bibfnamefont{M.}~\bibnamefont{Kramer}} \bibnamefont{and}
  \bibinfo{author}{\bibfnamefont{B.}~\bibnamefont{Stappers}},
  \bibinfo{journal}{Proceedings of the ISKAF2010 Science Meeting}
  (\bibinfo{year}{2010}).

\bibitem[{\citenamefont{{Dunkley} et~al.}(2011)\citenamefont{{Dunkley},
  {Hlozek}, {Sievers}, {Acquaviva}, {Ade}, {Aguirre}, {Amiri}, {Appel},
  {Barrientos}, {Battistelli} et~al.}}]{dhs+11}
\bibinfo{author}{\bibfnamefont{J.}~\bibnamefont{{Dunkley}}},
  \bibinfo{author}{\bibfnamefont{R.}~\bibnamefont{{Hlozek}}},
  \bibinfo{author}{\bibfnamefont{J.}~\bibnamefont{{Sievers}}},
  \bibinfo{author}{\bibfnamefont{V.}~\bibnamefont{{Acquaviva}}},
  \bibinfo{author}{\bibfnamefont{P.~A.~R.} \bibnamefont{{Ade}}},
  \bibinfo{author}{\bibfnamefont{P.}~\bibnamefont{{Aguirre}}},
  \bibinfo{author}{\bibfnamefont{M.}~\bibnamefont{{Amiri}}},
  \bibinfo{author}{\bibfnamefont{J.~W.} \bibnamefont{{Appel}}},
  \bibinfo{author}{\bibfnamefont{L.~F.} \bibnamefont{{Barrientos}}},
  \bibinfo{author}{\bibfnamefont{E.~S.} \bibnamefont{{Battistelli}}},
  \bibnamefont{et~al.}, \bibinfo{journal}{ApJ} \textbf{\bibinfo{volume}{739}},
  \bibinfo{pages}{52} (\bibinfo{year}{2011}).

\bibitem[{\citenamefont{{Urrestilla} et~al.}(2011)\citenamefont{{Urrestilla},
  {Bevis}, {Hindmarsh}, and {Kunz}}}]{ubhk11}
\bibinfo{author}{\bibfnamefont{J.}~\bibnamefont{{Urrestilla}}},
  \bibinfo{author}{\bibfnamefont{N.}~\bibnamefont{{Bevis}}},
  \bibinfo{author}{\bibfnamefont{M.}~\bibnamefont{{Hindmarsh}}},
  \bibnamefont{and} \bibinfo{author}{\bibfnamefont{M.}~\bibnamefont{{Kunz}}},
  \bibinfo{journal}{JCAP} \textbf{\bibinfo{volume}{12}}, \bibinfo{pages}{21}
  (\bibinfo{year}{2011}).

\bibitem[{\citenamefont{Christiansen et~al.}(2008)\citenamefont{Christiansen,
  Albin, James, Goldman, Maruyama, and Smoot}}]{cajg+08}
\bibinfo{author}{\bibfnamefont{J.~L.} \bibnamefont{Christiansen}},
  \bibinfo{author}{\bibfnamefont{E.}~\bibnamefont{Albin}},
  \bibinfo{author}{\bibfnamefont{K.}~\bibnamefont{James}},
  \bibinfo{author}{\bibfnamefont{J.}~\bibnamefont{Goldman}},
  \bibinfo{author}{\bibfnamefont{D.}~\bibnamefont{Maruyama}}, \bibnamefont{and}
  \bibinfo{author}{\bibfnamefont{G.~F.} \bibnamefont{Smoot}},
  \bibinfo{journal}{Phys. Rev. D} \textbf{\bibinfo{volume}{77}},
  \bibinfo{pages}{123509} (\bibinfo{year}{2008}).

\bibitem[{\citenamefont{{Christiansen}
  et~al.}(2011)\citenamefont{{Christiansen}, {Albin}, {Fletcher}, {Goldman},
  {Teng}, {Foley}, and {Smoot}}}]{caf+11}
\bibinfo{author}{\bibfnamefont{J.~L.} \bibnamefont{{Christiansen}}},
  \bibinfo{author}{\bibfnamefont{E.}~\bibnamefont{{Albin}}},
  \bibinfo{author}{\bibfnamefont{T.}~\bibnamefont{{Fletcher}}},
  \bibinfo{author}{\bibfnamefont{J.}~\bibnamefont{{Goldman}}},
  \bibinfo{author}{\bibfnamefont{I.~P.~W.} \bibnamefont{{Teng}}},
  \bibinfo{author}{\bibfnamefont{M.}~\bibnamefont{{Foley}}}, \bibnamefont{and}
  \bibinfo{author}{\bibfnamefont{G.~F.} \bibnamefont{{Smoot}}},
  \bibinfo{journal}{Phys. Rev. D} \textbf{\bibinfo{volume}{83}},
  \bibinfo{pages}{122004} (\bibinfo{year}{2011}).

\bibitem[{\citenamefont{{Tauber} et~al.}(2010)\citenamefont{{Tauber},
  {Mandolesi}, {Puget}, {Banos}, {Bersanelli}, {Bouchet}, {Butler}, {Charra},
  {Crone}, {Dodsworth} et~al.}}]{tmp+10}
\bibinfo{author}{\bibfnamefont{J.~A.} \bibnamefont{{Tauber}}},
  \bibinfo{author}{\bibfnamefont{N.}~\bibnamefont{{Mandolesi}}},
  \bibinfo{author}{\bibfnamefont{J.-L.} \bibnamefont{{Puget}}},
  \bibinfo{author}{\bibfnamefont{T.}~\bibnamefont{{Banos}}},
  \bibinfo{author}{\bibfnamefont{M.}~\bibnamefont{{Bersanelli}}},
  \bibinfo{author}{\bibfnamefont{F.~R.} \bibnamefont{{Bouchet}}},
  \bibinfo{author}{\bibfnamefont{R.~C.} \bibnamefont{{Butler}}},
  \bibinfo{author}{\bibfnamefont{J.}~\bibnamefont{{Charra}}},
  \bibinfo{author}{\bibfnamefont{G.}~\bibnamefont{{Crone}}},
  \bibinfo{author}{\bibfnamefont{J.}~\bibnamefont{{Dodsworth}}},
  \bibnamefont{et~al.}, \bibinfo{journal}{A\&A} \textbf{\bibinfo{volume}{520}},
  \bibinfo{pages}{A1} (\bibinfo{year}{2010}).

\bibitem[{\citenamefont{{Battye} et~al.}(2008)\citenamefont{{Battye},
  {Garbrecht}, {Moss}, and {Stoica}}}]{bgms08}
\bibinfo{author}{\bibfnamefont{R.~A.} \bibnamefont{{Battye}}},
  \bibinfo{author}{\bibfnamefont{B.}~\bibnamefont{{Garbrecht}}},
  \bibinfo{author}{\bibfnamefont{A.}~\bibnamefont{{Moss}}}, \bibnamefont{and}
  \bibinfo{author}{\bibfnamefont{H.}~\bibnamefont{{Stoica}}},
  \bibinfo{journal}{JCAP} \textbf{\bibinfo{volume}{1}}, \bibinfo{pages}{20}
  (\bibinfo{year}{2008}).

\bibitem[{\citenamefont{{Foreman} et~al.}(2011)\citenamefont{{Foreman}, {Moss},
  and {Scott}}}]{fms11}
\bibinfo{author}{\bibfnamefont{S.}~\bibnamefont{{Foreman}}},
  \bibinfo{author}{\bibfnamefont{A.}~\bibnamefont{{Moss}}}, \bibnamefont{and}
  \bibinfo{author}{\bibfnamefont{D.}~\bibnamefont{{Scott}}},
  \bibinfo{journal}{Phys. Rev. D} \textbf{\bibinfo{volume}{84}},
  \bibinfo{eid}{043522} (\bibinfo{year}{2011}).

\bibitem[{\citenamefont{{Dvorkin} et~al.}(2011)\citenamefont{{Dvorkin},
  {Wyman}, and {Hu}}}]{dwh11}
\bibinfo{author}{\bibfnamefont{C.}~\bibnamefont{{Dvorkin}}},
  \bibinfo{author}{\bibfnamefont{M.}~\bibnamefont{{Wyman}}}, \bibnamefont{and}
  \bibinfo{author}{\bibfnamefont{W.}~\bibnamefont{{Hu}}},
  \bibinfo{journal}{Phys. Rev. D} \textbf{\bibinfo{volume}{84}},
  \bibinfo{eid}{123519} (\bibinfo{year}{2011}).

\bibitem[{\citenamefont{{Hobbs} et~al.}(2010)\citenamefont{{Hobbs},
  {Archibald}, {Arzoumanian}, {Backer}, {Bailes}, {Bhat}, {Burgay},
  {Burke-Spolaor}, {Champion}, {Cognard} et~al.}}]{haa+10}
\bibinfo{author}{\bibfnamefont{G.}~\bibnamefont{{Hobbs}}},
  \bibinfo{author}{\bibfnamefont{A.}~\bibnamefont{{Archibald}}},
  \bibinfo{author}{\bibfnamefont{Z.}~\bibnamefont{{Arzoumanian}}},
  \bibinfo{author}{\bibfnamefont{D.}~\bibnamefont{{Backer}}},
  \bibinfo{author}{\bibfnamefont{M.}~\bibnamefont{{Bailes}}},
  \bibinfo{author}{\bibfnamefont{N.~D.~R.} \bibnamefont{{Bhat}}},
  \bibinfo{author}{\bibfnamefont{M.}~\bibnamefont{{Burgay}}},
  \bibinfo{author}{\bibfnamefont{S.}~\bibnamefont{{Burke-Spolaor}}},
  \bibinfo{author}{\bibfnamefont{D.}~\bibnamefont{{Champion}}},
  \bibinfo{author}{\bibfnamefont{I.}~\bibnamefont{{Cognard}}},
  \bibnamefont{et~al.}, \bibinfo{journal}{Class. Quant Grav.}
  \textbf{\bibinfo{volume}{27}}, \bibinfo{pages}{084013}
  (\bibinfo{year}{2010}).

\bibitem[{\citenamefont{{Dewdney} et~al.}(2009)\citenamefont{{Dewdney}, {Hall},
  {Schilizzi}, and {Lazio}}}]{dhsl09}
\bibinfo{author}{\bibfnamefont{P.~E.} \bibnamefont{{Dewdney}}},
  \bibinfo{author}{\bibfnamefont{P.~J.} \bibnamefont{{Hall}}},
  \bibinfo{author}{\bibfnamefont{R.~T.} \bibnamefont{{Schilizzi}}},
  \bibnamefont{and} \bibinfo{author}{\bibfnamefont{T.~J.~L.~W.}
  \bibnamefont{{Lazio}}}, \bibinfo{journal}{IEEE Proceedings}
  \textbf{\bibinfo{volume}{97}}, \bibinfo{pages}{1482} (\bibinfo{year}{2009}).

\bibitem[{\citenamefont{{Sesana} and {Vecchio}}(2010)}]{sv10}
\bibinfo{author}{\bibfnamefont{A.}~\bibnamefont{{Sesana}}} \bibnamefont{and}
  \bibinfo{author}{\bibfnamefont{A.}~\bibnamefont{{Vecchio}}},
  \bibinfo{journal}{Class. Quant Grav.} \textbf{\bibinfo{volume}{27}},
  \bibinfo{pages}{084016} (\bibinfo{year}{2010}).

\end{thebibliography}
\end{document}